\documentclass[twocolumn, tighten]{aastex62}
\graphicspath{{./}{figures/}}

\received{, 2021}
\revised{2021}
\accepted{2021}
\submitjournal{ApJ}

\shorttitle{Ce abundances in Open Clusters}
\shortauthors{Sales-Silva et al.}

\begin{document}

\title{Exploring the s-process history in the Galactic disk: Cerium abundances and gradients in Open Clusters from the OCCAM/APOGEE sample}

\correspondingauthor{J. V. Sales-Silva}
\email{joaovictor@on.br, joaovsaless@gmail.com}

\author{J. V. Sales-Silva}
\affil{Observat\'orio Nacional/MCTIC, R. Gen. Jos\'e Cristino, 77,  20921-400, Rio de Janeiro, Brazil}

\author{S. Daflon}
\affiliation{Observat\'orio Nacional/MCTIC, R. Gen. Jos\'e Cristino, 77,  20921-400, Rio de Janeiro, Brazil}

\author{K. Cunha}
\affil{Observat\'orio Nacional/MCTIC, R. Gen. Jos\'e Cristino, 77,  20921-400, Rio de Janeiro, Brazil}
\affiliation{Steward Observatory, University of Arizona Tucson AZ 85719} 

\author{D. Souto}
\affiliation{Departamento de F\'isica, Universidade Federal de Sergipe, Av. Marechal Rondon, S/N, 49000-000 S\~ao Crist\'ov\~ao, SE, Brazil}

\author{V. V. Smith}
\affiliation{Noirlab, Tucson AZ 85719}

\author{C. Chiappini}
\affil{Leibniz-Institut f{\"u}r Astrophysik Potsdam (AIP), An der Sternwarte 16, 14482 Potsdam, Germany}

\author{J. Donor}
\affiliation{Department of Physics \& Astronomy, Texas Christian University, TCU Box 298840, Fort Worth, TX 76129, USA}

\author{P. M. Frinchaboy}
\affil{Department of Physics \& Astronomy, Texas Christian University, TCU Box 298840, Fort Worth, TX 76129, USA}

\author{D. A. García-Hernández}
\affil{Instituto de Astrofísica de Canarias (IAC), E-38205 La Laguna, Tenerife, Spain}
\affil{Universidad de La Laguna (ULL), Departamento de Astrofísica, 38206 La Laguna, Tenerife, Spain}

\author{C. Hayes}
\affil{Department of Astronomy, University of Washington, Box 351580, Seattle, WA 98195, USA}

\author{S. R. Majewski}
\affil{Department of Astronomy, University of Virginia, Charlottesville, VA 22904-4325, USA}

\author{T. Masseron}
\affil{Instituto de Astrofísica de Canarias (IAC), E-38205 La Laguna, Tenerife, Spain}
\affil{Universidad de La Laguna (ULL), Departamento de Astrofísica, 38206 La Laguna, Tenerife, Spain}

\author{R. P. Schiavon}
\affil{Astrophysics Research Institute, Liverpool John Moores University, Liverpool, L3 5RF, UK}

\author{D. H. Weinberg}
\affil{Department of Astronomy \& Center for Cosmology and AstroParticle Physics, The Ohio State University, Columbus, OH 43210, USA}
\affil{Institute for Advanced Study, Princeton, NJ 08540, USA}

\author{R. L. Beaton}
\affil{The Carnegie Observatories, 813 Santa Barbara Street, Pasadena, CA 91101, USA}

\author{J. G. Fernández-Trincado}
\affil{Instituto de Astronom\'ia, Universidad Cat\'olica del Norte, Av. Angamos 0610, Antofagasta, Chile}
\affil{Instituto de Astronom\'ia y Ciencias Planetarias, Universidad de Atacama, Copayapu 485, Copiap\'o, Chile}

\author{H. J{\"o}nsson}
\affil{Materials Science and Applied Mathematics, Malm{\"o} University, SE-205 06 Malm{\"o}, Sweden}

\author{R. R. Lane}
\affil{Centro de Investigación en Astronomía, Universidad Bernardo O’Higgins, Avenida Viel 1497, Santiago, Chile}

\author{D. Minniti}
\affil{Departamento de Ciencias Fisicas, Facultad de Ciencias Exactas, Universidad Andres Bello, Av. Fernandez Concha 700, Las Condes, Santiago, Chile}
\affil{Vatican Observatory, V00120 Vatican City State, Italy}

\author{A. Manchado}
\affil{Instituto de Astrofísica de Canarias (IAC), E-38205 La Laguna, Tenerife, Spain}
\affil{Universidad de La Laguna (ULL), Departamento de Astrofísica, 38206 La Laguna, Tenerife, Spain}
\affil{Consejo Superior de Investigaciones Cient\'{\i}ficas, Spain}

\author{C. Moni Bidin}
\affil{Instituto de Astronom\'ia, Universidad Cat\'olica del Norte, Av. Angamos 0610, Antofagasta, Chile}

\author{C. Nitschelm}
\affil{Centro de Astronomía (CITEVA), Universidad de Antofagasta, Avenida Angamos 601, Antofagasta 1270300, Chile}

\author{J. O’Connell}
\affil{Departamento de Astronomía, Casilla 160-C, Universidad de Concepción, Concepción, Chile}

\author{S. Villanova}
\affil{Departamento de Astronomía, Casilla 160-C, Universidad de Concepción, Concepción, Chile}



\begin{abstract}

The APOGEE Open Cluster Chemical Abundances and Mapping (OCCAM) survey is used to probe the chemical evolution of the s-process element cerium in the Galactic disk. Cerium abundances were derived from measurements of Ce II lines in the APOGEE spectra using the Brussels Automatic Code for Characterizing High Accuracy Spectra (BACCHUS) in 218 stars belonging to 42 open clusters. Our results indicate that, in general, for Ages $<$ 4 Gyr, younger open clusters have higher [Ce/Fe] and [Ce/$\alpha$-element] ratios than older clusters. In addition, metallicity segregates open clusters in the [Ce/X]-Age plane (where X can be H, Fe, and the $\alpha$-elements O, Mg, Si, or Ca). These metallicity-dependant relations result in [Ce/Fe] and [Ce/$\alpha$] ratios with age that are not universal clocks. Radial gradients of [Ce/H] and [Ce/Fe] ratios in open clusters, binned by age, were derived for the first time, with d[Ce/H]dR$_{GC}$ being negative, while d[Ce/Fe]/dR$_{GC}$ is positive. [Ce/H] and [Ce/Fe] gradients are approximately constant over time, with the [Ce/Fe] gradient becoming slightly steeper, changing by $\sim$+0.009 dex-kpc$^{-1}$-Gyr$^{-1}$. Both the [Ce/H] and [Ce/Fe] gradients are shifted to lower values of [Ce/H] and [Ce/Fe] for older open clusters. The chemical pattern of Ce in open clusters across the Galactic disk is discussed within the context of s-process yields from AGB stars, $\sim$Gyr time delays in Ce enrichment of the interstellar medium, and the strong dependence of Ce nucleosynthesis on the metallicity of its AGB stellar sources.

\end{abstract}



\section{Introduction}

The elements heavier than the iron-peak elements are produced mostly via neutron captures onto atomic nuclei; in general, two distinct processes can account for most of the heavy element abundances, with one process involving slow neutron capture rates, the s-process, and one driven by very rapid neutron capture rates, the r-process \citep[][]{Burbidge1957, Kappeler2011}. The r-process elements are mainly synthesized in merging neutron stars \citep[][]{Thielemann2017}.  The production of s-process elements, on the other hand, can be divided into three components, based upon analyses of solar and solar system abundances: a weak s-process component (nickel to strontium, 60$\lesssim$A$\lesssim$90) produced in massive stars \citep[][]{Pignatari2010}; a strong s-process component, terminating in $^{208}$Pb, synthesized in low metallicity AGB stars \citep[][]{Gallino1998}; and finally, a main s-process component (A$>$90, which includes cerium) from low- and intermediate-mass AGB stars \citep[][]{Lugaro2003}. Although the s- and r-processes contribute in parallel to the abundance of a given heavy element, the study of s- or r-process dominated elements enables the isolated analysis of each of these processes. In particular, the cerium (Ce) abundance in the solar system has been produced mainly by the s-process \citep[83.5 $\pm$ 5.9\% contributed by the s-process,][]{Bisterzo2014} making Ce one of the ideal elements to explore the s-process history in stellar populations.

Understanding AGB yields is essential to correctly interpret the s-process chemical evolution of the Galaxy, as these stars are its principal producers. The s-process production in AGB stars depends on the efficiency of the formation of a $^{13}$C-pocket in thermally-pulsing (TP) AGB stars, as the main-component s-process neutrons are produced by $^{13}$C($\alpha$,n)$^{16}$O.\footnote{In the more massive AGB stars (M $>$ 4 solar mass), the free neutrons are produced instead by the $^{22}$Ne($\alpha$,n)$^{25}$Mg neutron source but only elements from the first-peak around Rb are overproduced \citep[e.g.,][and references therein]{Garcia-Hernandez2006, Garcia-Hernandez2013, van-Raai2012}} These $^{13}$C nuclei result from the mixing of H (protons) from the convective envelope of a TP-AGB star into the shell H- and $^{4}$He-burning regions, where the reaction $^{12}$C(p,$\gamma$)$^{13}$N($\beta^{+}$,$\nu$)$^{13}$C occurs.  The mixing of protons from the convective envelope into the shell-burning regions of TP-AGB stars depends upon such quantitites as the metallicity, initial stellar mass, rotation, or magnetic fields \citep[][for a review]{Gallino2006, Piersanti2013, Bisterzo2014, Cristallo2015, Battino2019, Vescovi2020, Karakas2014}. This production site, with many variables, highlights the complexity involved in understanding the chemical evolution of the s-process elements produced by AGB stars. In this context, open clusters are crucial to unraveling this complicated panorama because these objects provide well-determined distances and ages.

In the last decade, large spectroscopic surveys \citep[APOGEE, Gaia-ESO, and GALAH:][]{Majewski2017, Gilmore2012, DaSilva2015} have provided chemical abundances in large samples of open clusters through high-resolution spectroscopy, revealing details about the chemical evolution of the Galaxy. Along this line, \citet{Donor2020} performed a chemical analysis of Fe, $\alpha$, K, Na, Al, iron-peak elements using the 128 open cluster sample from the Open Cluster Chemical Abundances and Mapping (OCCAM) survey from the Sloan Digital Sky Survey (SDSS) IV \citep[]{Blanton2017} Apache Point Observatory Galactic Evolution Experiment 2 \citep[APOGEE 2;][]{Majewski2017} Data Release 16 \citep[DR16; ][]{Nidever2015,Holtzman2015,Jonsson2020}.  Some of their main results included reliable Galactic abundance gradients for sixteen elements and the evolution of [X/Fe] gradients as a function of age for some elements, although the analysis of Ce was not included in \citet{Donor2020} due to the imprecision of Ce abundances in the APOGEE DR16 database. This larger uncertainty in the Ce abundances occurred due to the use of only one Ce II absorption line in DR16.  In this study we extend the list of elements analyzed in the OCCAM sample to include cerium, using seven 
Ce II absorption lines in the APOGEE spectra, as discovered and studied by \citet{Cunha2017}.

The abundance of s-process elements in the Galactic disk has been the subject of intense study in recent years (e.g., using open clusters: \citet{DOrazi2009, Maiorca2011, Yong2012, Magrini2018, Spina2021}; as well as field stars: \citet{BattistiniBensby2016, Spina2018, Tautvaisiene2021} among others); 
the relationship between abundance and age may not be the same for all s-process elements \citep[][]{Yong2012, JacobsonFriel2013}.
Some studies found an increase of the [X/Fe] abundance ratio of s-process elements, mainly Ba, with decreasing age of the open clusters and field disk stars \citep[][]{DOrazi2009, Maiorca2012, Spina2018, Spina2021}.
However, \citet{Spina2020} proposed that the overabundance of Ba in young stars could be related to activity and magnectic enhancements. \citet{Baratella2021} explored different possible scenarios to explain Ba overabundances that they found in young open clusters (up to +0.6 dex in open clusters with ages $<$ 200 Myr), including chromospheric activity, but found these scenarios were not sufficient to explain the Ba overabundances. It is noted that their [La/Fe] abundances were approximately solar, leading them to suggest Ce (or La) to be a better tracer of the s-process and their temporal evolution (especially at younger ages).

For cerium, \citet{Maiorca2011}, \citet{Spina2018} and \citet{Casamiquela2021} found a clear growth of the [Ce/Fe] ratio with decreasing age for open clusters and field stars. 
\citet{Magrini2018} and \citet{Delgado-Mena2019} derived a lower correlation of the [Ce/Fe] ratio with age for open clusters and field stars respectively, while \citet{Tautvaisiene2021} determined a flat trend for thin disk stars. 
In addition, there was a metallicity dependence in the relationship between [Ce/Fe] ratio and age for field dwarf stars and open clusters (\citet{Delgado-Mena2019} and \citet{Casamiquela2021}. 
Here, we investigate the s-process history using the Ce abundances of the large and homogeneous sample of OCCAM to further probe trends in the Ce ratios with age. In addition, we also analyze the dependence of this relationship with metallicity, as this may be an important observational constraint for models of chemical evolution.

Meanwhile, large uncertainties in the ages of field stars raise interest in finding chemical ratios that can serve as universal clocks for these stars. The abundance ratios between the s-process and $\alpha$-elements are one target in this search, as they are produced by stars having very different lifetimes (Gyrs for stars producing s-process elements and Myrs for those producing $\alpha$-elements). The [Y/Mg] ratio appeared as one major candidate for such a universal chemical clock in stars \citep[][]{DaSilva2012, Nissen2015, Feltzing2017, Jofre2020}; however, recent studies indicate that the relationship between the [Y/Mg] ratio and age is not universal and varies throughout the Galactic disk \citep[][]{Delgado-Mena2019, Casali2020, Magrini2021, Casamiquela2021}. 
From a sample of 80 solar twins in the solar neighborhood, \citet{Jofre2020} found trends for Ce/Mg and Si with age, indicating that these were good chemical clocks. In this study, we examine the relationship between age and the abundance ratio of Ce with various $\alpha$-elements ( [Ce/$\alpha$] where $\alpha$ can be O, Mg, Si, or Ca) whose abundances have been determined in APOGEE/DR16.  This analysis allows us to further probe whether the correlation between age and [Ce/$\alpha$] is universal for open clusters.

In this paper, we also determine the [Ce/H] and [Ce/Fe] radial gradients (Section 4).  The large open cluster sample allows us to show the gradient binned by age, enabling us to explore the evolution of the Ce gradient over time. In Section 2, we present details of the sample, as well as the derivations of the Ce abundances. We also compare our results with those obtained in DR16 and the literature (Section 3). Concluding remarks about our results are found in the last Section.

\section{Sample and methodology}

APOGEE is a high-resolution, near-infrared \citep[1.514 to 1.696 microns,][]{Wilson2019} spectroscopic survey that in the latest public release (DR16) provided a detailed spectral analysis of approximately 430,000 stars \citep[][]{Zasowski2017, Jonsson2020}. The APOGEE observations occur on the 2.5m telescopes at Apache Point Observatory \citep[New Mexico, USA,][]{Gunn2006} and at Las Campanas Observatory \citep[La Serena, Chile,][]{Bowen1973}. The Open Cluster Chemical Abundances and Mapping (OCCAM) survey \citep[][]{Donor2020} used the atmospheric parameters and chemical abundances from DR16, which were obtained with the APOGEE Stellar Parameters and Abundances Pipeline \citep[ASPCAP,][]{GarciaPerez2016} and a customized line list \citep[][]{Smith2021}. Stars in the OCCAM sample were classified as open cluster members based on radial velocities, proper motions, spatial location, and derived metallicities \citep[][]{Donor2020}.

The OCCAM sample consists of 914 stars belonging to 128 open clusters. \citet{Donor2020} classified 71 of the APOGEE open clusters as high-quality clusters based on their color-magnitude diagram (CMD) and the reliability of the ASPCAP abundance results. \cite{Donor2020} investigated metallicity gradients for  Na, Al, K, $\alpha$, and iron-peak elements. In this study, we add to that list of elements and investigate the s-process dominated element Ce in the open clusters of the OCCAM sample.

\citet{Cunha2017} found eight measurable Ce II absorption lines in the APOGEE wavelength region. However, in DR16 the Ce abundances were estimated using only one Ce II absorption line (15784.8\AA) due to difficulties in fitting the other Ce II lines automatically with ASPCAP. In this study, we consider all Ce II lines from \citet{Cunha2017} to improve the reliability of the Ce abundances in the APOGEE open cluster stars. The $\log{gf}$ values of the Ce II lines used in our analysis are from \citet{Cunha2017} for the lines at 16376.5\AA\ and 16722.6\AA, and from \citet{Smith2021} for the lines at 15277.6\AA, 15784.8\AA, 15977.1\AA, 16327.3\AA, and 16595.2\AA.

The methodology adopted to derive Ce abundances relies on $\chi^2$-squared fits between observed and synthetic spectra made from the Brussels Automatic Code for Characterizing High Accuracy Spectra \citep[BACCHUS,][]{Masseron2016}. BACCHUS uses MARCS model atmospheres \citep[][]{Gustafsson2008} and the radiative transfer code Turbospectrum \citep[][]{AlvarezPlez1998, Plez2012}, which is exactly the machinery used in ASPCAP for DR16, and so makes the analysis self-consistent. Because C, N, and O abundances influence Ce II absorption lines in the APOGEE spectral region, we adopted the DR16 (uncalibrated) C, N, O abundances along with the atmospheric parameters ($T_{\rm eff}$, $\log{g}$, $\xi$, and [Fe/H]) to compute the syntheses in the Ce II line regions. The ASPCAP pipeline determines the atmospheric parameters and abundances automatically through best-fits between the synthetic and observed spectra for the entire APOGEE region \citep[][]{Jonsson2020}.

\begin{figure*}
	\includegraphics[width=15cm]{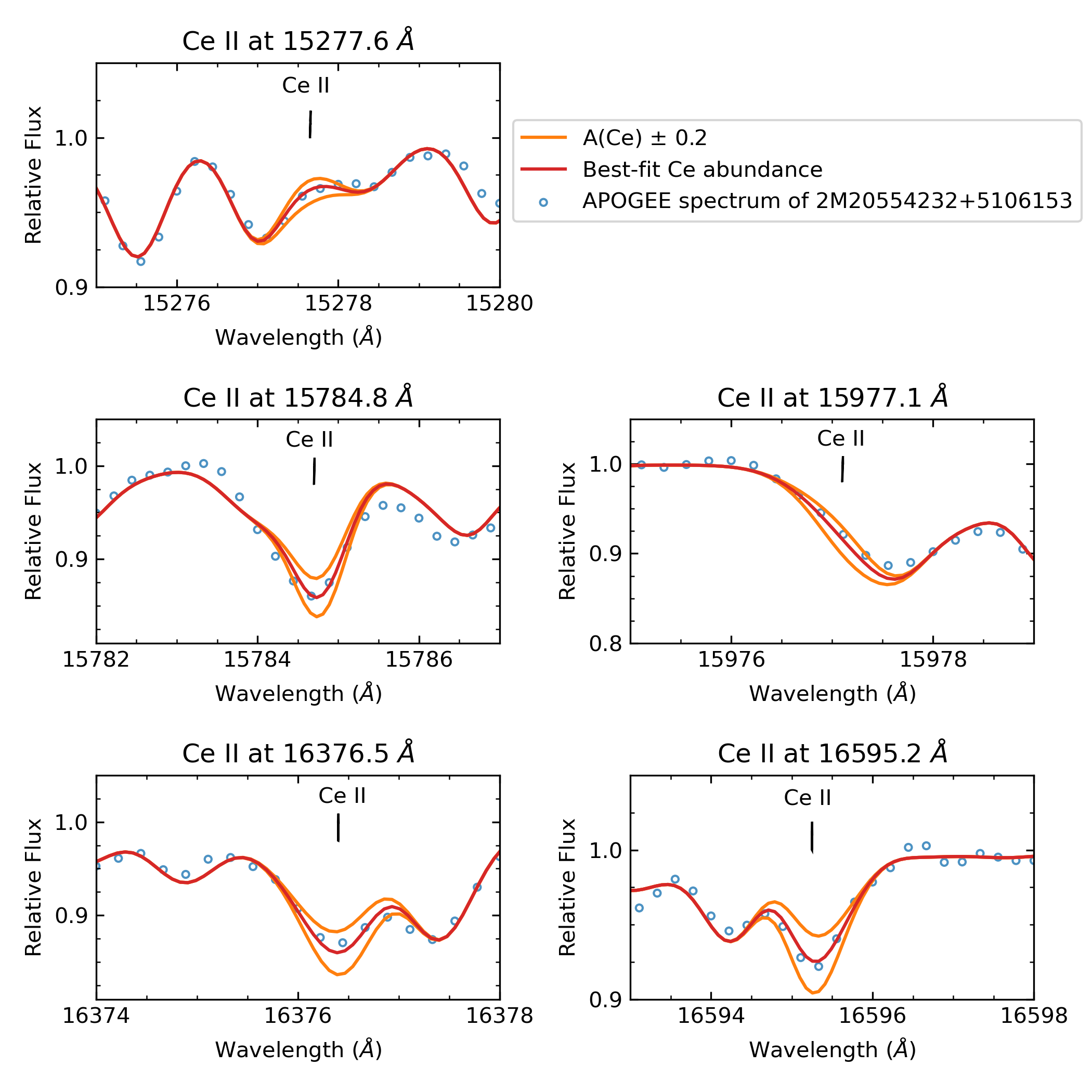}
    \caption{Observed (open blue circles) and synthetic spectra (solid lines) in the region of the five Ce II lines used to determine the Ce abundance of the Berkeley 53 red giant 2M20554232+5106153. Each panel shows one Ce II line and three synthetic spectra, with one synthesis representing the best fit Ce abundance (red lines) and the others with A(Ce) $\pm$ 0.2 dex (orange lines).}
    \label{sinteseCe}
\end{figure*}

\begin{figure}
	\includegraphics[width=\columnwidth]{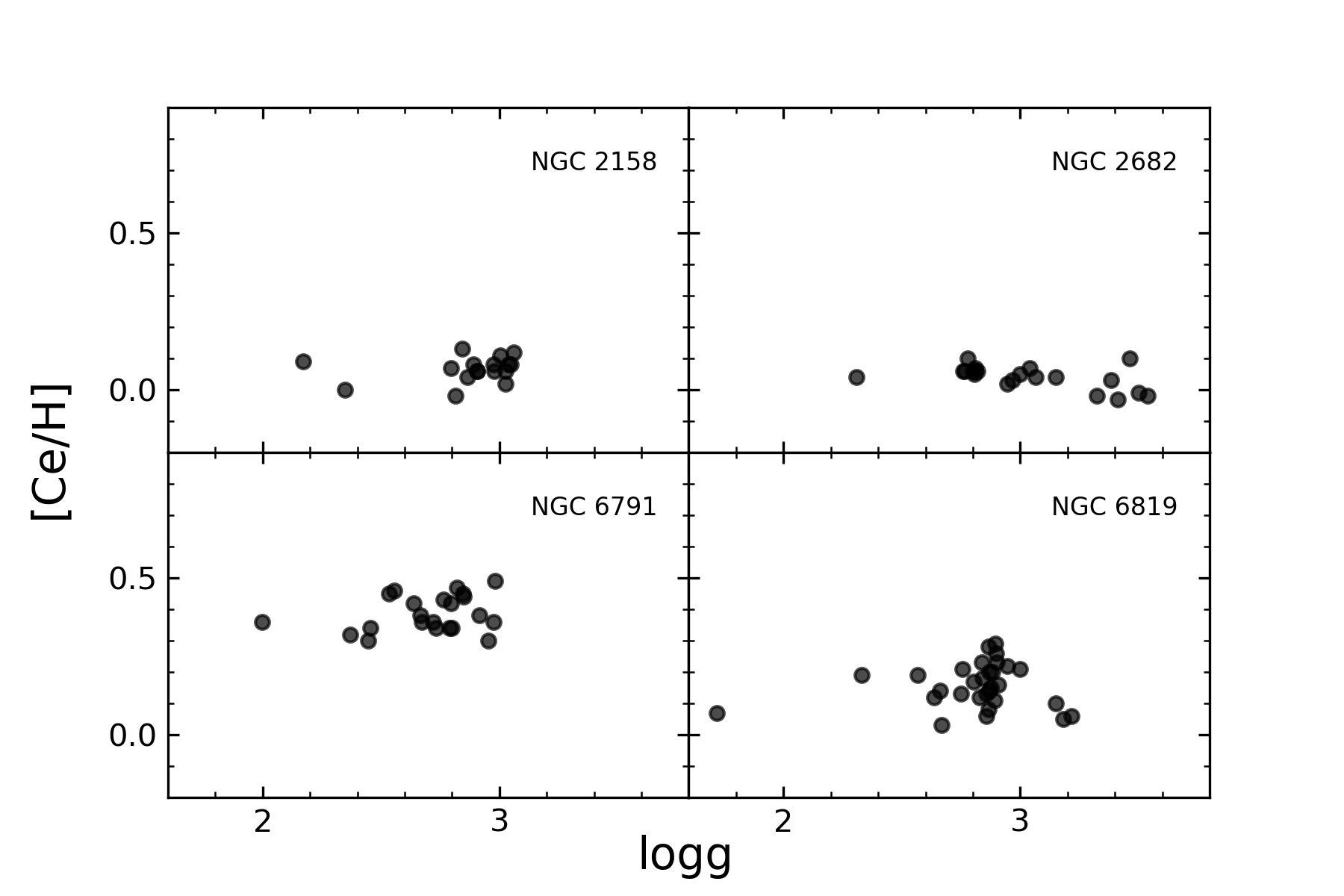}
    \caption{[Ce/H] versus surface gravity ($\log{g}$) for the stars in the open clusters NGC 2158, NGC 2682, NGC 6791, and NGC 6819, which are the ones having the largest numbers of stars analyzed. The OCCAM targets were  selected to have $\log{g}$ less than 3.70; dwarf stars were not considered.}
    \label{Ce_H_vs_logg}
\end{figure}

\begin{figure}
	\includegraphics[width=\columnwidth]{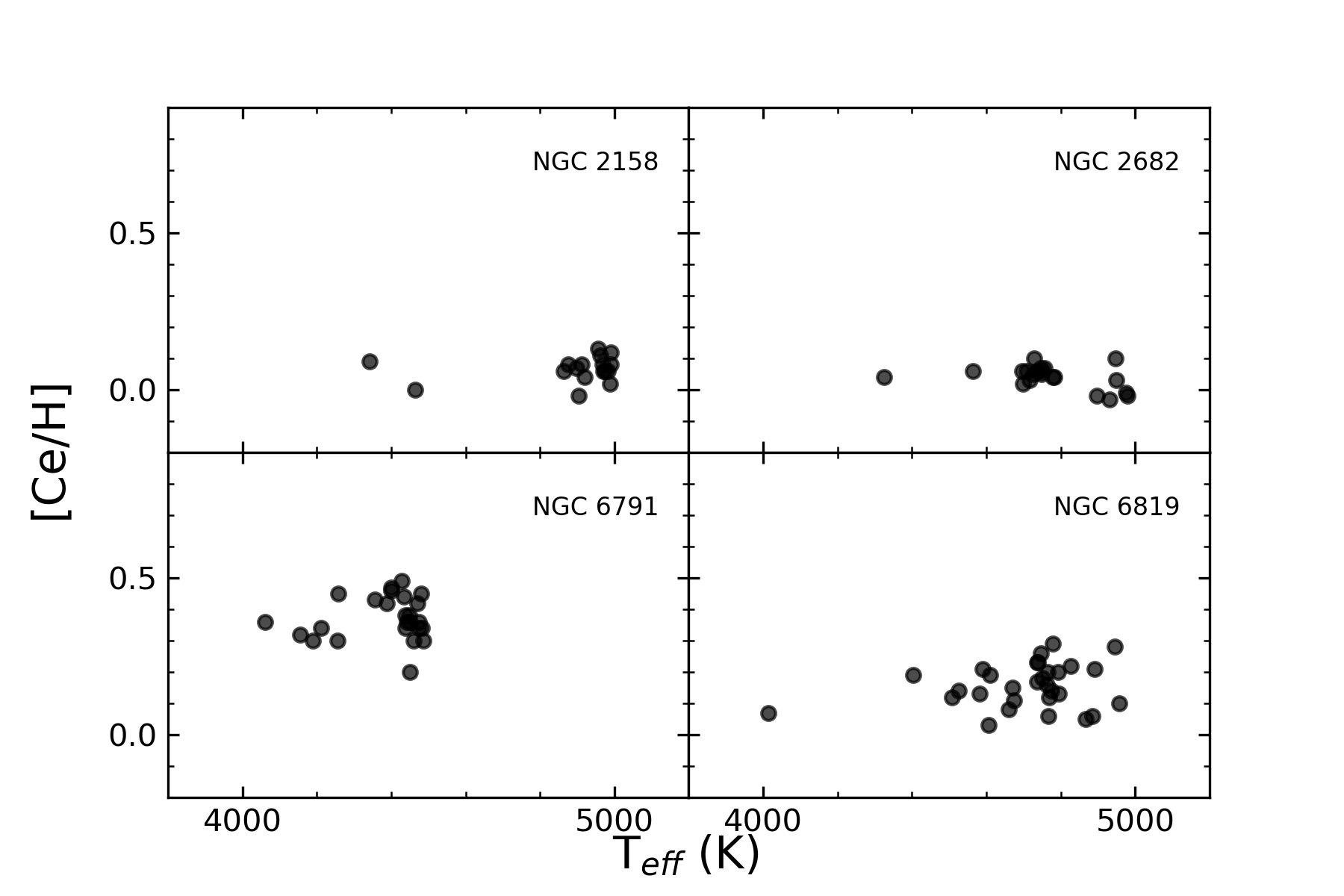}
    \caption{[Ce/H] versus $T_{\rm eff}$ for the studied stars members of the open clusters NGC 2158, NGC 2682, NGC 6791, and NGC 6819. Ce abundances were measured in the effective temperature interval between roughly 4000 and 5000 K. In general, there are no significant trends in A(Ce) with$T_{\rm eff}$.}
    \label{Ce_H_vs_teff}
\end{figure}

The Ce II lines in the APOGEE spectra can be weak and quite blended depending on the atmospheric parameters and chemical abundances of the studied stars \citep[][]{Cunha2017}. We verified that giants hotter than 5000 K show Ce II lines with small equivalent widths, as also indicated by \citet{Cunha2017}, while giants cooler than 4000 K present Ce II lines strongly blended with molecular bands. Thus, we selected only giant stars ($\log{g}$ $<$ 3.70 dex) with 4000$<$T$_{\rm eff}<$5000 K from the OCCAM sample. 
After a careful inspection of the seven Ce II lines (Table \ref{tab:star_abundances_table}) in each of the spectra of all targets, the final sample analyzed was reduced to 218 stars belonging to 42 open clusters; those were the stars from the high-quality OCCAM sample for which we could derive good Ce abundances with BACCHUS. We show in Figure \ref{sinteseCe} an example of the spectral syntheses and best-fit abundances for the five Ce II lines used to determine the Ce abundance of the star 2M20554232+5106153 from Berkeley 53. The individual Ce II lines and derived abundances for the sample stars are in Table \ref{tab:star_abundances_table}. For completion, we also present the atmospheric parameters used in the computations of the spectral syntheses, which are the uncalibrated values from DR16. All cluster stars in our sample are red-giants to minimize systematic errors and any possible effects of atomic diffusion  \citep[][]{Souto2018, Souto2019} in the abundances. To verify trends due to possible non-LTE effects or other systematic errors in the analysis, we show in Figures \ref{Ce_H_vs_logg} and \ref{Ce_H_vs_teff} [Ce/Fe] as functions of $\log{g}$ and $T_{\rm eff}$, respectively, for the four open clusters with the largest numbers of stars analyzed (NGC 2158, NGC 2682, NGC 6791, and NGC 6819). 
The general behavior of the Ce abundance with $T_{\rm eff}$ or $\log{g}$ is reasonably constant, with the modulus of the slope $\leq$ 0.06 dex dex$^{-1}$ in the [Ce/H]-$\log{g}$ plane and $\approx$ 0.00 dex K$^{-1}$ in the [Ce/H]-T$_{\rm eff}$ plane.

 The parameters of the open clusters adopted in this study are in Table \ref{tab:clusters_table}.  We present the average Ce abundances and the respective [Ce/H] and [Ce/Fe] ratios obtained for each studied cluster; we adopted the solar abundance of \citet{Grevesse2007}, which is the same abundance scale used in DR16 chemical abundances. The ages and galactocentric distances for the open clusters were taken from \citet{Cantat-Gaudin2020}, except for Berkeley 43 and FSR 0394, for which there were no estimates in that study. We use the values 
 from  \citet{Kharchenko2013}, instead, given that \citet{Cantat-Gaudin2020} find generally good agreement with the age and distance estimates from \citet{Kharchenko2013}.

We selected two sample stars (2M05240941+2937217 and 2M20554232+5106153) with different atmospheric parameters as references to estimate the uncertainties in Ce abundances derived in this study.
The errors in the Ce abundances were derived by varying each atmospheric parameter independently by their typical estimated uncertainties: $T_{\rm eff}$ by +90 K, $\log{g}$ by +0.2 dex, $\xi$ by +0.25\ ,km\,s$^{-1}$ and [Fe/H] by +0.1. In addition, we estimated the abundance uncertainties due to the synthetic fits of the Ce lines. We estimate the final errors by adding quadratically the uncertainties relative to each atmospheric parameter and synthetic fit. We show the uncertainties regarding the two stars in Table \ref{tab:error}.

\section{Comparison with DR16 and optical results from the literature}

\begin{figure}
	\includegraphics[width=\columnwidth]{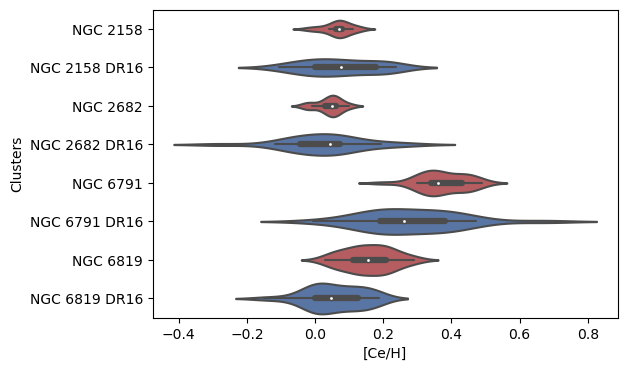}
    \caption{[Ce/H] distribution for the open clusters NGC 2158, NGC 2682, NGC 6791, and NGC 6819. The red sequences represents our [Ce/H] results while the blue sequences shows the DR16 values for the four clusters. White dots in the distribution indicate the median while the thick bar represents the interquartile range, and the thin bar shows the 95\% confidence interval. Wider regions of the distribution represent a higher probability that a star will have that [Ce/H] value.}
    \label{distribuicao_violino_Ce_H}
\end{figure}

We compare the Ce abundance results obtained in this study with those estimated in DR16. The reader is reminded that the Ce abundances from DR16 were derived with the ASPCAP pipeline and were based on a single Ce II line at 15784.8 \AA{}. 
The average difference between our results and those from DR16 has a small offset but has a significant standard deviation, $\Delta$ ([Ce/H]$_{this study}$ $-$ [Ce/H]$_{DR16}$) = 0.05$\pm$0.16, possibly indicating that the DR16 results have larger internal errors \citet{Jonsson2020}. In Figure \ref{distribuicao_violino_Ce_H}, we present violin plot distributions for the [Ce/H] abundances of the open clusters NGC 2158, NGC 2682, NGC 6791 and NGC 6819 obtained in this study (shown in red) and in DR16 (shown in blue). It is clear that, although the median abundance value for each cluster is not significantly offset, the Ce abundances derived here show much less internal dispersion than in DR16. Such small dispersions are expected under the paradigm that stars in open clusters do not show variations in chemical content.

\begin{figure}
	\includegraphics[width=\columnwidth]{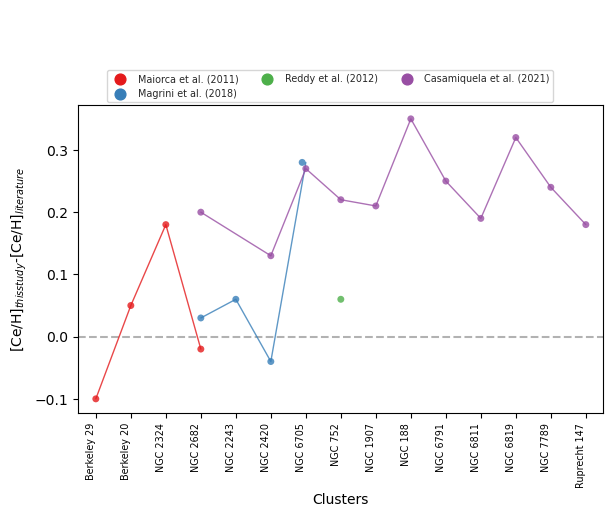}
    \caption{Comparison of the mean [Ce/H] abundances for the clusters obtained here from the near-infrared APOGEE spectra with  results from high-resolution optical spectroscopy from \citet{Maiorca2011} (red circles), \citet{Reddy2012} (green circle), \citet{Magrini2018} (blue circles) and \citet{Casamiquela2021} (purple circles) for the open clusters in common.}
    \label{diferenca_Ce_H}
\end{figure}

Most previous studies in the literature derived Ce abundances for relatively small (less than 10 clusters) open cluster samples using high-resolution optical spectra \citep[e. g.,][]{Reddy2012, Reddy2013, Santrich2013, Mishenina2015, PenaSuarez2018}. 
Larger samples have been investigated in \citet{Maiorca2011}, \citet{Magrini2018} and \citet{Casamiquela2021}. 
A comparison of the Ce abundance results for the open clusters in this study with the literature is presented in Figure \ref{diferenca_Ce_H}, where we show the differences of the mean [Ce/H] values obtained (This Study - Other studies) for the open clusters labeled in the x-axis of the figure.  The average Ce abundance difference with the studies of \citet{Maiorca2011}, \citet{Magrini2018} and \citet{Casamiquela2021} are, respectively, $\Delta$[Ce/H]= 0.03$\pm$0.10, 0.08$\pm$0.12, and  0.23$\pm$0.06. These indicate that there are small, or not significant, abundance offsets with the results from the first two studies when compared to ours, while with the recent study of \citet{Casamiquela2021} there is a more significant offset. 
For the well studied solar metallicity open cluster NGC 2682 (or M67), for example, the Ce abundance obtained by \citet{Maiorca2011} ([Ce/H]=0.06$\pm$0.05) and \citet{Magrini2018} ([Ce/H]=0.01$\pm$0.03) are roughly solar or ever so slightly enhanced, and, quite similar to our [Ce/H] value of 0.04$\pm$0.04, while the result in Casamiquela et al. (2021) is cerium-poor, [Ce/H]=$-$0.16 $\pm$ 0.03.
On the other hand, there are significant differences between our Ce abundances and those obtained for NGC 2324 in \citet{Maiorca2011} and NGC 6705 in \citet{Magrini2018}, with our [Ce/H] values being higher by 0.17 and 0.28 dex, respectively. For NGC 2324 there are no other Ce abundance determinations available in the literature for further comparisons, although it is worth noting that \citet{DOrazi2009} found a high Ba abundance (another heavy s-process dominated element) in the open cluster NGC 2324 ([Ba/H]=0.49), which would be generally in line with our Ce enrichment result for this cluster.

\begin{table*}
	\centering
	\caption{Clusters parameters and abundances ratios used in this study. The A(Ce) column shows the mean Ce abundance obtained in our study using BACCHUS. The galactocentric distance and age of the open clusters were obtained from \citet{Cantat-Gaudin2020}. R$_{GC}$ is in kpc. We used DR16 values to determine the average metallicity for each cluster.}
	\label{tab:clusters_table}
	\begin{tabular}{lccccccccc} 
		\hline
		Cluster & log Age & R$_{GC}$ & n$_{star}$ &  [Fe/H] & $\sigma_{Fe}$ & A(Ce) & $\sigma_{Ce}$ & [Ce/H] & [Ce/Fe] \\
		\hline
        Basel 11b &    8.36 &  10.121 &    1 & $-$0.004 &      --- &  1.940 &    --- &  0.240 &  0.244 \\
        Berkeley 17 &    9.86 &  11.668 &    7 & $-$0.164 &    0.026 &  1.633 &  0.051 & $-$0.067 &  0.097 \\
        Berkeley 19 &    9.34 &  14.890 &    1 & $-$0.323 &      --- &  1.640 &    --- & $-$0.060 &  0.263 \\
        Berkeley 20 &    9.68 &  16.320 &    1 & $-$0.398 &      --- &  1.440 &    --- & $-$0.260 &  0.138 \\
        Berkeley 29 &    9.49 &  20.577 &    1 & $-$0.450 &      --- &  1.460 &    --- & $-$0.240 &  0.210 \\
        Berkeley 43 &    8.79 &   7.120 &    1 &  0.026 &      --- &  2.020 &    --- &  0.320 &  0.294 \\
        Berkeley 53 &    8.99 &   9.026 &    6 & $-$0.084 &    0.023 &  1.875 &  0.041 &  0.175 &  0.259 \\
        Berkeley 66 &    9.49 &  12.349 &    2 & $-$0.159 &    0.031 &  1.770 &  0.014 &  0.070 &  0.229 \\
        Berkeley 98 &    9.39 &   9.788 &    1 &  0.004 &      --- &  1.780 &    --- &  0.080 &  0.076 \\
        BH 211 &    8.63 &   6.520 &    1 &  0.187 &      --- &  2.070 &    --- &  0.370 &  0.183 \\
        Collinder 220 &    8.37 &   8.080 &    1 & $-$0.077 &      --- &  2.000 &    --- &  0.300 &  0.377 \\
        Czernik 21 &    9.41 &  12.349 &    2 & $-$0.322 &    0.008 &  1.630 &  --- & $-$0.070 &  0.252 \\
        Czernik 30 &    9.46 &  13.779 &    2 & $-$0.396 &    0.008 &  1.505 &  0.021 & $-$0.195 &  0.201 \\
        FSR 0394 &    9.20 &  10.500 &    2 & $-$0.096 &    0.003 &  1.860 &  --- &  0.160 &  0.256 \\
        IC 1369 &    8.46 &   8.948 &    3 & $-$0.079 &    0.037 &  1.917 &  0.032 &  0.217 &  0.296 \\
        IC 166 &    9.12 &  12.418 &    1 & $-$0.086 &      --- &  1.860 &    --- &  0.160 &  0.246 \\
        King 2 &    9.61 &  13.264 &    1 & $-$0.359 &      --- &  1.530 &    --- & $-$0.170 &  0.189 \\
        King 5 &    9.01 &  10.526 &    1 & $-$0.156 &      --- &  1.820 &    --- &  0.120 &  0.276 \\
        King 7 &    8.35 &  11.194 &    4 & $-$0.160 &    0.024 &  1.978 &  0.052 &  0.278 &  0.438 \\
        NGC 1193 &    9.71 &  12.705 &    2 & $-$0.334 &    0.004 &  1.510 &  0.071 & $-$0.190 &  0.144 \\
        NGC 1245 &    9.08 &  11.118 &    1 & $-$0.139 &      --- &  1.810 &    --- &  0.110 &  0.249 \\
        NGC 1798 &    9.22 &  13.266 &    6 & $-$0.262 &    0.013 &  1.710 &  0.046 &  0.010 &  0.272 \\
        NGC 188 &    9.85 &   9.285 &   10 &  0.100 &    0.015 &  1.814 &  0.085 &  0.114 &  0.014 \\
        NGC 1907 &    8.77 &   9.947 &    1 & $-$0.078 &      --- &  1.930 &    --- &  0.230 &  0.308 \\
        NGC 2158 &    9.19 &  12.617 &   17 & $-$0.211 &    0.023 &  1.766 &  0.040 &  0.066 &  0.277 \\
        NGC 2204 &    9.32 &  11.344 &    6 & $-$0.267 &    0.017 &  1.707 &  0.069 &  0.007 &  0.274 \\
        NGC 2243 &    9.64 &  10.584 &    8 & $-$0.462 &    0.033 &  1.416 &  0.078 & $-$0.284 &  0.178 \\
        NGC 2304 &    8.96 &  12.019 &    1 & $-$0.142 &      --- &  1.850 &    --- &  0.150 &  0.292 \\
        NGC 2324 &    8.73 &  12.075 &    2 & $-$0.181 &    0.027 &  1.845 &  0.021 &  0.145 &  0.326 \\
        NGC 2420 &    9.24 &  10.683 &   10 & $-$0.190 &    0.033 &  1.698 &  0.051 & $-$0.002 &  0.188 \\
        NGC 2682 &    9.63 &   8.964 &   21 &  0.021 &    0.018 &  1.741 &  0.036 &  0.041 &  0.020 \\
        NGC 4337 &    9.16 &   7.454 &    6 &  0.240 &    0.039 &  2.010 &  0.054 &  0.310 &  0.070 \\
        NGC 6705 &    8.49 &   6.464 &   10 &  0.121 &    0.039 &  2.028 &  0.032 &  0.328 &  0.207 \\
        NGC 6791 &    9.80 &   7.942 &   25 &  0.355 &    0.034 &  2.072 &  0.069 &  0.372 &  0.017 \\
        NGC 6811 &    9.03 &   8.203 &    1 & $-$0.020 &      --- &  1.860 &    --- &  0.160 &  0.180 \\
        NGC 6819 &    9.35 &   8.027 &   30 &  0.055 &    0.030 &  1.857 &  0.070 &  0.157 &  0.102 \\
        NGC 752 &    9.07 &   8.669 &    1 & $-$0.041 &      --- &  1.850 &    --- &  0.150 &  0.191 \\
        NGC 7789 &    9.19 &   9.432 &   14 & $-$0.008 &    0.024 &  1.879 &  0.046 &  0.179 &  0.187 \\
        Ruprecht 147 &    9.48 &   8.046 &    1 &  0.138 &      --- &  1.840 &    --- &  0.140 &  0.002 \\
        SAI 116 &    8.10 &   7.528 &    2 &  0.161 &    0.011 &  2.040 &  0.071 &  0.340 &  0.179 \\
        Teutsch 84 &    9.02 &   6.018 &    1 &  0.214 &      --- &  2.000 &    --- &  0.300 &  0.086 \\
        Trumpler 5 &    9.63 &  11.211 &    3 & $-$0.439 &    0.006 &  1.380 &  0.026 & $-$0.320 &  0.119 \\
		\hline\hline
	\end{tabular}
\end{table*}

\begin{table*}
\caption{Ce abundance uncertainties for 2M05240941+2937217 and 2M20554232+5106153.} 
\centering
\label{tab:error}
\begin{tabular}{lcccccc}\\\hline\hline
Star & $\Delta T_{\rm eff}$ & $\Delta\log g$ & $\Delta\xi$ & $\Delta [Fe/H]$ & $\Delta$ A(Ce)$_{synth}$ & $\left( \sum \sigma^2 \right)^{1/2}$ \\
$_{\rule{0pt}{8pt}}$ & $+$90~K & $+$0.2 & $+$0.25 km\,s$^{-1}$ & $+$0.1 &   &  \\
\hline     
2M05240941+2937217    & $+$0.06  & $+$0.09 & $-$0.05 & $+$0.06 & $+$0.06 & 0.15 \\ 
2M20554232+5106153    & $+$0.05  & $+$0.09 & $-$0.03 & $+$0.02 & $+$0.05 & 0.12 \\
\hline
\end{tabular}

\par Notes. Each column gives the variation of the abundance caused by the change in $T_{\rm eff}$, $\log g$, $\xi$, \\and [Fe/H]. $\Delta$ A(Ce)$_{synth}$ indicate the abundance uncertainties due to the synthetic fits\\ of the Ce lines. The last column gives the compounded uncertainty.

\end{table*}

\section{Ce abundance trends}

\subsection{The [Ce/Fe]-[Fe/H] plane}

\begin{figure*}
\centering
	\includegraphics[width=15cm]{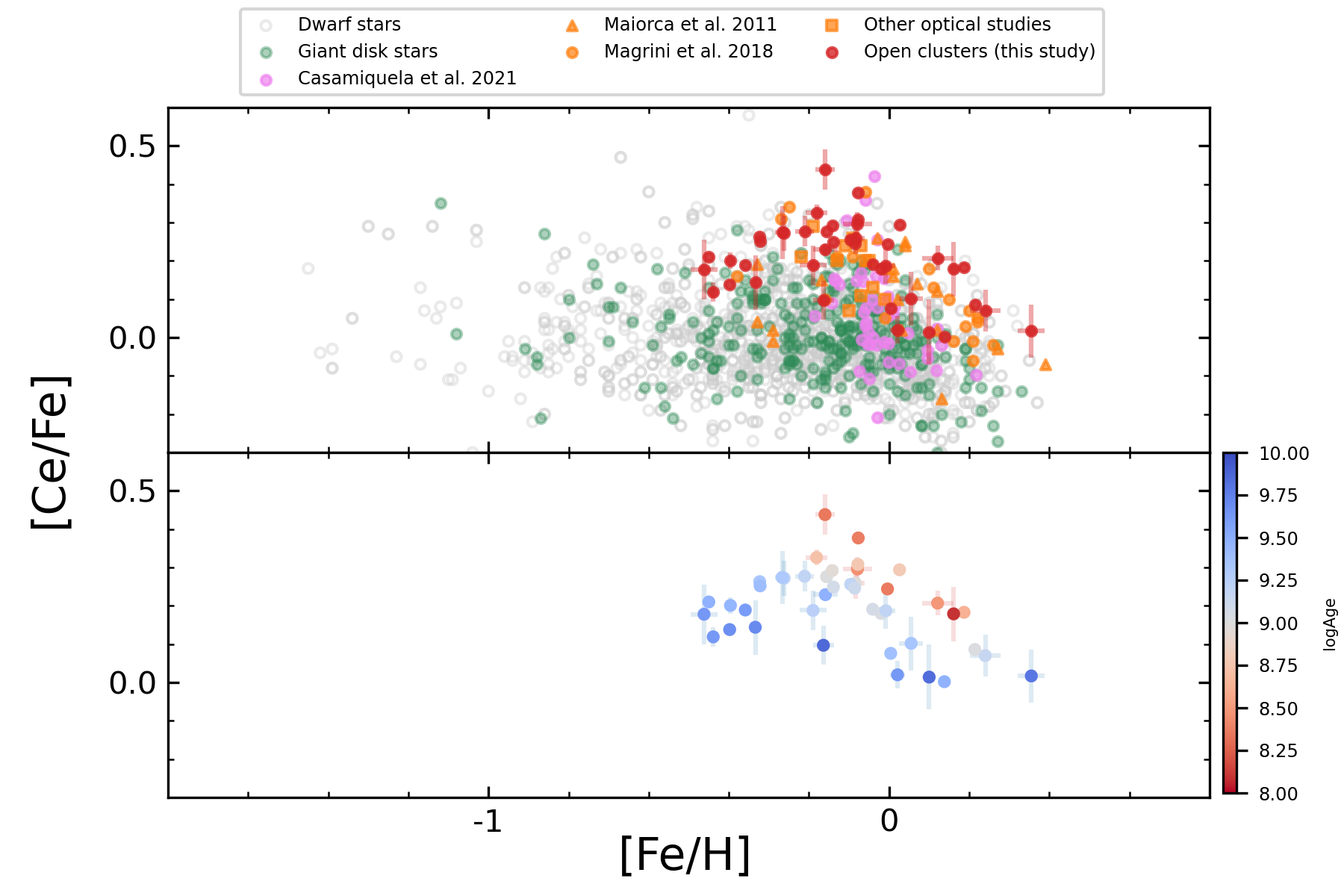}
    \caption{[Ce/Fe]-[Fe/H] plane for various stars in the Galactic disk. The top panel shows results for open clusters and field red-giants and dwarfs. The filled red circles represent the open clusters from the OCCAM/APOGEE sample, while the orange symbols represent results for open cluster from optical studies in the literature: triangles from \citet{Maiorca2011}, circles from \citet{Magrini2018}, squares from \citet{Reddy2012, Reddy2013, Santrich2013, Mishenina2015, PenaSuarez2018}. The violet circles represent open clusters results from \citet{Casamiquela2021} which shows an offset compared to our Ce abundances. The green symbols represent literature results for the disk giant stars from \citep[]{Forsberg2019} while gray symbols refer to dwarf disk stars from \citet[]{Reddy2003, Reddy2006, Mishenina2013, Bensby2014, BattistiniBensby2016, Fishlock2017}. The bottom panel shows the [Ce/Fe] results for open clusters in this study again, but with colors representing the cluster ages. Young open clusters (log Age $<$ 9.0; or Age $<$ 1 Gyr; red circles) present a higher [Ce/Fe] ratio than old open clusters in the same metallicity range.
    }
    \label{Ce_Fe_vs_Fe_H}
\end{figure*}

In Figure \ref{Ce_Fe_vs_Fe_H}, we show the results for [Ce/Fe] ratio as a function of [Fe/H] for the open cluster sample studied here. In the top panel of Figure \ref{Ce_Fe_vs_Fe_H}, we present our results as filled red circles while the results from the literature are shown as orange symbols (triangles, \citealt{Maiorca2011}; circles, \citealt{Magrini2018}; squares, other optical open cluster studies: \citealt{Reddy2012, Reddy2013, Santrich2013, Mishenina2015, PenaSuarez2018}). As violet circles, we show open cluster results from \citet{Casamiquela2021}, which have a significant offset when compared to our Ce abundances (see Figure \ref{diferenca_Ce_H}). We also show, for comparison, results for dwarf stars  \citep[gray circles,][]{Reddy2003, Reddy2006, Mishenina2013, Bensby2014, BattistiniBensby2016, Fishlock2017}, and red-giant stars from the Galactic disk \citep[green circles,][]{Forsberg2019}. 
The main feature is that the Ce abundance results for open clusters in all studies generally overlap in the [Ce/Fe]-[Fe/H] plane; [Ce/Fe] increases as the metallicity decreases, with a possible downturn in the trend at roughly $-$0.2 in [Fe/H]. 

The chemical pattern for the open clusters is found to be generally overabundant in the [Ce/Fe] ratio when compared to most disk stars (giants and dwarfs) in the same metallicity range. Ba, another s-process dominated element, also shows an overabundance in open clusters when compared to field disk stars \citep[][]{Yong2012}. The field stars being systematically older than the open clusters may contribute to this difference, as pointed out by \citet{Yong2012}.

In the lower panel of Figure \ref{Ce_Fe_vs_Fe_H}, we show the studied open cluster sample but now with color representing the log Age (the age the color bar shown on the right side of the plot). It is apparent that younger open clusters show [Ce/Fe] ratios greater than older open clusters in the same metallicity range. It is the older clusters in our sample that show a change of slope in [Ce/Fe] at roughly [Fe/H]$\approx-$0.2; we note, however, that the open clusters with the lowest metallicities in our sample are all older. 

The chemical evolution model of \citet{Prantzos2018}, which considers the yields from low- and intermediate-mass stars, rotating massive stars, and an r-process component, finds a [Ce/Fe] ratio $\sim$0.03 at solar metallicity and a maximum [Ce/Fe] value of 0.20 around [Fe/H]=$-$0.3, followed by a drop of [Ce/Fe] ratio for lower metallicities range \citep[see Figure 16 in][]{Prantzos2018}; this evolution reproduces well the relation between the [Ce/Fe] ratio and metallicity shown by the old open clusters in our sample (dark blue circles in the lower panel of Figure \ref{Ce_Fe_vs_Fe_H}), while young open clusters and the bulk of field giant stars from \citet{Forsberg2019} present, respectively, higher and lower [Ce/Fe] values than that in the \citet{Prantzos2018} model.

In the metallicity range spanned by the Galactic open cluster population, AGB stellar model calculations indicate an increase in the production of heavy s-process elements (like Ce) with a decrease in [Fe/H] \citep[][]{Gallino2006, Cristallo2015, KarakasLugaro2016, Battino2019}. The dependence of the s-process on [Fe/H] is due to the $^{56}$Fe acting as the seed nucleus for the synthesis of the s-process elements coupled with the reaction $^{13}$C($\alpha$,n)$^{16}$O being a primary source of neutrons. At low metallicities, the ratio of neutrons to Fe-seed increases as [Fe/H] decreases, resulting in larger neutron exposures with decreasing metallicity \citep[][]{Cristallo2009, Cristallo2011, Cristallo2015,Karakas2014,KarakasLugaro2016}.

\subsection{The chemical evolution of Ce}

\begin{figure}
\centering
	\includegraphics[width=\columnwidth]{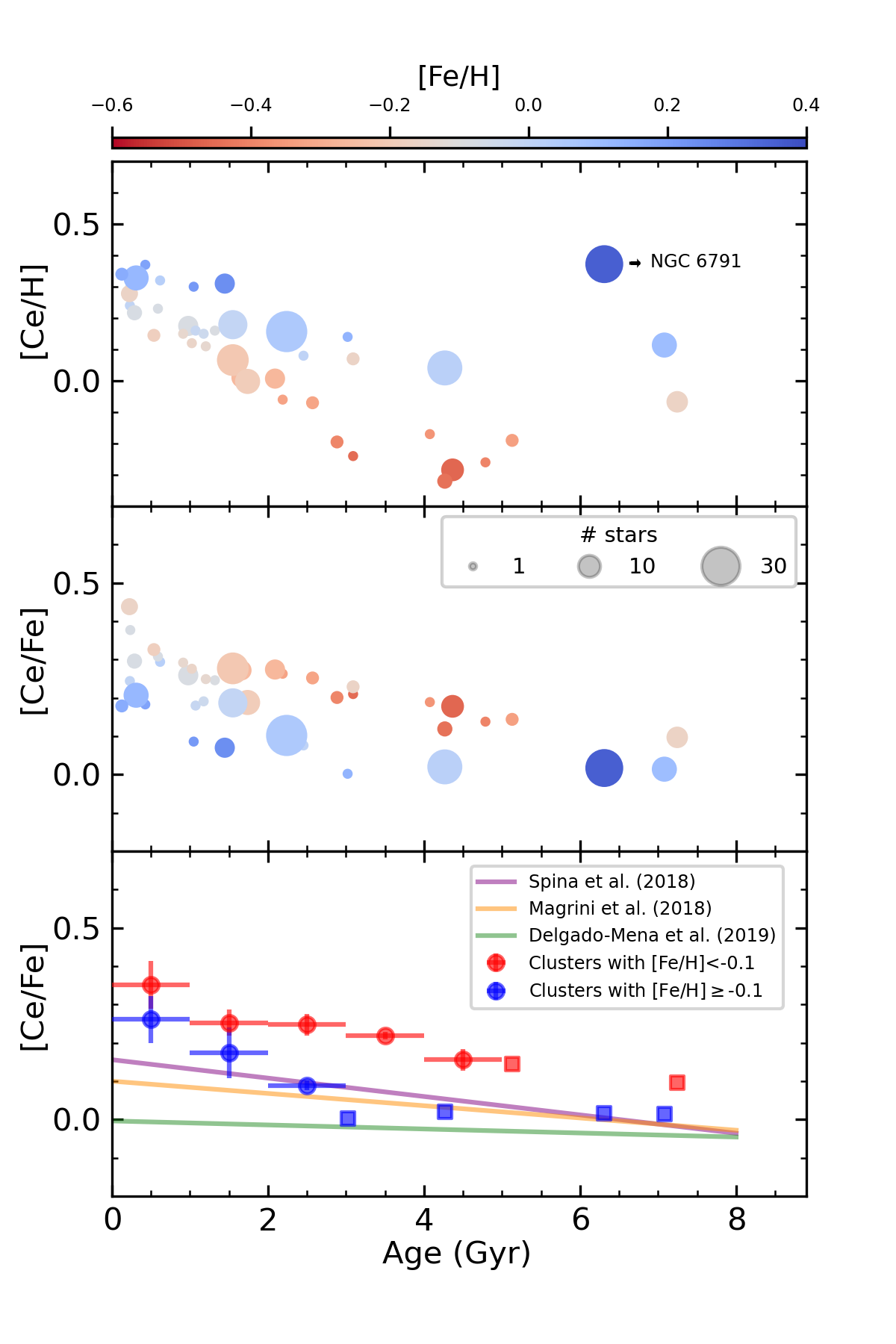}
    \caption{The evolution of [Ce/H] and [Ce/Fe] ratios for the studied open clusters as a function of their age. Upper and middle panels: circles represent the mean [Ce/H] and [Ce/Fe] obtained for the open clusters and circle colors indicate their mean metallicities, while circle sizes symbolize the number of stars analyzed in each open cluster (see Table \ref{tab:clusters_table}). Bottom Panel: circles represent the average of the [Ce/Fe] ratio for the open clusters at 1 Gyr bin and in two metallicity regimes. Red circles represent the mean using only open clusters with [Fe/H]$<-$0.1, while blue circles refer to the average for clusters with [Fe/H]$\geq-$0.1. Squares represent 1 Gyr bin regions where we have only one open cluster. The purple, orange and green lines represent the best linear fits obtained by \citet{Spina2018}, \citet{Magrini2018} and \citet{Delgado-Mena2019}, respectively.
    }
    \label{Ce_Fe_vs_age_feh}
\end{figure}


\subsubsection{[Ce/H]}

To further explore trends in the cerium abundances with metallicity and age, in Figure \ref{Ce_Fe_vs_age_feh}, we plot the time evolution of [Ce/H] and [Ce/Fe] for the studied open cluster sample; the circles in the upper and middle panels of Figure \ref{Ce_Fe_vs_age_feh} are now color coded by metallicity, according to the color bar at the top of the figure. 
It is apparent that metallicity segregates the open clusters in the [Ce/X]-Age plane. In the case of [Ce/H] (top panel), open clusters with larger metallicities (shown in blue) have larger [Ce/H] values for the same age. In addition, [Ce/H] in open clusters with similar metallicities exhibit a correlation with age; open clusters with metallicities around 0.0 (light blue circles) show an increase in [Ce/H] with decreasing age for clusters with Age $\lessapprox$ 4.0 Gyr. A similar evolution of [Ce/H] with age is found for open clusters of lower metallicities $<-$0.1 dex (red circles in Figure \ref{Ce_Fe_vs_age_feh}), but shifted to smaller [Ce/H] values.

The oldest open clusters in this sample, with ages $>$ than 6 Gyr old, are only three in number, making it challenging to reach meaningful conclusions about the [Ce/H] evolution at early times in Galactic disk history. The behavior of the three oldest open clusters (age $>$ 6 Gyr) in this sample (NGC 6791,  NGC 188 and Berkeley 17) indicates that these do not follow the same [Ce/H] sequence as the younger clusters. 
Open clusters with Age $>$ 6 Gyr and [Fe/H]$\leq$0.1, for example, have [Ce/H] values similar to the clusters aged between 2 and 3 Gyr old. The [Ce/H] values result ultimately from the combination of AGB yields and the star formation rate in the birthplace of open clusters. As will be discussed below, the old clusters seem to follow the same lower sequence in the [Ce/Fe] as shown in the middle panel of the figure.

Being the most metal-rich ([Fe/H]=+0.36), as well as one of the oldest open clusters, NGC 6791 is worth discussing in comparison to the much younger, but also metal-rich ([Fe/H]=+0.19) open cluster, BH211.  Both clusters have the highest [Ce/H] abundances ([Ce/H]=+0.37 in both), yet have very different ages. BH211 is a very young \citep[Age=0.42 Gyrs;][]{Cantat-Gaudin2020}, metal-rich open cluster, whereas NGC 6791 is an old open cluster \citep[Age=6.31 Gyrs;][]{Cantat-Gaudin2020} which is even more metal rich.
This simple comparison of these two clusters makes it evident that clusters having very different ages can reach similar [Ce/H] values. The old age and chemical enrichment of NGC 6791 (very high [Fe/H] and high [Ce/H]) suggest that this cluster was probably formed in a region with a high star formation rate.
Observational studies of the interstellar medium indicate that regions of the inner disk are characterized by a higher star formation rate than regions of the outer disk \citep[e. g.][]{Misiriotis2006, Djordjevic2019}. 
Our results for the [Ce/H] gradient support an outward radial migration scenario of NGC 6791 from the innermost regions of the disk, as shown in the next subsection. Chemical abundance and dynamic studies of NGC 6791 also indicate radial migration to its current position \citep[][]{Jilkova2012, Martinez-Medina2018, Villanova2018, ChenZhao2020}.

\subsubsection{[Ce/Fe]}

In addition to having a dependence on metallicity, stellar evolution models show that AGB yields are heavily dependent on stellar mass, with low-mass stars (around 2$\pm$1M$_{\odot}$ for [Fe/H]$\approx-$1.0) having the largest yields\footnote{The nucleosynthesis predictions for low-mass AGB stars ($<$ 3-4 M$_{\odot}$) generally agree quite well among the different AGB nucleosynthesis models/codes (e.g., FRUITY, Monash, ATON, NuGrid/MESA). The situation is very different for the higher mass ($>$ 4 M$_{\odot}$) AGBs, where the model predictions are very dependent on the nucleosynthesis code used \citep[see][for a detailed discussion about the model predictions of different codes]{KarakasLugaro2016}.}\citep[][]{Cristallo2015}. Low-mass stars may require a few Gyr to add their chemical products, such as the s-process elements, to the interstellar medium, resulting in a delay for the interstellar medium enrichment of the s-process elements from this source. The evolution of the [Ce/Fe] ratio observed is in line with a delay time between the enrichment caused by AGB stars and SN Ia. In Figure \ref{Ce_Fe_vs_age_feh} (middle and bottom panel), are shown the [Ce/Fe] ratio versus age for this sample. Open clusters with lower metallicities (red circles) have a higher [Ce/Fe] ratio than clusters with higher metallicities (blue circles); this behavior is reversed when considering only [Ce/H]. 
The strong dependence of [Ce/H] on the star formation rate (as shown by its gradient plotted in the next subsection) may explain the different behaviors for [Ce/H] and [Ce/Fe]. For Ages $<$ 4 Gyr, the [Ce/Fe] ratio increases with decreasing age of the open clusters in all metallicity regimes. This result is in agreement with previous studies from optical high-resolution spectroscopy for open clusters and field disk stars \citep[e.g.][]{Maiorca2011, Spina2018}.

The open clusters can be segregated into two groups in the [Ce/Fe]- Age plane, as those having [Fe/H]$\geq-$0.1 and [Fe/H]$<-$0.1, and we compute the average [Ce/Fe] at each Gyr bin for each group. In the lower panel of Figure \ref{Ce_Fe_vs_age_feh}, we show the average [Ce/Fe] per bin along with their respective standard deviations, 
with the horizontal bars corresponding to the bin size. We note that when having only one cluster in a bin we use the result for that cluster (square symbols in the lower panel of Figure \ref{Ce_Fe_vs_age_feh}). 
There seems to be an overall similar relation between the [Ce/Fe] ratio and age in the two groups but shifted in [Ce/Fe]. For Ages $<$ 4 Gyr, we see an increase in the [Ce/Fe] ratio with decreasing age.
However, for Ages $>$ 4 Gyr, we have roughly a constant value of the [Ce/Fe] ratio with age, with [Ce/Fe] $\approx$ 0.0 and $\approx$ 0.13 for the metal-rich ([Fe/H]$\geq-$0.1) and metal-poor ([Fe/H]$<-$0.1) open clusters, respectively. 

In the bottom panel of Figure \ref{Ce_Fe_vs_age_feh}, we include linear fits in the [Ce/Fe]-Age plane as presented in studies by \citet{Spina2018}, \citet{Magrini2018} and \citet{Delgado-Mena2019}.
\citet{Spina2018} used a sample of 79 solar twin stars (having, by definition, [Fe/H]$\approx$0.0) and derived an increase in the [Ce/Fe] ratio with decreasing age, similar to the trend derived by \citet{Magrini2018}, who used a sample of solar neighborhood open clusters (10) and a sample of thin disk field stars, with ages less than 8 Gyr. 
Both of the trends from \citet{Spina2018} and \citet{Magrini2018} are similar, and track our overall results for the open clusters with [Fe/H]$\geq$-0.1, but for the older clusters we find a flat behavior of [Ce/Fe] with age, with a transition at around 3-4 Gyr to increasing [Ce/Fe] for younger clusters.  At young ages, our results do diverge slightly
from the simplified linear trends from \citet{Spina2018} and \citet{Magrini2018}.

The linear trend from \citet{Delgado-Mena2019} is flatter than those from both \citet{Spina2018} and \citet{Magrini2018}, as well as the open cluster trends found here. For the thin disk stars in their sample with ages less than 8 Gyr there is significant scatter that increases at young ages (their Figure 7), but \citet{Delgado-Mena2019} find a distribution of [Ce/Fe] that is approximately flat as a function of age with a mean [Ce/Fe]$\sim$0.0 (see green line in the bottom panel of Figure \ref{Ce_Fe_vs_age_feh}). Their best fit line to their data is not in good agreement with our results for the open clusters with [Fe/H] $<$ -0.1 (red circles) and, although in better agreement with the open clusters with [Fe/H] $>$ -0.1 (blue circles), there is more divergence for the younger clusters, as [Ce/Fe] rises for younger open cluster ages. We note that one difference between the field dwarfs from \citet{Delgado-Mena2019} and the open clusters studied here is the lack of young open clusters with low values of [Ce/Fe] ($\sim$-0.1 to -0.2) that is seen in the disk population. \citet{Delgado-Mena2019} also investigate the linear fits to thin disk stars with different [Fe/H] ranges. They find that the relationship between s-process abundances (including Ce) and Age varies with metallicity, as also indicated by our results.



The change in the time evolution of the [Ce/Fe] ratio with metallicity in open clusters highlights the dependence of the s-process element synthesis on the metallicity. Stellar evolution models indicate lower [X/Fe] ratios of s-process elements in higher metallicity AGB stars \citep[][]{Cristallo2015, KarakasLugaro2016, Battino2019}, corroborating our results. 


\subsubsection{[Ce/$\alpha$]}

\begin{figure*}
\centering
	\includegraphics[width=15cm]{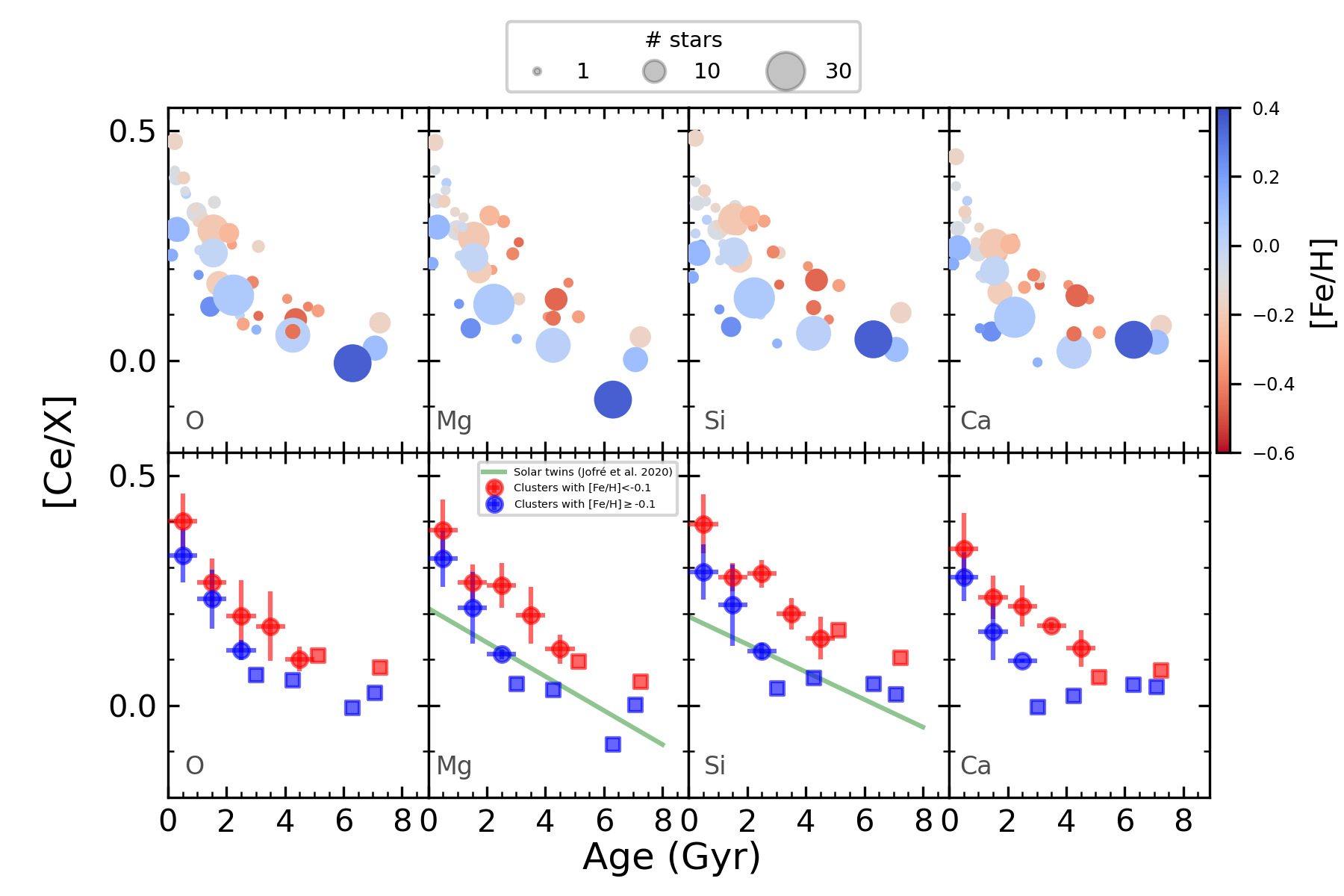}
    \caption{The evolution of the abundances of Ce relative to the $\alpha$ elements O, Mg, Si, and Ca abundances as a function of cluster age. Top panels: Circles represent the mean bracket abundance values for the open clusters. Circle colors indicate their mean metallicities. Circle sizes symbolize the number of stars analyzed in each open cluster (see Table \ref{tab:clusters_table}). Bottom panels: The average of the Ce abundance relative to the $\alpha$ elements as a function of open cluster age. The average values were computed for cluster ages within 1 Gyr bins and segregating the sample into two metallicity regimes. Red circles represent the mean using only open clusters with [Fe/H]$<-$0.1, while blue circles refer to the average for clusters with [Fe/H]$\geq-$0.1. Squares represent 1 Gyr bin regions where we have only one open cluster. The green lines represent the best linear fits obtained by \citet{Jofre2020} for solar twins.
    }
    \label{Ce_X_vs_age_feh}
\end{figure*}

The s-process elements are mainly produced by low and intermediate-mass stars that have longer lifetimes than high mass stars, which are the source of the $\alpha$-elements. The difference in their lifetimes has led to previous suggestions that s-process to alpha-element ratios could serve as good chemical chronometers \citep[e. g.][]{DaSilva2012, Nissen2015, Feltzing2017}. In this study, we seek to investigate whether the Ce-to-$\alpha$ element ratios are universal clocks.
In Figure \ref{Ce_X_vs_age_feh}, we show the evolution of [Ce/$\alpha$] ratios with cluster age, using the O, Mg, Si and Ca uncalibrated abundances from APOGEE DR16. Metallicity segregates the open clusters in the [Ce/$\alpha$]-age plane (Figure \ref{Ce_X_vs_age_feh}). In general, clusters with lower metallicities have higher [Ce/$\alpha$] ratios, although the trend of the [Ce/$\alpha$] ratio with age is similar across the entire metallicity range: there is an increase in the [Ce/$\alpha$] ratio with decreasing open cluster age.

As done previously for Fe (bottom panel of Figure \ref{Ce_Fe_vs_age_feh}), we also computed the average Ce to $\alpha$-element abundance ratios in the same Gyr bins and for the same two metallicity regimes (above and below [Fe/H]=$-$0.1); the behavior of [Ce/$\alpha$] versus cluster age is shown in  Figure \ref{Ce_X_vs_age_feh} for the 'low' (in red)  and 'high' (in blue) metallicity cluster groups. The average [Ce/$\alpha$] ratios for the lower metallicity ([Fe/H]$<$-0.1) open cluster group are shifted to higher values when compared to the high metallicity group ([Fe/H]$\geq$-0.1), with average differences between these two metallicity groups of 0.07$\pm$0.02, 0.09$\pm$0.04, 0.11$\pm$0.04 and 0.10$\pm$0.05 for [Ce/O], [Ce/Mg], [Ce/Si], and [Ce/Ca] ratios, respectively. These differences are approximately equal for all [Ce/$\alpha$] abundance ratios.

Overall, the [Ce/$\alpha$] ratio for all studied $\alpha$ elements (O, Mg, Si and Ca) shows a similar dependence with age in both metallicity groups. For old open clusters with ages $>$ 4 Gyr, there is an approximately constant relation of the [Ce/$\alpha$] ratio with age also for both metallicity groups. (As previously mentioned, our sample of open clusters with Age $>$ 4 Gyr is small, however, there is a need to analyze a more robust open cluster sample to confirm (or not) the constant evolution of [Ce/$\alpha$] with time.) On the other hand, open clusters with Age $<$ 4 Gyr show an increase in the [Ce/X] ratio with decreasing age in the two metallicity groups, a behavior that is reminiscent of that found for the [Ce/Fe] ratio. 
\citet{Jofre2020} analyzed trends in the [Ce/Mg] and [Ce/Si]-age planes for the same sample of solar twin stars from \citet{Spina2018}. 
In the bottom panel of Figure \ref{Ce_X_vs_age_feh}, we show their best-fit line obtained for their sample of solar neighborhood solar twins (solid green line); the overall trend is the same as the one found for the open clusters in general; there is a better agreement with those open clusters having [Fe/H]$\geq-$0.1, which is reasonable, as the solar twins have near-solar metallicity; however, although the general behavior is similar, our data may suggest a more complex behavior.

The increase of [Ce/$\alpha$] and [Ce/Fe] ratios is interpreted as a signature of the late chemical evolution of Ce. We note that the [Fe/$\alpha$] values are generally within $\sim$ 0.1 and show no significant trend with age for either the metal-rich or metal-poor subset. Therefore, the [Ce/$\alpha$]-age trend provides information that cannot be obtained from the [$\alpha$/Fe] ratio.

Finally, the results here indicate that the [Ce/$\alpha$]-Age ratio is not universal for open clusters, but, rather, is strongly dependent on metallicity, as also shown for the [Y/$\alpha$] ratio \citep[e.g.][]{Delgado-Mena2019, Magrini2021}. Recently, \citet{Casali2020} and \citet{Magrini2021} indicated that the non-universality of the [s-process/$\alpha$]-age-[Fe/H] relation is caused by star formation history and s-process yields, with their metallicity dependence. In particular, \citet{Magrini2021} used models of the chemical evolution of the Galaxy that consider magnetic-buoyancy-induced mixing in AGB stars to explain the change in the relationship between [Y/Mg]-Age with metallicity observed in the open clusters from \citet{Magrini2018}. \citet{Magrini2021} pointed out that the mixing triggered by magnetic fields may cause a change of s-process production and in its relationship with metallicity by changing the $^{13}$C pocket (main source of s-process neutrons) inside the TP AGB stars. This scenario presented by \citet{Magrini2021} exposes the complexity that involves s-process nucleosynthesis and [Fe/H]. Homogeneous studies with a significant sample of objects with well-defined ages, such as the one presented here, are essential in testing the evolutionary models of AGBs and unveiling the complex relationship between the s-process production and metallicity. 

\subsection{The Ce abundance gradient}

\begin{figure}
\centering
	\includegraphics[width=\columnwidth]{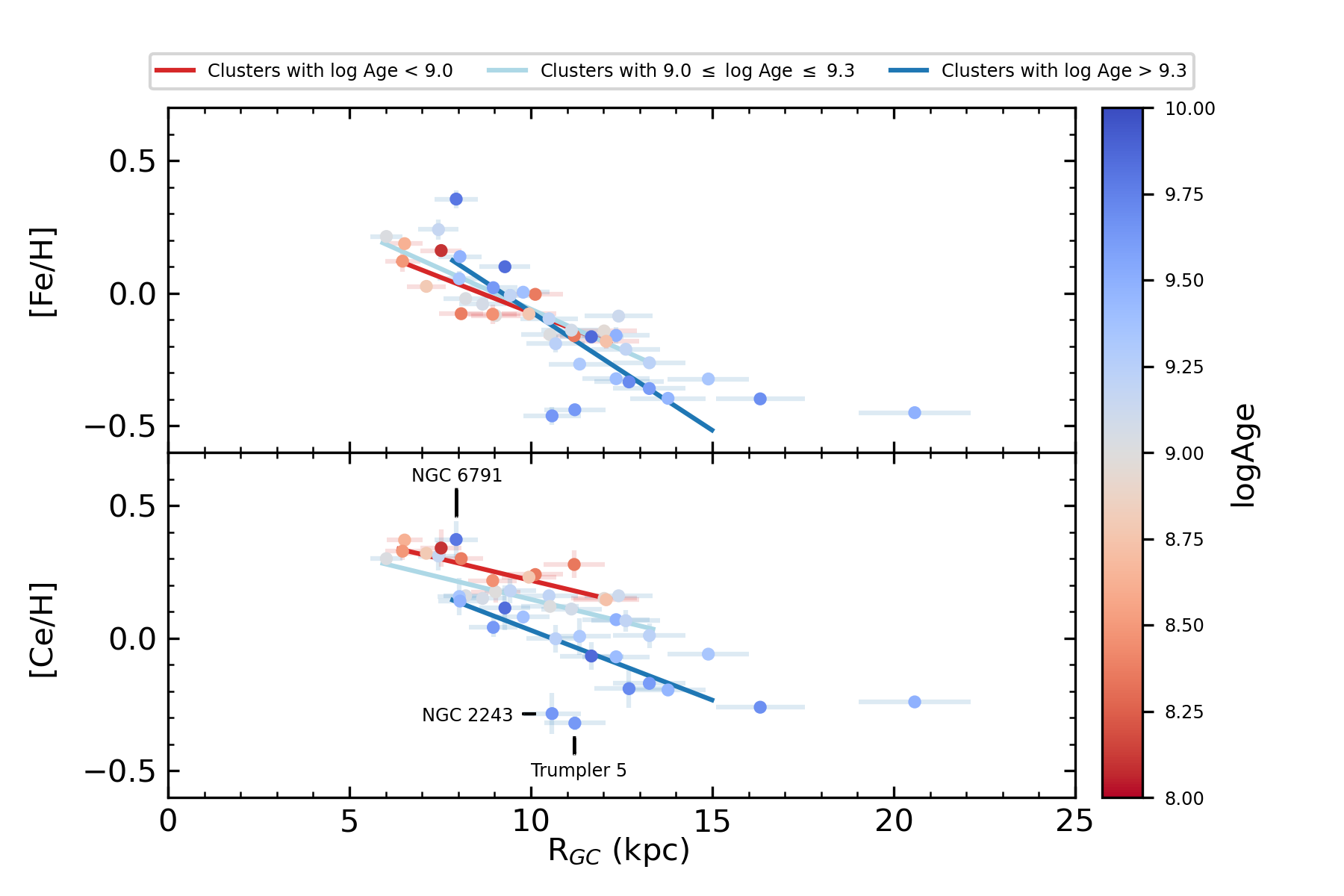}
    \caption{The [Fe/H] (top panel) and [Ce/H] (bottom panel) cluster mean abundances as a function of cluster galactocentric distance; the colors represent cluster age. Three gradients were computed for samples of open clusters segregated according to their age. In both panels, the linear fits shown as red, light blue, and dark blue lines correspond to open clusters with Age$<$1.0 Gyr, 1 Gyr$\leq$Age$\leq$2 Gyr, and Age$>$2.0 Gyr, respectively. The fits were computed for R$_{GC}<$15kpc. There is an age segregation in [Ce/H] as a function of galactocentric distance. In general, younger open clusters (red circles) show higher [Ce/H] ratios than older clusters (blue circles) at a given galactocentric distance. The behavior for [Fe/H] is different than that from [Ce/H]. 
    }
    \label{Ce_H_Fe_H_vs_rgc_logage}
\end{figure}

\begin{figure}
\centering
	\includegraphics[width=\columnwidth]{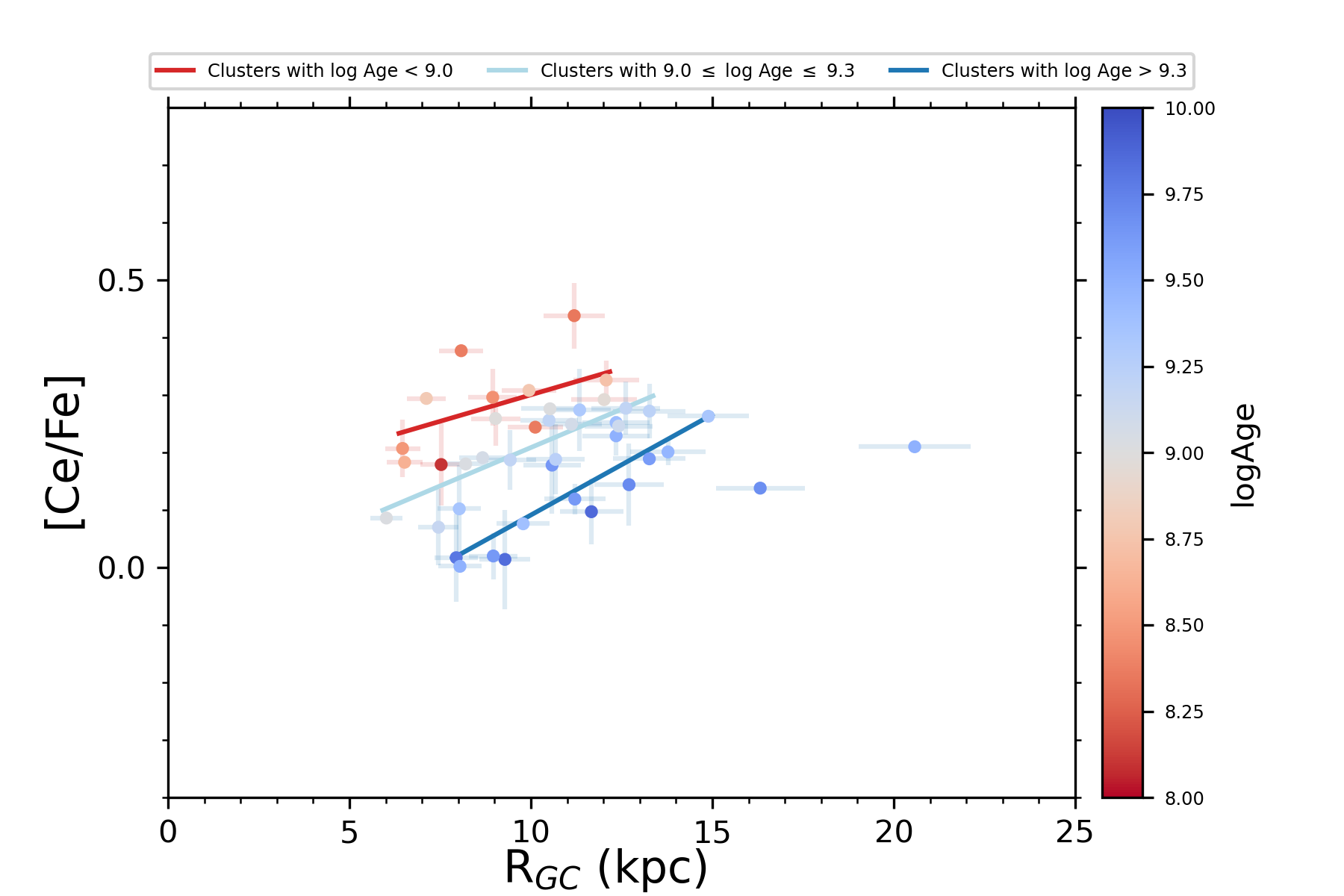}
    \caption{The [Ce/Fe] gradients obtained for the studied open cluster sample. Three gradients were computed as in Figure \ref{Ce_H_Fe_H_vs_rgc_logage}. In general, younger open clusters (red circles) show higher [Ce/Fe] ratios than older clusters (blue circles) at a given galactocentric distances.
    }
    \label{Ce_Fe_vs_rgc_logage}
\end{figure}

\begin{table*}
	\centering
	\scriptsize
	\caption{Radial abundance gradients (dex kpc$^{-1}$) and intercept coeficient of the best linear fits binned by age for open clusters with R$_{GC} <$15 kpc. We also show the number of open clusters (\#) and galactocentric distance range covered by each age sample.}
	\label{tab:gradients_table}
	\begin{tabular}{lcc|cc|cc|cc} \hline\hline
     & & & \multicolumn{2}{c}{[Fe/H]} & \multicolumn{2}{c}{[Ce/H]}  &  \multicolumn{2}{c}{[Ce/Fe]}\\
      & \#  & R$_{GC}$ (kpc) range & d[Fe/H]/dR$_{GC}$  & intercept & d[Ce/H]/dR$_{GC}$ & intercept & d[Ce/Fe]/dR$_{GC}$ & intercept  \\
\hline     
All open clusters       & 40 &  6.02 $\leq$R$_{GC}\leq$ 14.89   & $-$0.071 $\pm$ 0.008 &  0.634 $\pm$ 0.082  &  $-$0.070 $\pm$ 0.007   & 0.614 $\pm$ 0.077  & 0.014 $\pm$ 0.007  & 0.061 $\pm$ 0.071  \\ 
Age $<$ 1 Gyr           & 12 &  6.46 $\leq$R$_{GC}\leq$ 12.07   & $-$0.054 $\pm$ 0.011 &  0.462 $\pm$ 0.104  &  $-$0.033 $\pm$ 0.006   & 0.544 $\pm$ 0.061  & 0.018 $\pm$ 0.008  & 0.116 $\pm$ 0.079 \\
1 $\leq$Age$\leq$2 Gyr  & 12 &  6.02 $\leq$R$_{GC}\leq$ 13.27   & $-$0.061 $\pm$ 0.010 &  0.552 $\pm$ 0.107  &  $-$0.033 $\pm$ 0.007   & 0.479 $\pm$ 0.078  & 0.027 $\pm$ 0.007  & -0.058 $\pm$ 0.077 \\
Age $>$ 2.0 Gyr         & 16 &  7.94 $\leq$R$_{GC}\leq$ 14.89   & $-$0.089 $\pm$ 0.018 &  0.825 $\pm$ 0.205  &  $-$0.053 $\pm$ 0.018   & 0.557 $\pm$ 0.198  & 0.035 $\pm$ 0.007  & -0.258 $\pm$ 0.077 \\
\hline
	\end{tabular}
\end{table*}

Radial abundance gradients in the Galaxy provide information on the star formation rate and serve as observational constraints to models of chemical evolution. In particular, open clusters are essential pieces in studies of abundance gradients as their distances and ages are more accurate than those for field stars. In Figure \ref{Ce_H_Fe_H_vs_rgc_logage}, we show the [Fe/H] (top panel) and [Ce/H] (bottom panel) gradients for the studied open cluster sample, using the homogeneous distance estimates from \citet{Cantat-Gaudin2020}. 

The studied open clusters cover the galactocentric distance range between 6.0 and 20.6 kpc, with all but two clusters having galactocentric distances less than 15 kpc (Table \ref{tab:clusters_table}). We realized the best linear fits for the gradients using the maximum likelihood with associated uncertainties being estimated through the Markov-Chain Monte Carlo (MCMC) routine from the emcee python package \citep[][]{Foreman-Mackey2013}, as previously done in \citet{Donor2020} for [Fe/H] gradients.

When considering the cluster sample having galactocentric distances R$_{GC} <$ 15 kpc, we obtain a [Ce/H] gradient of $-$0.070$\pm$0.007 dex kpc$^{-1}$,
approximately equal to our estimate for the [Fe/H] gradient ($-$0.071$\pm$0.008 dex kpc$^{-1}$),
which, as expected, is close to the metallicity gradient found in \citet[$-$0.068 dex kpc $^{-1}$]{Donor2020} from the analysis of a larger number of open clusters from the OCCAM sample. This is also in agreement with the metallicity gradient from \citet[$-$0.076 dex kpc$^{-1}$]{Spina2021} obtained using a combination of APOGEE and GALAH results put on the same reference scale. Concerning [Ce/H], the earlier study by \citet{Maiorca2011} found that [Ce/H] decreases with galactocentric distance from the analysis of a sample of 19 open clusters, but that study did not compute a [Ce/H] gradient.

In Figure \ref{Ce_H_Fe_H_vs_rgc_logage}, we now focus on gradients as a function of open cluster age. In both panels of the figure, the filled circle colors represent the log Age, according to the color bar. [Fe/H] (top panel) and [Ce/H] (bottom panel) gradients are obvious. First, it seems clear that there is an age segregation in the [Ce/H]-R$_{GC}$ plane: the younger the open cluster, the higher the [Ce/H] value throughout the thin disk (at the entire R$_{GC}$ range probed). A similar behavior is not seen for the [Fe/H]-R$_{GC}$ plane (top panel of the figure), as the populations do not segregate in this parameter space.
In general, as discussed previously, the age segregation of Ce abundances may be due to late enrichment of Ce to the natal cloud.

Given the age segregation in the [Ce/H]-R$_{GC}$ plane, we now divide the open cluster sample into three populations and compute the gradients considering R$_{GC} <$ 15 kpc: very young clusters with ages less than 1 Gyr, clusters with ages within the narrow range between 1 and 2 Gyr, and older clusters having ages larger than 2 Gyr. The linear fits to the data in each case are shown in Figure \ref{Ce_H_Fe_H_vs_rgc_logage} as red, light blue and dark blue lines respectively for the  Age$<$1.0 Gyr, 1 Gyr$\leq$Age$\leq$2 Gyr, and Age$>$2.0 Gyr samples.
In Table \ref{tab:gradients_table}, we present the radial abundance gradients (dex kpc$^{-1}$) binned by age for our open cluster sample with R$_{GC} <$ 15 kpc. We find that the slopes of the linear fits in the [Ce/H]-R$_{GC}$ plane are approximately equal for the different age groups of the open clusters. Open clusters with Age $<$ 1Gyr and 1 Gyr$\leq$Age$\leq$2 Gyr show a gradient of $-$0.033 $\pm$ 0.007 dex kpc$^{-1}$, just slightly flatter but similar within the uncertainties, to the gradient obtained for clusters with ages $>$2.0 Gyr ($-$0.053 $\pm$ 0.018 dex kpc$^{-1}$). 

The larger uncertainty in the best fit for the older group is caused by greater dispersion of this sample. We observed that three old open clusters (NGC 6791, Trumpler 5, and NGC 2243) present the highest residuals\footnote{The residuals (the difference between the observed value and the predicted by the linear fit, [Ce/H]-[Ce/H]$_{fit}$, in the cluster R$_{GC}$) for each open cluster using the fits from their respective age group.} of our sample ($>$ 0.22 dex), contributing to the imprecision of the linear fit of old clusters. The [Fe/H] and [Ce/H] gradients for the old open clusters without NGC 6791, Trumpler 5, and NGC 2243 are -0.081 $\pm$ 0.010 and -0.043 $\pm$ 0.011, respectively, being these gradients less uncertain but equal within the uncertainties to the gradients obtained considering the entire sample of old clusters. Probably these open clusters underwent a radial migration process. NGC 6791 is known to exhibit significant radial migration \citep[][]{Martinez-Medina2018, Villanova2018, ChenZhao2020}. \citet{Miglio2021} have identified a population of metal rich red giant stars ([Fe/H]$>$0.2) at the solar galactocentric distance with a chemical pattern and age similar to NGC 6791, indicating that likely these stars also suffered a radial migration from their birthplace to the solar neighborhood. NGC 2243 (Age= 4.4 Gyr) and Trumpler 5 (Age= 4.3 Gyr) have a very low value [Ce/H] $\sim$ $-$0.3, which, by sharing a very low [Ce/H], would seem to follow a potentially flatter gradient of the two old clusters that are found beyond 15 kpc. In general, radial migration of old open clusters is expected to be more significant than that of young open clusters due to longer exposure to the bar and spiral arm perturbations \citep[][]{Jilkova2012, ChenZhao2020}.

Concerning iron, the metallicities and gradients obtained for the R$_{GC}$ $<$ 15 kpc open cluster sample do not exhibit clear segregation with age. The [Fe/H] gradients become slightly steeper with the increasing age of open clusters, as previously discussed in \citet{Donor2020}. [Fe/H] gradients are steeper than [Ce/H] gradients in all age bins (see Table \ref{tab:gradients_table}).

[Ce/Fe] ratio gradients are shown in Figure \ref{Ce_Fe_vs_rgc_logage}; the best fit slopes were computed segregating the cluster sample according to the same age bins as done for [Ce/H] and [Fe/H]. In general, younger open clusters show higher [Ce/Fe] ratios than older clusters (blue circles) at a given galactocentric distance. For R$_{GC}$ $<$ 15 kpc, we find an increasing [Ce/Fe] ratio with increasing R$_{GC}$, which is opposite to the behavior of the [Ce/H] gradient. The [Ce/Fe] gradients for the open clusters with Age$<$1Gyr, 1 Gyr$\leq$Age$\leq$2 Gyr and Age$>$2Gyr are 0.018 $\pm$ 0.007, 0.027 $\pm$ 0.007 and 0.035 $\pm$ 0.007 dex kpc$^{-1}$, respectively. We find that the slopes of the linear fits in the [Ce/Fe]-R$_{GC}$ plane are very similar, becoming just modestly steeper with increasing open cluster age, with the [Ce/Fe] gradient changing by $\sim$+0.009 dex-kpc$^{-1}$-Gyr$^{-1}$. This change is approximately equal to the gradient uncertainties ($\approx$ 0.007, see Table \ref{tab:gradients_table}). We noted that the [Ce/Fe] ratios for the open clusters NGC 6791, NGC 2243, and Trumpler 5 are consistent with the [Ce/Fe] linear gradient shown by the old open clusters.

\citet{Magrini2018} also found an overall increase in the [Ce/Fe] ratio with galactocentric distance using the open cluster sample from the GAIA-ESO survey, but they do not present a gradient. More recently, \citet{Tautvaisiene2021} estimated the [Ce/Fe] gradient for a sample of 424 thin disk stars spanning ages from 0.1 to 9.5 Gyr, and R$_{GC}$ from 5.5 to 11.8 kpc. For all thin disk stars, they found a [Ce/Fe] gradient of +0.015$\pm$0.007 dex kpc$^{-1}$, a very similar gradient to our estimate using the entire open cluster sample (+0.014$\pm$0.007 dex kpc$^{-1}$). \citet{Tautvaisiene2021} did not calculate the [Ce/Fe] gradient binned by age. The [Ce/Fe] gradient obtained here is also in line with the recent results for Ba from \citet{Spina2021}. The latter study found that the [Ba/Fe] ratio increases with galactocentric distance for 5$<$R$_{GC}$ $<$ 12 kpc. The dependence of the production of heavy s-process elements (such as Ce) on metallicity can explain the increase in the [Ce/Fe] ratio with increasing R$_{GC}$. AGBs from regions with lower metallicity (outer disk) show greater Ce yields than AGBs with high metallicity (inner disk) \citep[][]{Cristallo2015, KarakasLugaro2016}.

Only two clusters in our sample (Berkeley 20 and Berkeley 29) have galactocentric distances greater than 15 kpc; gradient determinations for the outer disk using this sample would not be meaningful. However, these two distant open clusters have similar ages, and [Ce/H] and [Ce/Fe] ratios, which may indicate a constant [Ce/H] and [Ce/Fe] gradient for the old open clusters in the outer disk. 

\section{Conclusions}

The evolution and gradient of s-process elements in the Galactic disk are still not well defined due to considerable distance and age uncertainties for field stars and small and heterogeneous open cluster samples. Large spectroscopic surveys such as GAIA-ESO \citep[][]{Gilmore2012}, GALAH \citep[][]{DaSilva2015}, as well as APOGEE \citep[][]{Majewski2017} are changing this scenario by increasing significantly the number of homogeneous chemical abundance measurements for s-process elements for field stars and to a lesser extent open clusters.  Along these lines, we determined the abundance of the s-process dominated element cerium, Ce, for 218 stars belonging to 42 open clusters from the OCCAM/APOGEE DR16 survey. The Ce abundances obtained in this study allowed us to determine details of the chemical evolution of Ce in the Galactic disk and its relationship with metallicity. In addition, we estimated the Ce gradient and its change over time for the studied open cluster sample. Our results can be summarized as follows:

\textbf{[Ce/Fe]-[Fe/H] plane:} The [Ce/Fe] ratio increases as the metallicity decrease for the different age sets of the open clusters, with a possible change in the trend for [Fe/H]$<-$0.2 for the old open clusters. Our results also indicate that older open clusters have lower [Ce/Fe] ratio values than the young open clusters in the same metallicity range. The [Ce/Fe] ratios of our sample, which was derived from the APOGEE spectra in the near-infrared, are slightly overabundant if one compares with literature Ce abundances from high-resolution optical spectroscopy obtained for dwarf and giant stars in the field. This overabundance of Ce in the open cluster population relative to field stars is in line with results from the literature from other s-process elements that also find similar behavior. 
Age may contribute to such differences, field stars being systematically older than open clusters.  

\textbf{Chemical evolution of Ce:} Metallicity segregates open clusters in the [Ce/X]-Age plane, with X being H, Fe, or $\alpha$ elements (O, Mg, Si, or Ca). Open clusters with lower metallicity show [Ce/Fe] and [Ce/$\alpha$] ratios higher than those with high metallicity at a given age. For Ages $<$ 4 Gyr, the Ce abundance increases with decreasing age of the open clusters. In other words, younger open clusters show higher [Ce/Fe] and [Ce/$\alpha$] ratios than older open clusters with similar metallicities. For Ages $>$ 4 Gyr, the trend of the [Ce/Fe] and [Ce/$\alpha$] ratios with age are approximately constant, but our sample is small in this age range.

\textbf{[Ce/$\alpha$] as a stellar chemical clock:} The abundance ratio between s-process and $\alpha$ elements has emerged in the literature as the main candidate for the universal chemical clock for stars. The examination of such a ratio in the open clusters provides an excellent opportunity to test this hypothesis. Our results indicate that the relationship of the [Ce/$\alpha$] ratio with age is not the same across the Galactic disk, which is possibly due to the dependence of AGB yields on metallicity. 

\textbf{Ce abundance gradients:} For clusters with R$_{GC} <$15 kpc, we find a negative ($-$0.070 $\pm$ 0.007 dex  kpc$^{-1}$) and positive (0.014 $\pm$ 0.007 dex  kpc$^{-1}$) gradient for the [Ce/H] and [Ce/Fe] ratios, respectively. Age segregates the open clusters in the [Ce/H]-R$_{GC}$ and [Ce/Fe]-R$_{GC}$ planes, a different behavior when compared to the metallicity gradient, which does not show this separation. The linear gradients in [Ce/H] and [Ce/Fe] shift to smaller values in [Ce/H] and [Ce/Fe] for the older open clusters. We also find that the [Ce/H] and [Ce/Fe] gradients are approximately constant with cluster age. The [Ce/Fe] gradient becomes slightly steeper over time, changing by $\sim$+0.009 dex-kpc$^{-1}$-Gyr$^{-1}$, marginally greater than the gradient uncertainties ($\sim$ 0.007). 


Overall, our results indicate a strong dependence of the Ce abundance with metallicity and age. Iron ($^{56}$Fe) nuclei work as seeds for the s-process, hence the close relationship between metallicity and the s-process. The production of heavy s-process elements, like Ce, is lower in high metallicity AGB stars, due to the lower number of neutrons per iron-56 seed nucleus, which favors the production of the light s-process elements (Sr, Y, and Zr) \citep[][]{Cristallo2009, Cristallo2011, Cristallo2015,Karakas2014,KarakasLugaro2016}. 
The nature of the behavior of increasing abundance of heavy s-process elements with decreasing age for open clusters is not fully understood \citep[][]{Baratella2021}. The relationship between [Ce/Fe] and age for open clusters may be related to the delay in the enrichment of some Ce-producing stars, such as 1.5$M_{\odot}$ stars, which takes a few Gyr to add AGB products to the interstellar medium. However, AGB models indicate low yields of heavy s-process elements for very low mass stars \citep[][]{Cristallo2015, KarakasLugaro2016}. The formation of an extended $^{13}$C pocket induced by mixing processes can increase Ce production \citep[e.g.][]{Battino2021} and may explain its overabundance in the young open clusters, as pointed out by \citet{Maiorca2011}.

\acknowledgments

JVSS acknowledges the PCI/CNPQ programme under the grant 301863/2021-0.


P.M.F., J.D., and J.O. acknowledge support for this research from the National Science Foundation (AST-1311835 \& AST-1715662).  P.M.F. also acknowledges some of this work was performed at the Aspen Center for Physics, which is supported by National Science Foundation grant PHY-1607611.

SRM acknowledges support from NSF grant AST-1909497.

DAGH acknowledges support from the State Research Agency (AEI) of the Spanish Ministry of Science, Innovation and Universities (MCIU) and the European Regional Development Fund (FEDER) under grant AYA2017-88254-P.

Funding for the Sloan Digital Sky Survey IV has been provided by the Alfred P. Sloan Foundation, the U.S. Department of Energy Office of Science, and the Participating Institutions. SDSS-IV acknowledges support and resources from the Center for High-Performance Computing at the University of Utah. The SDSS web site is www.sdss.org.

SDSS-IV is managed by the Astrophysical Research consortium for the Participating Institutions of the SDSS Collaboration including the Brazilian Participation Group, the Carnegie Institution for Science, Carnegie Mellon University, the Chilean Participation Group, the French Participation Group, HarvardSmithsonian Center for Astrophysics, Instituto de Astrofísica de Canarias, The Johns Hopkins University, Kavli Institute for the Physics and Mathematics of the Universe (IPMU)/University of Tokyo, Lawrence Berkeley National Laboratory, Leibniz Institut fur Astrophysik Potsdam (AIP), Max-Planck-Institut fur Astronomie (MPIA Heidelberg), Max-Planck Institut fur Astrophysik (MPA Garching), Max-Planck-Institut fur Extraterrestrische Physik (MPE), National Astronomical Observatory of China, New Mexico State University, New York University, University of Notre Dame, Observatório Nacional/MCTI, The Ohio State University, Pennsylvania State University, Shanghai Astronomical Observatory, United Kingdom Participation Group, Universidad Nacional Autónoma de México, University of Arizona, University of Colorado Boulder, University of Oxford, University of Portsmouth, University of Utah, University of Virginia, University of Washington, University of Wisconsin, Vanderbilt University, and Yale University.

This research made use of following Python packages: {\tt matplotlib} \citep[][]{Hunter2007}, {\tt Numpy} \citep[][]{Harris2020}, {\tt Scipy} \citep[][]{Virtanen2020}.

%

\vspace{5mm}
\facilities{Sloan (APOGEE)}


\software{\tt matplotlib} \citep[][]{Hunter2007}, {\tt Numpy} \citep[][]{Harris2020}, {\tt Scipy} \citep[][]{Virtanen2020}



\appendix

\section{Ce abundances for all cluster stars} 
\label{sec:ceabundances}

\begin{table*}
	\centering
	\tiny
	\caption{Line by line Ce abundances and atmospheric parameters ($T_{\rm eff}$, $\log{g}$, $\xi$, and [Fe/H]) for all cluster stars. In the third column, we present the signal-to-noise of the spectra.}
	\label{tab:star_abundances_table}
	\begin{tabular}{ccccccccccccccc} 
		\hline\hline
		\multicolumn{7}{c}{} & \multicolumn{7}{c}{Ce II absorption lines (\AA)} & \\
        \cline{8-14}
		Cluster & ID & SNR & $T_{\rm eff}$ (K) & $\log{g}$ & $\xi$ (km\,s$^{-1}$) & [Fe/H] & 15277.6 & 15784.8 & 15977.1 & 16327.3 & 16376.5 & 16595.2 & 16722.6 & $<$A(Ce)$>$ $\pm \sigma$ \\
		\hline
Basel 11b  &  2M05581816+2158437 &   296 &  4759 &  2.67 &   1.23 & $-$0.00 &  ---  &  ---  &  1.88  &  ---  &  1.97  &  1.97  &  ---  &  1.94$\pm$0.05  \\
Berkeley 17  &  2M05202118+3035544 &   166 &  4746 &  2.77 &   1.14 &$-$0.13 &  ---  &  1.69  &  ---  &  ---  &  ---  &  1.74  &  ---  &  1.72$\pm$0.04  \\
Berkeley 17  &  2M05202905+3032414 &    98 &  4708 &  2.99 &   0.83 &$-$0.15 &  ---  &  1.61  &  ---  &  ---  &  ---  &  1.70  &  ---  &  1.66$\pm$0.06  \\
Berkeley 17  &  2M05203121+3035067 &   172 &  4773 &  2.79 &   1.13 & $-$0.16 &  ---  &  1.63  &  ---  &  ---  &  1.55  &  1.75  &  ---  &  1.64$\pm$0.10  \\
Berkeley 17  &  2M05203650+3030351 &   396 &  4340 &  2.20 &   1.18 & $-$0.15 &  ---  &  ---  &  ---  &  ---  &  1.61  &  1.59  &  ---  &  1.60$\pm$0.01  \\
Berkeley 17  &  2M05203799+3034414 &   333 &  4202 &  2.10 &   1.15 & $-$0.15 &  ---  &  1.63  &  ---  &  ---  &  1.53  &  1.57  &  ---  &  1.58$\pm$0.05  \\
Berkeley 17  &  2M05204143+3036042 &   168 &  4772 &  2.77 &   1.16 & $-$0.20 &  ---  &  1.56  &  ---  &  ---  &  1.51  &  1.68  &  ---  &  1.58$\pm$0.09  \\
Berkeley 17  &  2M05204488+3038020 &   180 &  4753 &  2.72 &   1.18 & $-$0.20 &  ---  &  1.64  &  ---  &  ---  &  ---  &  1.66  &  ---  &  1.65$\pm$0.01  \\
Berkeley 19  &  2M05240941+2937217 &   110 &  4381 &  2.03 &   1.23 & $-$0.32 &  ---  &  1.73  &  1.51  &  ---  &  1.67  &  1.65  & ---   &  1.64$\pm$0.09  \\
Berkeley 20  &  2M05323895+0011203 &    68 &  4313 &  1.99 &   1.21 & $-$0.40 &  ---  & 1.51   &  ---  &  ---  &  1.36  &  1.45  &  ---  &  1.44$\pm$0.08  \\
Berkeley 29  &  2M06531569+1656176 &    56 &  4635 &  2.62 &   1.11 & $-$0.45 &  ---  &  1.62  &  1.35  &  ---  &  1.42  &  ---  &  ---  &  1.46$\pm$0.14  \\
Berkeley 43 &  2M19152201+1115544 &   544 &  4715 &  2.85 &   1.13 &  0.03 &  2.00  &  ---  &  ---  &  ---  &  ---  &  2.04  &  ---  &  2.02$\pm$0.03  \\
Berkeley 53 &  2M20554232+5106153 &   217 &  4695 &  2.66 &   1.14 & $-$0.06 &  1.74  &  1.95  &  1.74  &  ---  &  1.89  &  1.93  &  ---  &  1.85$\pm$0.10  \\
Berkeley 53 &  2M20554936+5106545 &   367 &  4362 &  2.17 &   1.19 & $-$0.09 &  1.79  &  ---  &  1.81  &  ---  &  ---  &  1.91  &  ---  &  1.84$\pm$0.06  \\
Berkeley 53 &  2M20554998+5102175 &    94 &  4669 &  2.64 &   1.18 & $-$0.08 &  ---  &  ---  &  ---  &  ---  &  1.90  &  1.88  &  ---  &  1.89$\pm$0.01  \\
Berkeley 53 &  2M20555767+5103206 &   276 &  4915 &  2.88 &   1.19 & $-$0.12 &  ---  &  1.99  &  1.79  &  ---  &  1.97  &  1.89  &  ---  &  1.91$\pm$0.09  \\
Berkeley 53 &  2M20555959+5100466 &    55 &  4937 &  3.00 &   1.12 & $-$0.10 &  ---  &  ---  &  ---  &  ---  &  1.85  &  1.80  &  ---  &  1.83$\pm$0.04  \\
Berkeley 53 &  2M20561018+5102389 &   320 &  4820 &  2.71 &   1.21 & $-$0.06 &  ---  &  1.94  &  1.88  &  ---  &  ---  &  1.97  &  ---  &  1.93$\pm$0.05  \\
Berkeley 66  &  2M03040128+5846422 &    60 &  4893 &  2.78 &   1.24 & $-$0.18 &  ---  &  ---  &  ---  &  ---  &  1.71  &  1.81  &  ---  &  1.76$\pm$0.07  \\
Berkeley 66  &  2M03042797+5845042 &    59 &  4907 &  2.84 &   1.22 & $-$0.14 &  ---  &  1.72  &  1.85  &  ---  &  1.77  &  ---  &  ---  &  1.78$\pm$0.07  \\
Berkeley 98 &  2M22423502+5222084 &    93 &  4495 &  2.67 &   1.12 &  0.00 &  ---  &  ---  &  1.80  &  ---  &  1.81  &  1.76  &  1.76  &  1.78$\pm$0.03  \\
BH 211 &  2M17021851-4109170 &   398 &  4789 &  2.86 &   1.16 &  0.19 &  ---  &  2.09  &  2.03  &  ---  &  ---  &  2.09  &  ---  &  2.07$\pm$0.03  \\
Collinder 220 &  2M10260294-5755255 &   836 &  4804 &  2.55 &   1.35 & $-$0.08 &  1.92  &  ---  &  ---  &  ---  &  ---  &  2.08  &  ---  &  2.00$\pm$0.11  \\
Czernik 21  &  2M05263726+3600404 &   150 &  4978 &  2.93 &   1.15 & $-$0.32 &  ---  &  1.74  &  ---  &  ---  &  1.56  &  1.58  &  ---  &  1.63$\pm$0.10  \\
Czernik 21  &  2M05264047+3602191 &   114 &  4878 &  2.86 &   1.12 & $-$0.33 &  ---  &  1.57  &  ---  &  ---  &  1.61  &  1.70  &  ---  &  1.63$\pm$0.07  \\
Czernik 30  &  2M07310830-0956359 &   170 &  4286 &  1.94 &   1.23 & $-$0.39 &  ---  &  1.62  &  ---  &  ---  &  1.47  &  1.46  &  ---  &  1.52$\pm$0.09  \\
Czernik 30  &  2M07311590-0955415 &   113 &  4440 &  2.22 &   1.24 & $-$0.40 &  ---  &  ---  &  ---  &  ---  &  1.42  &  1.56  &  ---  &  1.49$\pm$0.10  \\
FSR 0394 &  2M22545788+5844048 &   157 &  4728 &  2.76 &   1.17 & $-$0.10 &  ---  &  1.76  &  ---  &  ---  &  ---  &  1.93  &  1.88  &  1.86$\pm$0.09  \\
FSR 0394 &  2M22550718+5842026 &   149 &  4918 &  2.77 &   1.22 & $-$0.09 &  ---  &  1.80  &  ---  &  ---  &  1.93  &  ---  &  ---  &  1.86$\pm$0.09  \\
IC 166  &  2M01522953+6151427 &   126 &  4807 &  2.83 &   1.16 & $-$0.09 &  ---  &  ---  &  ---  &  ---  &  1.86  &  ---  &  ---  &  1.86  \\
IC 1369 &  2M21115265+4744571 &   238 &  4953 &  2.78 &   1.29 & $-$0.04 &  ---  &  ---  &  1.91  &  ---  &  ---  &  1.96  &  ---  &  1.94$\pm$0.04  \\
IC 1369 &  2M21120996+4744158 &   267 &  4919 &  2.83 &   1.22 & $-$0.07 &  ---  &  ---  &  1.89  &  ---  &  ---  &  1.97  &  ---  &  1.93$\pm$0.06  \\
IC 1369 &  2M21121345+4745256 &   385 &  4968 &  2.55 &   1.50 & $-$0.12 &  1.79  &  ---  &  ---  &  ---  &  ---  &  1.98  &  ---  &  1.88$\pm$0.13  \\
King 2   &  2M00510072+5810562 &   232 &  4062 &  1.66 &   1.24 & $-$0.36 &  ---  &  1.61  &  1.50  &  1.51  &  ---  &  1.53  &  ---  &  1.53$\pm$0.05  \\
King 5  &  2M03142548+5247355 &   619 &  4209 &  1.79 &   1.23 & $-$0.16 &  ---  &  1.86  &  1.76  &  ---  &  ---  &  1.85  &  ---  &  1.82$\pm$0.06  \\
King 7  &  2M03590443+5148003 &   519 &  4895 &  2.45 &   1.50 & $-$0.15 &  2.01  &  1.98  &  ---  &  ---  &  ---  &  2.02  &  ---  &  2.00$\pm$0.02  \\
King 7  &  2M03591013+5145193 &   291 &  4706 &  2.17 &   1.48 & $-$0.18 &  1.99  &  ---  &  1.90  &  ---  &  ---  &  1.93  &  ---  &  1.94$\pm$0.05  \\
King 7  &  2M03591747+5147014 &   522 &  4317 &  1.86 &   1.33 & $-$0.13 &  1.94  &  ---  &  1.89  &  ---  &  ---  &  1.96  &  ---  &  1.93$\pm$0.04  \\
King 7  &  2M03592828+5148425 &   409 &  4848 &  2.40 &   1.51 & $-$0.18 &  2.04  &  ---  &  ---  &  ---  &  ---  &  2.04  &  ---  &  2.04$\pm$0.00  \\
NGC 188   &  2M00415197+8527070 &   409 &  4609 &  2.74 &   1.12 &  0.11 &  1.87  &  ---  &  ---  &  ---  &  ---  &  1.86  &  ---  &  1.87$\pm$0.01  \\
NGC 188   &  2M00422570+8516219 &   272 &  4562 &  2.91 &   1.02 &  0.09 &  1.86  &  ---  &  ---  &  ---  &  ---  &  1.79  &  ---  &  1.83$\pm$0.05  \\
NGC 188   &  2M00444460+8532163 &   243 &  4791 &  3.31 &   1.00 &  0.11 &  1.56  &  ---  &  ---  &  ---  &  1.69  &  ---  &  ---  &  1.62$\pm$0.09  \\
NGC 188   &  2M00472975+8524140 &   362 &  4661 &  2.97 &   1.03 &  0.13 &  1.59  &  ---  &  ---  &  ---  &  ---  &  1.86  &  ---  &  1.72$\pm$0.19  \\
NGC 188   &  2M00512176+8512377 &   207 &  4665 &  3.04 &   1.05 &  0.08 &  1.86  &  ---  &  ---  &  ---  &  ---  &  ---  &  1.89  &  1.88$\pm$0.02  \\
NGC 188   &  2M00533497+8511145 &   373 &  4650 &  3.02 &   0.94 &  0.09 &  ---  &  ---  &  ---  &  ---  &  ---  &  1.81  &  ---  &  1.81  \\
NGC 188   &  2M00533572+8520583 &   280 &  4517 &  2.81 &   1.06 &  0.10 &  1.87  &  ---  &  ---  &  ---  &  ---  &  1.76  &  ---  &  1.82$\pm$0.08  \\
NGC 188   &  2M00541152+8515231 &   460 &  4621 &  2.72 &   1.12 &  0.09 &  1.82  &  ---  &  ---  &  ---  &  ---  &  1.82  &  1.82  &  1.82$\pm$0.00  \\
NGC 188   &  2M00543664+8501152 &   483 &  4637 &  2.78 &   1.14 &  0.10 &  ---  &  ---  &  ---  &  ---  &  ---  &  1.87  &  1.91  &  1.89$\pm$0.03  \\
NGC 188  &  2M00571844+8510288 &   373 &  4573 &  2.72 &   1.13 &  0.10 &  ---  &  ---  &  ---  &  ---  &  ---  &  1.80  &  1.95  &  1.88$\pm$0.11  \\
NGC 752  &  2M01562163+3736084 &  1137 &  4814 &  3.00 &   1.06 & $-$0.04 &  ---  &  ---  &  ---  &  ---  &  1.78  &  1.92  &  ---  &  1.85$\pm$0.10  \\
NGC 1193  &  2M03060593+4421203 &   112 &  4660 &  2.57 &   1.16 & $-$0.33 &  ---  &  1.38  &  ---  &  ---  &  ---  &  ---  &  1.54  &  1.46$\pm$0.11  \\
NGC 1193  &  2M03060808+4423347 &    90 &  4718 &  2.69 &   1.10 & $-$0.34 &  ---  &  ---  &  ---  &  ---  &  1.56  &  ---  &  ---  &  1.56  \\
NGC 1245  &  2M03141134+4709173 &   388 &  4481 &  2.28 &   1.21 & $-$0.14 &  ---  &  ---  &  1.71  &  ---  &  1.90  &  1.81  &  ---  &  1.81$\pm$0.10  \\
NGC 1798  &  2M05112446+4740027 &   259 &  4399 &  2.13 &   1.20 & $-$0.27 &  1.65  &  ---  &  ---  &  ---  &  1.73  &  1.74  &  ---  &  1.71$\pm$0.05  \\
NGC 1798  &  2M05113666+4741482 &   200 &  4656 &  2.45 &   1.22 & $-$0.27 &  ---  &  1.82  &  ---  &  ---  &  1.66  &  1.76  &  ---  &  1.75$\pm$0.08  \\
NGC 1798  &  2M05113768+4742329 &   138 &  4762 &  2.55 &   1.21 & $-$0.27 &  ---  &  1.62  &  ---  &  ---  &  1.69  &  ---  &  ---  &  1.66$\pm$0.05  \\
NGC 1798  &  2M05114006+4739238 &   182 &  4692 &  2.52 &   1.18 & $-$0.24 &  ---  &  1.80  &  ---  &  ---  &  1.73  &  1.80  &  ---  &  1.78$\pm$0.04  \\
NGC 1798  &  2M05114134+4740406 &   115 &  4821 &  2.70 &   1.19 & $-$0.25 &  ---  &  1.66  &  1.59  &  ---  &  1.80  &  ---  &  ---  &  1.68$\pm$0.11  \\
NGC 1798  &  2M05114626+4743422 &   181 &  4603 &  2.40 &   1.22 & $-$0.26 &  1.70  &  ---  &  1.62  &  ---  &  1.75  &  1.67  &  ---  &  1.68$\pm$0.05  \\
NGC 1907  &  2M05280420+3519163 &   366 &  4941 &  2.89 &   1.22 & $-$0.08 &  ---  &  1.93  &  ---  &  ---  &  1.92  &  1.94  &  ---  &  1.93$\pm$0.01  \\
NGC 2158  &  2M06070155+2401470 &    79 &  4898 &  2.80 &   1.26 & $-$0.21 &  1.76  &  1.86  &  ---  &  ---  &  1.74  &  1.71  &  ---  &  1.77$\pm$0.07  \\
NGC 2158  &  2M06070415+2409180 &    56 &  4957 &  2.84 &   1.26 & $-$0.18 &  ---  &  ---  &  ---  &  ---  &  1.81  &  1.84  &  ---  &  1.83$\pm$0.02  \\
NGC 2158  &  2M06071494+2407517 &   143 &  4464 &  2.35 &   1.22 & $-$0.22 &  1.70  &  ---  &  1.59  &  ---  &  1.76  &  1.76  &  ---  &  1.70$\pm$0.08  \\
NGC 2158  &  2M06071696+2402007 &    72 &  4920 &  2.86 &   1.23 & $-$0.19 &  1.76  &  1.81  &  ---  &  ---  &  1.66  &  ---  &  ---  &  1.74$\pm$0.08  \\
NGC 2158  &  2M06071787+2405542 &   164 &  4342 &  2.17 &   1.22 & $-$0.20 &  ---  &  ---  &  1.69  &  ---  &  1.83  &  1.80  &  1.84  &  1.79$\pm$0.07  \\
NGC 2158  &  2M06071913+2400148 &    71 &  4974 &  2.90 &   1.21 & $-$0.21 &  1.78  &  1.87  &  ---  &  ---  &  1.65  &  1.75  &  ---  &  1.76$\pm$0.09  \\
NGC 2158  &  2M06072041+2407463 &    62 &  4903 &  2.81 &   1.21 & $-$0.24 &  ---  &  1.75  &  ---  &  ---  &  1.62  &  1.68  &  ---  &  1.68$\pm$0.07  \\
NGC 2158  &  2M06072443+2400524 &    65 &  4877 &  2.89 &   1.22 & $-$0.23 &  ---  &  ---  &  ---  &  ---  &  1.71  &  1.85  &  ---  &  1.78$\pm$0.10  \\
NGC 2158  &  2M06072624+2409568 &    75 &  4990 &  3.04 &   1.14 & $-$0.21 &  ---  &  1.80  &  1.64  &  ---  &  1.88  &  1.80  &  ---  &  1.78$\pm$0.10  \\
NGC 2158  &  2M06072907+2402151 &    87 &  4912 &  2.98 &   1.20 & $-$0.18 &  1.66  &  ---  &  1.81  &  ---  &  1.76  &  1.88  &  ---  &  1.78$\pm$0.09  \\
NGC 2158  &  2M06072918+2408185 &    81 &  4989 &  3.03 &   1.21 & $-$0.21 &  1.73  &  ---  &  ---  &  ---  &  1.69  &  1.75  &  ---  &  1.72$\pm$0.03  \\
NGC 2158  &  2M06073636+2405001 &    74 &  4989 &  3.06 &   1.20 & $-$0.19 &  1.77  &  1.88  &  ---  &  ---  &  1.81  &  ---  &  ---  &  1.82$\pm$0.06  \\
NGC 2158  &  2M06073917+2409098 &    73 &  4970 &  2.98 &   1.21 & $-$0.17 &  ---  &  1.79  &  ---  &  ---  &  1.76  &  1.74  &  ---  &  1.76$\pm$0.03  \\
NGC 2158  &  2M06073998+2403546 &    79 &  4962 &  3.00 &   1.18 & $-$0.22 &  ---  &  1.74  &  ---  &  ---  &  1.88  &  ---  &  ---  &  1.81$\pm$0.10  \\
NGC 2158  &  2M06074162+2405540 &    67 &  4864 &  2.91 &   1.21 & $-$0.23 &  ---  &  1.82  &  ---  &  ---  &  1.70  &  ---  &  ---  &  1.76$\pm$0.08  \\
NGC 2158  &  2M06074272+2402514 &    74 &  4982 &  3.03 &   1.26 & $-$0.26 &  ---  &  1.78  &  ---  &  ---  &  1.73  &  ---  &  ---  &  1.76$\pm$0.04  \\
NGC 2158  &  2M06075243+2403561 &    77 &  4968 &  3.05 &   1.17 & $-$0.22 &  ---  &  ---  &  ---  &   --- &  1.79  &  1.77  &  ---  &  1.78$\pm$0.01  \\
NGC 2204  &  2M06151360-1841498 &   148 &  4937 &  2.83 &   1.25 & $-$0.28 &  ---  &  ---  &  ---  &  ---  &  1.70  &  ---  &  ---  &  1.70  \\
NGC 2204  &  2M06152142-1835512 &   264 &  4655 &  2.54 &   1.25 & $-$0.24 &  ---  &  ---  &  ---  &  1.76  &  1.81  &  1.79  &  ---  &  1.79$\pm$0.03  \\
NGC 2204  &  2M06153043-1838239 &   163 &  4695 &  2.67 &   1.18 & $-$0.26 &  ---  &  ---  &  ---  &  ---  &  1.70  &  1.61  &  ---  &  1.66$\pm$0.06  \\
NGC 2204  &  2M06153192-1839369 &   322 &  4428 &  2.18 &   1.24 & $-$0.28 &  ---  &  ---  &  1.61  &  1.68  &  1.78  &  1.75  &  ---  &  1.70$\pm$0.08  \\
NGC 2204  &  2M06153696-1836091 &   133 &  4999 &  2.92 &   1.25 & $-$0.26 &  ---  &  ---  &  ---  &  ---  &  ---  &  1.61  &  ---  &  1.61  \\
NGC 2204  &  2M06154970-1837393 &   271 &  4473 &  2.29 &   1.24 & $-$0.27 &  ---  &  ---  &  ---  &  ---  &  1.76  &  1.80  &  ---  &  1.78$\pm$0.03  \\
		\hline
	\end{tabular}
\end{table*}

\begin{table*}
	\centering
	\tiny
	\begin{tabular}{ccccccccccccccc} 
	    \multicolumn{15}{c}{Table \ref{tab:star_abundances_table} (continued).}\\
		\hline\hline
		\multicolumn{7}{c}{} & \multicolumn{7}{c}{Ce II absorption lines (\AA)} & \\
        \cline{8-14}
		Cluster & ID & SNR & $T_{\rm eff}$ (K) & $\log{g}$ & $\xi$ (km\,s$^{-1}$) & [Fe/H] & 15277.6 & 15784.8 & 15977.1 & 16327.3 & 16376.5 & 16595.2 & 16722.6 & $<$A(Ce)$>$ $\pm \sigma$ \\
		\hline
NGC 2243  &  2M06292300-3117299 &   189 &  4967 &  2.77 &   1.32 & $-$0.51 &  1.32  &  ---  &  1.31  &  ---  &  1.38  &  ---  &  ---  &  1.32$\pm$0.05  \\
NGC 2243  &  2M06292939-3115459 &   140 &  4979 &  2.80 &   1.30 & $-$0.46 &  1.47  &  ---  &  ---  &  ---  &  1.48  &  1.38  &  ---  &  1.44$\pm$0.06  \\
NGC 2243  &  2M06293009-3116587 &   293 &  4576 &  2.31 &   1.24 & $-$0.51 &  1.34  &  ---  &  1.31  &  ---  &  1.40  &  1.41  &  ---  &  1.37$\pm$0.05  \\
NGC 2243  &  2M06293525-3115470 &    60 &  4951 &  3.07 &   1.26 & $-$0.45 &  ---  &  ---  &  ---  &  ---  &  1.36  &  ---  &  ---  &  1.36  \\
NGC 2243  &  2M06293565-3117110 &   124 &  4969 &  2.78 &   1.28 & $-$0.45 &  ---  &  ---  &  1.38  &  ---  &  1.46  &  ---  &  ---  &  1.42$\pm$0.06  \\
NGC 2243  &  2M06294150-3114360 &   172 &  4678 &  2.56 &   1.16 & $-$0.44 &  1.35  &  ---  &  ---  &  ---  &  1.45  &  1.52  &  ---  &  1.44$\pm$0.09  \\
NGC 2243  &  2M06294583-3115382 &   231 &  4928 &  2.74 &   1.30 & $-$0.46 &  1.34  &  ---  &  ---  &  ---  &  1.37  &  1.49  &  ---  &  1.40$\pm$0.08  \\
NGC 2243  &  2M06295100-3114428 &   100 &  4886 &  3.36 &   0.71 & $-$0.41 &  ---  &  ---  &  ---  &  ---  &  1.58  &  ---  &  ---  &  1.58  \\
NGC 2304  &  2M06550345+1759521 &   217 &  4770 &  2.76 &   1.15 & $-$0.14 &  1.85  &  ---  &  1.82  &  ---  &  1.87  &  1.87  &  ---  &  1.85$\pm$0.02  \\
NGC 2324  &  2M07035166+0106381 &   350 &  4902 &  2.74 &   1.24 & $-$0.20 &  1.73  &  ---  &  1.82  &  ---  & 1.86   &  1.91  &  ---  &  1.83$\pm$0.08  \\
NGC 2324  &  2M07040031+0058168 &   217 &  4431 &  2.06 &   1.29 & $-$0.16 &  1.78  &  ---  &  1.84  &  ---  &  1.94  &  1.86  &  ---  &  1.86$\pm$0.07  \\
NGC 2420  &  2M07380545+2136507 &   219 &  4842 &  2.97 &   1.12 & $-$0.19 &  1.75  &  ---  &  ---  &  ---  &  1.58  &  1.74  &  ----  &  1.69$\pm$0.10  \\
NGC 2420  &  2M07380627+2136542 &   413 &  4691 &  2.71 &   1.16 & $-$0.24 &  ---  &  ---  &  ---  &  ---  &  1.63  &  1.68  &  ----  &  1.65$\pm$0.04  \\
NGC 2420  &  2M07381507+2134589 &  1076 &  4091 &  1.69 &   1.22 & $-$0.24 &  ---  &  ---  &  1.63  &  ---  &  1.79  &  1.74  &  ----  &  1.72$\pm$0.08  \\
NGC 2420  &  2M07381549+2138015 &   322 &  4898 &  2.91 &   1.16 & $-$0.19 &  1.65  &  ---  &  ---  &  ---  &  1.63  &  1.74  &  ----  &  1.67$\pm$0.06  \\
NGC 2420  &  2M07382148+2135050 &   221 &  4870 &  3.00 &   1.14 & $-$0.18 &  ---  &  ---  &  1.67  &  ---  &  1.72  &  1.78  &  ----  &  1.72$\pm$0.06  \\
NGC 2420  &  2M07382195+2135508 &   270 &  4896 &  2.93 &   1.15 & $-$0.14 &  ---  &  ---  &  1.48  &  ---  &  1.83  &  ---  &  ----  &  1.66$\pm$0.25  \\
NGC 2420  &  2M07382670+2128514 &   130 &  4888 &  2.96 &   1.18 & $-$0.16 &  ---  &  1.76  &  1.76  &  ---  &  1.79  &  1.81  &  ----  &  1.78$\pm$0.02  \\
NGC 2420  &  2M07382696+2138244 &   315 &  4836 &  2.81 &   1.18 & $-$0.17 &  1.76  &  ---  &  1.71  &  ---  &  1.78  &  1.83  &  ----  &  1.77$\pm$0.05  \\
NGC 2420  &  2M07382984+2134509 &   229 &  4781 &  2.93 &   1.11 & $-$0.19 &  1.61  &  ---  &  ---  &  ---  &  1.64  &  1.61  &  ----  &  1.62$\pm$0.02  \\
NGC 2420  &  2M07383760+2134119 &   269 &  4912 &  2.94 &   1.15 & $-$0.18 &  1.66  &  ---  &  1.64  &  ---  &  1.66  &  1.85  &  ----  &  1.70$\pm$0.10  \\
NGC 2682  &  2M08504964+1135089 &   342 &  4726 &  3.00 &   1.08 &  0.04 &  ---  &  ---  &  1.71  &  ---  &  1.73  &  1.81  &  ---  &  1.75$\pm$0.05  \\
NGC 2682  &  2M08510839+1147121 &   171 &  4946 &  3.46 &   1.04 &  0.04 &  ---  &  ---  &  ---  &  ---  &  1.71  &  1.88  &  ---  &  1.80$\pm$0.12  \\
NGC 2682  &  2M08511269+1152423 &   768 &  4736 &  2.82 &   1.11 &  0.01 &  1.77  &  ---  &  1.64  &  ---  &  1.74  &  1.87  &  ---  &  1.76$\pm$0.09  \\
NGC 2682  &  2M08511704+1150464 &   371 &  4698 &  2.95 &   1.08 & $-$0.03 &  ---  &  1.82  &  1.62  &  ---  &  1.65  &  1.78  &  ---  &  1.72$\pm$0.10  \\
NGC 2682  &  2M08511897+1158110 &   393 &  4948 &  3.38 &   1.09 &  0.02 &  ---  &  1.84  &  ---  &  ---  &  1.59  &  1.76  &  ---  &  1.73$\pm$0.13  \\
NGC 2682  &  2M08512156+1146061 &   645 &  4757 &  3.04 &   1.08 &  0.05 &  ---  &  1.79  &  ---  &  ---  &  1.72  &  1.81  &  ---  &  1.77$\pm$0.05  \\
NGC 2682  &  2M08512280+1148016 &  1001 &  4728 &  2.78 &   1.16 &  0.04 &  ---  &  1.92  &  1.65  &  ---  &  1.81  &  1.81  &  ---  &  1.80$\pm$0.11  \\
NGC 2682  &  2M08512618+1153520 &   785 &  4747 &  2.81 &   1.13 &  0.00 &  ---  &  ---  &  ---  &  ---  &  1.72  &  1.82  &  ---  &  1.77$\pm$0.07  \\
NGC 2682  &  2M08512898+1150330 &   949 &  4696 &  2.77 &   1.13 &  0.02 &  ---  &  ---  &  ---  &  ---  &  1.71  &  1.82  &  ---  &  1.76$\pm$0.08  \\
NGC 2682  &  2M08513577+1153347 &   205 &  4932 &  3.41 &   1.01 &  0.02 &  ---  &  1.80  &  ---  &  ---  &  1.52  &  1.69  &  ---  &  1.67$\pm$0.14  \\
NGC 2682  &  2M08513938+1151456 &   390 &  4898 &  3.32 &   1.06 &  0.03 &  ---  &  1.82  &  ---  &  ---  &  1.59  &  1.64  &  ---  &  1.68$\pm$0.12  \\
NGC 2682  &  2M08514234+1150076 &   270 &  4783 &  3.15 &   1.07 &  0.03 &  ---  &  ---  &  ---  &  ---  &  1.64  &  1.84  &  ---  &  1.74$\pm$0.14  \\
NGC 2682  &  2M08514235+1151230 &   668 &  4716 &  2.97 &   1.10 &  0.00 &  1.76  &  1.77  &  ---  &  ---  &  1.62  &  1.78  &  ---  &  1.73$\pm$0.08  \\
NGC 2682  &  2M08514388+1156425 &   950 &  4751 &  2.80 &   1.16 &  0.01 &  1.62  &  1.89  &  ---  &  ---  &  1.73  &  1.80  &  ---  &  1.76$\pm$0.11  \\
NGC 2682  &  2M08514507+1147459 &   466 &  4778 &  3.06 &   1.10 &  0.01 &  ---  &  ---  &  ---  &  ---  &  1.70  &  1.77  &  ---  &  1.74$\pm$0.05  \\
NGC 2682  &  2M08515952+1155049 &   996 &  4748 &  2.81 &   1.13 &  0.00 &  ---  &  ---  &  ---  &  ---  &  1.72  &  1.78  &  ---  &  1.75$\pm$0.04  \\
NGC 2682  &  2M08521097+1131491 &   667 &  4563 &  2.76 &   1.07 &  0.04 &  1.78  &  1.78  &  ---  &  ---  &  1.68  &  1.79  &  ---  &  1.76$\pm$0.05  \\
NGC 2682  &  2M08521656+1119380 &  1058 &  4326 &  2.31 &   1.10 &  0.01 &  ---  &  1.75  &  ---  &  ---  &  1.73  &  1.75  &  ---  &  1.74$\pm$0.01  \\
NGC 2682  &  2M08521856+1144263 &   503 &  4708 &  2.81 &   1.14 &  0.02 &  1.79  &  ---  &  1.67  &  ---  &  1.80  &  1.76  &  ---  &  1.76$\pm$0.06  \\
NGC 2682  &  2M08522003+1127362 &   260 &  4975 &  3.50 &   1.06 &  0.02 &  ---  &  1.76  &  ---  &  ---  &  1.50  &  1.82  &  ---  &  1.69$\pm$0.17  \\
NGC 2682  &  2M08522636+1141277 &   197 &  4980 &  3.54 &   0.85 &  0.03 &  ---  &  ---  &  ---  &  ---  &  1.58  &  1.78  &  ---  &  1.68$\pm$0.14  \\
NGC 4337 &  2M12235244-5806564 &   191 &  4885 &  3.14 &   1.12 &  0.23 &  2.06  &  ---  &  ---  &  ---  &  1.91  &  2.02  &  2.00  &  2.00$\pm$0.06  \\
NGC 4337 &  2M12235665-5807252 &   159 &  4857 &  3.17 &   1.09 &  0.26 &  ---  &  ---  &  ---  &  ---  &  2.00  &  2.10  &  ---  &  2.05$\pm$0.07  \\
NGC 4337 &  2M12240101-5807554 &   585 &  4286 &  2.26 &   1.19 &  0.22 &  1.86  &  ---  &  ---  &  ---  &  2.07  &  1.96  &  2.10  &  2.00$\pm$0.11  \\
NGC 4337 &  2M12240488-5805099 &   168 &  4880 &  3.23 &   1.09 &  0.31 &  ---  &  ---  &  ---  &  ---  &  1.99  &  2.17  &  ---  &  2.08$\pm$0.13  \\
NGC 4337 &  2M12240586-5807152 &   156 &  4906 &  3.18 &   1.09 &  0.19 &  ---  &  ---  &  ---  &  ---  &  ---  &  2.01  &  ---  &  2.01  \\
NGC 4337 &  2M12241575-5808502 &   200 &  4889 &  3.16 &   1.12 &  0.22 &  ---  &  ---  &  ---  &  ---  &  1.84  &  2.00  &  ---  &  1.92$\pm$0.11  \\
NGC 6705 &  2M18505494-0616182 &   394 &  4650 &  2.84 &   1.02 &  0.12 &  ---  &  ---  &  1.99  &  ---  &  2.05  &  2.00  &  ---  &  2.01$\pm$0.03  \\
NGC 6705 &  2M18505944-0612435 &   376 &  4870 &  3.15 &   0.83 &  0.09 &  2.09  &  ---  &  1.89  &  ---  &  ---  &  2.07  &  ---  &  2.02$\pm$0.11  \\
NGC 6705 &  2M18510092-0614564 &   457 &  4783 &  2.89 &   1.09 &  0.10 &  ---  &  ---  &  2.00  &  ---  &  ---  &  2.03  &  ---  &  2.01$\pm$0.02  \\
NGC 6705 &  2M18510399-0620414 &   455 &  4717 &  2.72 &   1.20 &  0.16 &  ---  &  1.99  &  1.94  &  ---  &  ---  &  2.07  &  ---  &  2.00$\pm$0.07  \\
NGC 6705 &  2M18510626-0615134 &   438 &  4775 &  2.80 &   1.18 &  0.16 &  ---  &  ---  &  1.93  &  ---  &  ---  &  2.07  &  ---  &  2.00$\pm$0.10  \\
NGC 6705 &  2M18510661-0612442 &   453 &  4758 &  2.88 &   1.14 &  0.17 &  ---  &  2.04  &  2.06  &  ---  &  ---  &  2.12  &  ---  &  2.07$\pm$0.04  \\
NGC 6705 &  2M18510786-0617119 &   350 &  4778 &  3.00 &   1.10 &  0.05 &  ---  &  ---  &  2.05  &  ---  &  ---  &  ---  &  ---  &  2.05  \\
NGC 6705 &  2M18511048-0615470 &   475 &  4749 &  2.87 &   1.12 &  0.09 &  2.09  &  ---  &  2.08  &  ---  &  ---  &  1.98  &  ---  &  2.05$\pm$0.06  \\
NGC 6705 &  2M18511452-0616551 &   297 &  4811 &  3.06 &   1.02 &  0.11 &  ---  &  ---  &  2.06  &  ---  &  ---  &  2.09  &  ---  &  2.08$\pm$0.02  \\
NGC 6705 &  2M18511571-0618146 &   465 &  4731 &  2.74 &   1.21 &  0.16 &  1.98  &  2.01  &  1.91  &  ---  &  2.07  &  1.98  &  ---  &  1.99$\pm$0.06  \\
NGC 6791 &  2M19203005+3750191 &    64 &  4470 &  2.80 &   1.08 &  0.35 &  2.09  &  ---  &  2.11  &  ---  &  2.15  &  ---  &  ---  &  2.12$\pm$0.03  \\
NGC 6791 &  2M19203485+3746298 &    66 &  4448 &  2.67 &   1.06 &  0.36 &  ---  &  ---  &  2.04  &  ---  &  2.10  &  ---  &  2.10  &  2.08$\pm$0.03  \\
NGC 6791 &  2M19203784+3745249 &    41 &  4400 &  2.82 &   0.92 &  0.34 &  ---  &  ---  &  2.17  &  ---  &  ---  &  ---  &  ---  &  2.17  \\
NGC 6791 &  2M19203934+3748048 &    77 &  4212 &  2.45 &   1.09 &  0.30 &  ---  &  2.03  &  2.06  &  ---  &  ---  &  2.05  &  2.03  &  2.04$\pm$0.02  \\
NGC 6791 &  2M19204517+3744339 &    45 &  4356 &  2.76 &   0.75 &  0.40 &  ---  &  ---  &  2.21  &  ---  &  ---  &  2.06  &  ---  &  2.13$\pm$0.11  \\
NGC 6791 &  2M19205287+3745331 &    67 &  4474 &  2.80 &   1.06 &  0.36 &  ---  &  ---  &  ---  &  ---  &  ---  &  ---  &  2.04  &  2.04  \\
NGC 6791 &  2M19205368+3750236 &    58 &  4483 &  2.79 &   1.05 &  0.41 &  ---  &  ---  &  ---  &  ---  &  ---  &  2.04  &  ---  &  2.04  \\
NGC 6791 &  2M19205530+3743152 &   124 &  4189 &  2.44 &   1.00 &  0.34 &  ---  &  ---  &  1.98  &  ---  &  ---  &  2.03  &  ---  &  2.00$\pm$0.04  \\
NGC 6791 &  2M19205629+3744334 &    63 &  4443 &  2.72 &   1.07 &  0.39 &  2.11  &  ---  &  2.06  &  ---  &  ---  &  2.00  &  ---  &  2.06$\pm$0.06  \\
NGC 6791 &  2M19210052+3750188 &    67 &  4429 &  2.98 &   0.75 &  0.28 &  ---  &  ---  &  2.15  &  ---  &  ---  &  ---  &  2.23  &  2.19$\pm$0.06  \\
NGC 6791 &  2M19210086+3745339 &    69 &  4387 &  2.64 &   1.10 &  0.39 &  ---  &  2.14  &  2.02  &  ---  &  ---  &  2.17  &  2.16  &  2.12$\pm$0.07  \\
NGC 6791 &  2M19210112+3742134 &   137 &  4156 &  2.37 &   1.08 &  0.32 &  ---  &  ---  &  1.97  &  ---  &  ---  &  1.93  &  2.15  &  2.02$\pm$0.12  \\
NGC 6791 &  2M19210426+3747187 &   134 &  4061 &  2.00 &   1.15 &  0.33 &  2.09  &  ---  &  1.95  &  2.15  &  ---  &  ---  &  2.04  &  2.06$\pm$0.08  \\
NGC 6791 &  2M19210483+3741036 &    98 &  4480 &  2.84 &   0.86 &  0.41 &  ---  &  ---  &  2.13  &  ---  &  ---  &  2.25  &  2.08  &  2.15$\pm$0.09  \\
NGC 6791 &  2M19210604+3752049 &    79 &  4474 &  2.98 &   0.99 &  0.35 &  ---  &  1.93  &  ---  &  ---  &  ---  &  2.20  &  ---  &  2.06$\pm$0.19  \\
NGC 6791 &  2M19210629+3744596 &    67 &  4438 &  2.73 &   1.02 &  0.36 &  ---  &  ---  &  1.98  &  ---  &  ---  &  2.07  &  2.07  &  2.04$\pm$0.05  \\
NGC 6791 &  2M19211007+3750008 &   103 &  4435 &  2.85 &   1.04 &  0.32 &  2.16  &  ---  &  ---  &  ---  &  2.12  &  ---  &  ---  &  2.14$\pm$0.03  \\
NGC 6791 &  2M19211300+3743005 &    63 &  4439 &  2.91 &   1.00 &  0.36 &  ---  &  ---  &  ---  &  ---  &  2.14  &  2.02  &  ---  &  2.08$\pm$0.08  \\
NGC 6791 &  2M19203266+3746221 &   113 &  4257 &  2.53 &   1.10 &  0.37 &  2.13  &  2.23  &  ---  &  ---  &  2.10  &  ---  &  ---  &  2.15$\pm$0.07  \\
NGC 6791 &  2M19204356+3747019 &   112 &  4255 &  2.41 &   1.14 &  0.30 &  2.05  &  ---  &  1.90  &  ---  &  2.01  &  1.91  &  2.12  &  2.00$\pm$0.09  \\
NGC 6791 &  2M19204965+3744077 &   118 &  4461 &  2.64 &   1.11 &  0.35 &  ---  &  ---  &  ---  &  ---  &  1.99  &  ---  &  2.00  &  2.00$\pm$0.01  \\
NGC 6791 &  2M19205784+3747067 &   102 &  4486 &  2.95 &   1.06 &  0.37 &  2.13  &  ---  &  ---  &  ---  &  ---  &  1.88  &  ---  &  2.00$\pm$0.18  \\
NGC 6791 &  2M19205874+3743130 &   109 &  4449 &  2.67 &   1.10 &  0.38 &  2.05  &  ---  &  2.03  &  ---  &  ---  &  2.10  &  ---  &  2.06$\pm$0.04  \\
NGC 6791 &  2M19210086+3746396 &   117 &  4450 &  2.60 &   1.14 &  0.35 &  ---  &  ---  &  1.79  &  ---  &  2.00  &  ---  &  ---  &  1.90$\pm$0.15  \\
NGC 6791 &  2M19211725+3743187 &   118 &  4400 &  2.56 &   1.12 &  0.40 &  ---  &  ---  &  2.01  &  ---  &  ---  &  2.25  &  2.22  &  2.16$\pm$0.13  \\
NGC 6811 &  2M19373462+4624098 &   435 &  4944 &  3.03 &   1.14 & $-$0.02 &  ---  &  1.85  &  1.82  &  ---  &  ---  &  1.92  &  ---  &  1.86$\pm$0.05  \\
		\hline
	\end{tabular}
\end{table*}

\begin{table*}
	\centering
	\tiny
	\begin{tabular}{ccccccccccccccc} 
	    \multicolumn{15}{c}{Table \ref{tab:star_abundances_table} (continued).}\\
		\hline\hline
		\multicolumn{7}{c}{} & \multicolumn{7}{c}{Ce II absorption lines (\AA)} & \\
        \cline{8-14}
		Cluster & ID & SNR & $T_{\rm eff}$ (K) & $\log{g}$ & $\xi$ (km\,s$^{-1}$) & [Fe/H] & 15277.6 & 15784.8 & 15977.1 & 16327.3 & 16376.5 & 16595.2 & 16722.6 & $<$A(Ce)$>$ $\pm \sigma$ \\
		\hline
NGC 6819 &  2M19404803+4008085 &   324 &  4507 &  2.64 &   1.11 &  0.02 &  ---  &  1.82  &  ---  &  ---  &  ---  &  1.82  &  ---  &  1.82$\pm$0.0  \\
NGC 6819 &  2M19404965+4014313 &   223 &  4675 &  2.89 &   1.11 &  0.05 &  ---  &  1.88  &  ---  &  ---  &  1.73  &  1.82  &  ---  &  1.81$\pm$0.08  \\
NGC 6819 &  2M19405020+4013109 &   237 &  4738 &  2.90 &   1.10 &  0.09 &  ---  &  ---  &  ---  &  ---  &  1.93  &  1.93  &  ---  &  1.93$\pm$0.0  \\
NGC 6819 &  2M19405601+4013395 &   141 &  4885 &  3.22 &   0.86 &  0.06 &  ---  &  1.69  &  ---  &  ---  &  1.84  &  ---  &  ---  &  1.76$\pm$0.11  \\
NGC 6819 &  2M19405797+4008174 &   288 &  4826 &  2.95 &   1.13 &  0.09 &  1.96  &  1.95  &  ---  &  ---  &  1.84  &  1.94  &  ---  &  1.92$\pm$0.06  \\
NGC 6819 &  2M19410524+4014042 &   130 &  4778 &  2.89 &   1.12 &  0.08 &  ---  &  ---  &  ---  &  ---  &  1.89  &  2.09  &  2.00  &  1.99$\pm$0.10  \\
NGC 6819 &  2M19410622+4010532 &   135 &  4867 &  3.18 &   1.04 &  0.07 &  ---  &  ---  &  ---  &  ---  &  1.75  &  1.75  &  ---  &  1.75$\pm$0.0  \\
NGC 6819 &  2M19410858+4013299 &   249 &  4751 &  2.84 &   1.13 &  0.08 &  ---  &  1.80  &  ---  &  ---  &  1.90  &  1.94  &  ---  &  1.88$\pm$0.07  \\
NGC 6819 &  2M19410926+4014436 &   257 &  4762 &  2.91 &   1.08 &  0.07 &  ---  &  ---  &  ---  &  ---  &  1.88  &  1.84  &  ---  &  1.86$\pm$0.03  \\
NGC 6819 &  2M19410991+4015495 &   148 &  4737 &  2.80 &   1.13 &  0.00 &  ---  &  ---  &  ---  &  ---  &  1.81  &  1.83  &  1.98  &  1.87$\pm$0.09  \\
NGC 6819 &  2M19411102+4011116 &   370 &  4944 &  2.87 &   1.24 &  0.06 &  1.88  &  ---  &  ---  &  ---  &  ---  &  2.01  &  2.06  &  1.98$\pm$0.09  \\
NGC 6819 &  2M19411115+4011422 &   305 &  4611 &  2.57 &   1.18 &  0.09 &  1.82  &  1.82  &  ---  &  ---  &  ---  &  2.02  &  ---  &  1.89$\pm$0.12  \\
NGC 6819 &  2M19411279+4012238 &   188 &  4769 &  2.83 &   1.15 &  0.07 &  ---  &  1.75  &  ---  &  ---  &  1.88  &  ---  &  ---  &  1.82$\pm$0.09  \\
NGC 6819 &  2M19411345+4011561 &   171 &  4767 &  2.86 &   1.14 &  0.03 &  ---  &  ---  &  ---  &  ---  &  1.76  &  ---  &  ---  &  1.76  \\
NGC 6819 &  2M19411355+4012205 &   260 &  4795 &  2.86 &   1.14 &  0.03 &  ---  &  1.91  &  ---  &  ---  &  1.77  &  1.82  &  ---  &  1.83$\pm$0.07  \\
NGC 6819 &  2M19411476+4011008 &   281 &  4892 &  3.00 &   1.15 &  0.10 &  ---  &  ---  &  ---  &  ---  &  1.90  &  1.92  &  ---  &  1.91$\pm$0.01  \\
NGC 6819 &  2M19411564+4010105 &   133 &  4747 &  2.90 &   1.12 &  0.09 &  ---  &  1.82  &  ---  &  ---  &  ---  &  2.10  &  ---  &  1.96$\pm$0.20  \\
NGC 6819 &  2M19411705+4010517 &   805 &  4014 &  1.72 &   1.21 & $-$0.01 &  ---  &  1.82  &  ---  &  1.70  &  ---  &  1.73  &  1.83  &  1.77$\pm$0.06  \\
NGC 6819 &  2M19411893+4011408 &   168 &  4659 &  2.87 &   1.09 &  0.03 &  ---  &  1.75  &  1.68  &  ---  &  1.84  &  1.85  &  ---  &  1.78$\pm$0.08  \\
NGC 6819 &  2M19412136+4011002 &   132 &  4589 &  2.75 &   1.08 &  0.04 &  ---  &  ---  &  ---  &  ---  &  1.91  &  ---  &  ---  &  1.91  \\
NGC 6819 &  2M19412147+4013573 &   251 &  4765 &  2.87 &   1.14 &  0.09 &  ---  &  1.94  &  ---  &  ---  &  1.80  &  1.95  &  1.93  &  1.90$\pm$0.07  \\
NGC 6819 &  2M19412176+4012111 &   260 &  4582 &  2.75 &   1.13 &  0.06 &  ---  &  1.84  &  ---  &  ---  &  1.78  &  1.88  &  ---  &  1.83$\pm$0.05  \\
NGC 6819 &  2M19412245+4012033 &   133 &  4957 &  3.15 &   1.08 &  0.06 &  ---  &  ---  &  ---  &  ---  &  1.80  &  ---  &  ---  &  1.80  \\
NGC 6819 &  2M19412658+4011418 &   347 &  4404 &  2.33 &   1.18 &  0.01 &  ---  &  1.89  &  ---  &  ---  &  1.87  &  1.90  &  ---  &  1.89$\pm$0.02  \\
NGC 6819 &  2M19412707+4012283 &   214 &  4525 &  2.66 &   1.10 &  0.04 &  ---  &  1.89  &  ---  &  ---  &  1.84  &  1.80  &  ---  &  1.84$\pm$0.05  \\
NGC 6819 &  2M19412915+4013040 &   185 &  4792 &  2.88 &   1.15 &  0.07 &  ---  &  1.83  &  ---  &  ---  &  1.93  &  1.93  &  ---  &  1.90$\pm$0.06  \\
NGC 6819 &  2M19412942+4014199 &   137 &  4670 &  2.88 &   1.08 &  0.05 &  ---  &  1.86  &  1.85  &  ---  &  1.83  &  1.86  &  ---  &  1.85$\pm$0.01  \\
NGC 6819 &  2M19412953+4012210 &   270 &  4737 &  2.84 &   1.16 &  0.07 &  ---  &  1.91  &  ---  &  ---  &  1.83  &  1.98  &  1.98  &  1.93$\pm$0.07  \\
NGC 6819 &  2M19413027+4015218 &   262 &  4774 &  2.87 &   1.15 &  0.06 &  ---  &  1.89  &  ---  &  ---  &  1.80  &  1.84  &  ---  &  1.84$\pm$0.05  \\
NGC 6819 &  2M19413330+4012349 &   261 &  4606 &  2.67 &   1.15 & $-$0.00 &  ---  &  1.73  &  ---  &  ---  &  1.70  &  1.75  &  ---  &  1.73$\pm$0.03  \\
NGC 7789 &  2M23554966+5639180 &   279 &  4424 &  2.35 &   1.21 & $-$0.00 &  1.83  &  1.96  &  1.74  &  ---  &  ---  &  1.84  &  ---  &  1.84$\pm$0.09  \\
NGC 7789 &  2M23562953+5648399 &   319 &  4948 &  3.12 &   1.09 & $-$0.03 &  ---  &  1.87  &  ---  &  ---  &  ---  &  1.88  &  ---  &  1.88$\pm$0.01  \\
NGC 7789 &  2M23563930+5645242 &   310 &  4966 &  3.13 &   1.10 &  0.01 &  1.80  &  ---  &  ---  &  ---  &  ---  &  1.92  &  ---  &  1.86$\pm$0.08  \\
NGC 7789 &  2M23564304+5650477 &   322 &  4929 &  3.09 &   1.12 &  0.02 &  ---  &  ---  &  ---  &  ---  &  1.91  &  1.93  &  ---  &  1.92$\pm$0.01  \\
NGC 7789 &  2M23565751+5645272 &   685 &  4531 &  2.61 &   1.11 & $-$0.00 &  1.87  &  1.92  &  ---  &  ---  &  1.88  &  1.90  &  ---  &  1.89$\pm$0.02  \\
NGC 7789 &  2M23570895+5648504 &   290 &  4981 &  3.13 &   1.15 &  0.04 &  1.82  &  1.92  &  ---  &  ---  &  1.91  &  1.88  &  ---  &  1.88$\pm$0.04  \\
NGC 7789 &  2M23571400+5640586 &   604 &  4472 &  2.49 &   1.15 & $-$0.01 &  1.80  &  ---  &  1.80  &  ---  &  ---  &  1.92  &  ---  &  1.84$\pm$0.07  \\
NGC 7789 &  2M23571728+5645333 &   128 &  4992 &  3.18 &   1.14 & $-$0.02 &  ---  &  1.94  &  ---  &  ---  &  ---  &  1.99  &  ---  &  1.96$\pm$0.04  \\
NGC 7789 &  2M23571847+5650271 &   326 &  4879 &  2.98 &   1.13 & $-$0.03 &  ---  &  1.86  &  1.84  &  ---  &  1.83  &  1.90  &  ---  &  1.86$\pm$0.03  \\
NGC 7789 &  2M23573184+5641221 &   934 &  4352 &  2.27 &   1.17 & $-$0.02 &  1.80  &  ---  &  1.83  &  1.91  &  ---  &  1.94  &  ---  &  1.87$\pm$0.07  \\
NGC 7789 &  2M23573563+5640000 &   139 &  4948 &  3.16 &   1.09 &  0.02 &  ---  &  1.99  &  ---  &  ---  &  ---  &  1.97  &  ---  &  1.98$\pm$0.01  \\
NGC 7789 &  2M23580015+5650125 &   687 &  4369 &  2.29 &   1.14 & $-$0.05 &  1.75  &  ---  &  ---  &  ---  &  ---  &  1.88  &  ---  &  1.82$\pm$0.09  \\
NGC 7789 &  2M23580275+5647208 &   289 &  4716 &  2.90 &   1.06 & $-$0.02 &  1.75  &  ---  &  ---  &  ---  &  1.91  &  1.96  &  ---  &  1.87$\pm$0.11  \\
NGC 7789 &  2M23581471+5651466 &   713 &  4251 &  2.10 &   1.18 & $-$0.03 &  1.78  &  1.83  &  1.82  &  1.81  &  ---  &  1.95  &  ---  &  1.84$\pm$0.07  \\
Ruprecht 147 &  2M19164574-1635226 &   999 &  4781 &  3.15 &   0.99 &  0.14 &  ---  &  1.76  &  ---  &  ---  &  1.81  &  1.95  &  ---  &  1.84$\pm$0.10  \\
SAI 116 &  2M11491181-6214125 &   400 &  4652 &  2.62 &   1.26 &  0.17 &  2.07  & ---   &  ---  &  ---  &  ---  &  2.11  &  ---  &  2.09$\pm$0.03  \\
SAI 116 &  2M11491918-6214038 &   358 &  4601 &  2.53 &   1.20 &  0.15 &  ---  &  ---  &  ---  &  ---  &  2.00  &  1.98  &  ---  &  1.99$\pm$0.01  \\
Teutsch 84 &  2M17041246-4206305 &   107 &  4934 &  3.18 &   1.12 &  0.21 &  ---  &  2.01  &  ---  &  ---  &  1.98  &  ---  &  ---  &  2.00$\pm$0.02  \\
Trumpler 5  &  2M06363859+0938525 &   127 &  4787 &  2.73 &   1.15 & $-$0.43 &  ---  &  ---  &  ---  &  ---  &  1.41  &  1.33  &  ---  &  1.37$\pm$0.06  \\
Trumpler 5  &  2M06364229+0925257 &   344 &  4286 &  1.96 &   1.21 & $-$0.44 &  1.39  &  ---  &  ---  &  ---  &  1.41  &  1.43  &  ---  &  1.41$\pm$0.02  \\
Trumpler 5  &  2M06364741+0919364 &    95 &  4830 &  2.80 &   1.19 & $-$0.44 &  ---  &  ---  &  ---  &  ---  &  1.36  &  ---  &  ---  &  1.36  \\
		\hline
	\end{tabular}
\end{table*}




\bibliography{ce_paper}

\begin{thebibliography}{}
\expandafter\ifx\csname natexlab\endcsname\relax\def\natexlab#1{#1}\fi
\providecommand{\url}[1]{\href{#1}{#1}}

\bibitem[{{Alvarez} \& {Plez}(1998)}]{AlvarezPlez1998}
{Alvarez}, R., \& {Plez}, B. 1998, \aap, 330, 1109

\bibitem[{{Baratella} {et~al.}(2021){Baratella}, {D'Orazi}, {Sheminova},
  {Spina}, {Carraro}, {Gratton}, {Magrini}, {Randich}, {Lugaro}, {Pignatari},
  {Romano}, {Biazzo}, {Bragaglia}, {Casali}, {Desidera}, {Frasca}, {de Silva},
  {Melo}, {Van der Swaelmen}, {Tautvai{\v{s}}ien\{\textbackslash. e\}},
  {Jim{\'e}nez-Esteban}, {Gilmore}, {Bensby}, {Smiljanic}, {Bayo},
  {Franciosini}, {Gonneau}, {Hourihane}, {Jofr{\'e}}, {Monaco}, {Morbidelli},
  {Sacco}, {Sbordone}, {Worley}, \& {Zaggia}}]{Baratella2021}
{Baratella}, M., {D'Orazi}, V., {Sheminova}, V., {et~al.} 2021, arXiv e-prints,
  arXiv:2107.12381

\bibitem[{{Battino} {et~al.}(2021){Battino}, {Lederer-Woods}, {Cseh},
  {Denissenkov}, \& {Herwig}}]{Battino2021}
{Battino}, U., {Lederer-Woods}, C., {Cseh}, B., {Denissenkov}, P., \& {Herwig},
  F. 2021, Universe, 7, 25

\bibitem[{{Battino} {et~al.}(2019){Battino}, {Tattersall}, {Lederer-Woods},
  {Herwig}, {Denissenkov}, {Hirschi}, {Trappitsch}, {den Hartogh}, {Pignatari},
  \& {NuGrid Collaboration}}]{Battino2019}
{Battino}, U., {Tattersall}, A., {Lederer-Woods}, C., {et~al.} 2019, \mnras,
  489, 1082

\bibitem[{{Battistini} \& {Bensby}(2016)}]{BattistiniBensby2016}
{Battistini}, C., \& {Bensby}, T. 2016, \aap, 586, A49

\bibitem[{{Bensby} {et~al.}(2014){Bensby}, {Feltzing}, \& {Oey}}]{Bensby2014}
{Bensby}, T., {Feltzing}, S., \& {Oey}, M.~S. 2014, \aap, 562, A71

\bibitem[{{Bisterzo} {et~al.}(2014){Bisterzo}, {Travaglio}, {Gallino},
  {Wiescher}, \& {K{\"a}ppeler}}]{Bisterzo2014}
{Bisterzo}, S., {Travaglio}, C., {Gallino}, R., {Wiescher}, M., \&
  {K{\"a}ppeler}, F. 2014, \apj, 787, 10

\bibitem[{{Blanton} {et~al.}(2017){Blanton}, {Bershady}, {Abolfathi},
  {Albareti}, {Allende Prieto}, {Almeida}, {Alonso-Garc{\'\i}a}, {Anders},
  {Anderson}, {Andrews}, {Aquino-Ort{\'\i}z}, {Arag{\'o}n-Salamanca},
  {Argudo-Fern{\'a}ndez}, {Armengaud}, {Aubourg}, {Avila-Reese}, {Badenes},
  {Bailey}, {Barger}, {Barrera-Ballesteros}, {Bartosz}, {Bates}, {Baumgarten},
  {Bautista}, {Beaton}, {Beers}, {Belfiore}, {Bender}, {Berlind}, {Bernardi},
  {Beutler}, {Bird}, {Bizyaev}, {Blanc}, {Blomqvist}, {Bolton}, {Boquien},
  {Borissova}, {van den Bosch}, {Bovy}, {Brandt}, {Brinkmann}, {Brownstein},
  {Bundy}, {Burgasser}, {Burtin}, {Busca}, {Cappellari}, {Delgado Carigi},
  {Carlberg}, {Carnero Rosell}, {Carrera}, {Chanover}, {Cherinka}, {Cheung},
  {G{\'o}mez Maqueo Chew}, {Chiappini}, {Choi}, {Chojnowski}, {Chuang},
  {Chung}, {Cirolini}, {Clerc}, {Cohen}, {Comparat}, {da Costa}, {Cousinou},
  {Covey}, {Crane}, {Croft}, {Cruz-Gonzalez}, {Garrido Cuadra}, {Cunha},
  {Damke}, {Darling}, {Davies}, {Dawson}, {de la Macorra}, {Dell'Agli}, {De
  Lee}, {Delubac}, {Di Mille}, {Diamond-Stanic}, {Cano-D{\'\i}az}, {Donor},
  {Downes}, {Drory}, {du Mas des Bourboux}, {Duckworth}, {Dwelly}, {Dyer},
  {Ebelke}, {Eigenbrot}, {Eisenstein}, {Emsellem}, {Eracleous}, {Escoffier},
  {Evans}, {Fan}, {Fern{\'a}ndez-Alvar}, {Fernandez-Trincado}, {Feuillet},
  {Finoguenov}, {Fleming}, {Font-Ribera}, {Fredrickson}, {Freischlad},
  {Frinchaboy}, {Fuentes}, {Galbany}, {Garcia-Dias},
  {Garc{\'\i}a-Hern{\'a}ndez}, {Gaulme}, {Geisler}, {Gelfand},
  {Gil-Mar{\'\i}n}, {Gillespie}, {Goddard}, {Gonzalez-Perez}, {Grabowski},
  {Green}, {Grier}, {Gunn}, {Guo}, {Guy}, {Hagen}, {Hahn}, {Hall}, {Harding},
  {Hasselquist}, {Hawley}, {Hearty}, {Gonzalez Hern{\'a}ndez}, {Ho}, {Hogg},
  {Holley-Bockelmann}, {Holtzman}, {Holzer}, {Huehnerhoff}, {Hutchinson},
  {Hwang}, {Ibarra-Medel}, {da Silva Ilha}, {Ivans}, {Ivory}, {Jackson},
  {Jensen}, {Johnson}, {Jones}, {J{\"o}nsson}, {Jullo}, {Kamble}, {Kinemuchi},
  {Kirkby}, {Kitaura}, {Klaene}, {Knapp}, {Kneib}, {Kollmeier}, {Lacerna},
  {Lane}, {Lang}, {Law}, {Lazarz}, {Lee}, {Le Goff}, {Liang}, {Li}, {Li},
  {Lian}, {Lima}, {Lin}, {Lin}, {Bertran de Lis}, {Liu}, {de Icaza Lizaola},
  {Long}, {Lucatello}, {Lundgren}, {MacDonald}, {Deconto Machado}, {MacLeod},
  {Mahadevan}, {Geimba Maia}, {Maiolino}, {Majewski}, {Malanushenko},
  {Malanushenko}, {Manchado}, {Mao}, {Maraston}, {Marques-Chaves}, {Masseron},
  {Masters}, {McBride}, {McDermid}, {McGrath}, {McGreer}, {Medina Pe{\~n}a},
  {Melendez}, {Merloni}, {Merrifield}, {Meszaros}, {Meza}, {Minchev},
  {Minniti}, {Miyaji}, {More}, {Mulchaey}, {M{\"u}ller-S{\'a}nchez}, {Muna},
  {Munoz}, {Myers}, {Nair}, {Nandra}, {Correa do Nascimento}, {Negrete},
  {Ness}, {Newman}, {Nichol}, {Nidever}, {Nitschelm}, {Ntelis}, {O'Connell},
  {Oelkers}, {Oravetz}, {Oravetz}, {Pace}, {Padilla}, {Palanque-Delabrouille},
  {Alonso Palicio}, {Pan}, {Parejko}, {Parikh}, {P{\^a}ris}, {Park}, {Patten},
  {Peirani}, {Pellejero-Ibanez}, {Penny}, {Percival}, {Perez-Fournon},
  {Petitjean}, {Pieri}, {Pinsonneault}, {Pisani}, {Poleski}, {Prada},
  {Prakash}, {Queiroz}, {Raddick}, {Raichoor}, {Barboza Rembold}, {Richstein},
  {Riffel}, {Riffel}, {Rix}, {Robin}, {Rockosi}, {Rodr{\'\i}guez-Torres},
  {Roman-Lopes}, {Rom{\'a}n-Z{\'u}{\~n}iga}, {Rosado}, {Ross}, {Rossi}, {Ruan},
  {Ruggeri}, {Rykoff}, {Salazar-Albornoz}, {Salvato}, {S{\'a}nchez}, {Aguado},
  {S{\'a}nchez-Gallego}, {Santana}, {Santiago}, {Sayres}, {Schiavon}, {da Silva
  Schimoia}, {Schlafly}, {Schlegel}, {Schneider}, {Schultheis}, {Schuster},
  {Schwope}, {Seo}, {Shao}, {Shen}, {Shetrone}, {Shull}, {Simon}, {Skinner},
  {Skrutskie}, {Slosar}, {Smith}, {Sobeck}, {Sobreira}, {Somers}, {Souto},
  {Stark}, {Stassun}, {Stauffer}, {Steinmetz}, {Storchi-Bergmann},
  {Streblyanska}, {Stringfellow}, {Su{\'a}rez}, {Sun}, {Suzuki}, {Szigeti},
  {Taghizadeh-Popp}, {Tang}, {Tao}, {Tayar}, {Tembe}, {Teske}, {Thakar},
  {Thomas}, {Thompson}, {Tinker}, {Tissera}, {Tojeiro}, {Hernandez Toledo}, {de
  la Torre}, {Tremonti}, {Troup}, {Valenzuela}, {Martinez Valpuesta},
  {Vargas-Gonz{\'a}lez}, {Vargas-Maga{\~n}a}, {Vazquez}, {Villanova}, {Vivek},
  {Vogt}, {Wake}, {Walterbos}, {Wang}, {Weaver}, {Weijmans}, {Weinberg},
  {Westfall}, {Whelan}, {Wild}, {Wilson}, {Wood-Vasey}, {Wylezalek}, {Xiao},
  {Yan}, {Yang}, {Ybarra}, {Y{\`e}che}, {Zakamska}, {Zamora}, {Zarrouk},
  {Zasowski}, {Zhang}, {Zhao}, {Zheng}, {Zheng}, {Zhou}, {Zhou}, {Zhu},
  {Zoccali}, \& {Zou}}]{Blanton2017}
{Blanton}, M.~R., {Bershady}, M.~A., {Abolfathi}, B., {et~al.} 2017, \aj, 154,
  28

\bibitem[{{Bowen} \& {Vaughan}(1973)}]{Bowen1973}
{Bowen}, I.~S., \& {Vaughan}, A.~H., J. 1973, \ao, 12, 1430

\bibitem[{{Burbidge} {et~al.}(1957){Burbidge}, {Burbidge}, {Fowler}, \&
  {Hoyle}}]{Burbidge1957}
{Burbidge}, E.~M., {Burbidge}, G.~R., {Fowler}, W.~A., \& {Hoyle}, F. 1957,
  Reviews of Modern Physics, 29, 547

\bibitem[{{Cantat-Gaudin} {et~al.}(2020){Cantat-Gaudin}, {Anders},
  {Castro-Ginard}, {Jordi}, {Romero-G{\'o}mez}, {Soubiran}, {Casamiquela},
  {Tarricq}, {Moitinho}, {Vallenari}, {Bragaglia}, {Krone-Martins}, \&
  {Kounkel}}]{Cantat-Gaudin2020}
{Cantat-Gaudin}, T., {Anders}, F., {Castro-Ginard}, A., {et~al.} 2020, \aap,
  640, A1

\bibitem[{{Casali} {et~al.}(2020){Casali}, {Spina}, {Magrini}, {Karakas},
  {Kobayashi}, {Casey}, {Feltzing}, {Van der Swaelmen}, {Tsantaki},
  {Jofr{\'e}}, {Bragaglia}, {Feuillet}, {Bensby}, {Biazzo}, {Gonneau},
  {Tautvai{\v{s}}ien{\.{e}}}, {Baratella}, {Roccatagliata}, {Pancino}, {Sousa},
  {Adibekyan}, {Martell}, {Bayo}, {Jackson}, {Jeffries}, {Gilmore}, {Randich},
  {Alfaro}, {Koposov}, {Korn}, {Recio-Blanco}, {Smiljanic}, {Franciosini},
  {Hourihane}, {Monaco}, {Morbidelli}, {Sacco}, {Worley}, \&
  {Zaggia}}]{Casali2020}
{Casali}, G., {Spina}, L., {Magrini}, L., {et~al.} 2020, \aap, 639, A127

\bibitem[{{Casamiquela} {et~al.}(2021){Casamiquela}, {Soubiran}, {Jofr{\'e}},
  {Chiappini}, {Lagarde}, {Tarricq}, {Carrera}, {Jordi},
  {Balaguer-N{\'u}{\~n}ez}, {Carbajo-Hijarrubia}, \&
  {Blanco-Cuaresma}}]{Casamiquela2021}
{Casamiquela}, L., {Soubiran}, C., {Jofr{\'e}}, P., {et~al.} 2021, \aap, 652,
  A25

\bibitem[{{Chen} \& {Zhao}(2020)}]{ChenZhao2020}
{Chen}, Y.~Q., \& {Zhao}, G. 2020, \mnras, 495, 2673

\bibitem[{{Cristallo} {et~al.}(2009){Cristallo}, {Straniero}, {Gallino},
  {Piersanti}, {Dom{\'\i}nguez}, \& {Lederer}}]{Cristallo2009}
{Cristallo}, S., {Straniero}, O., {Gallino}, R., {et~al.} 2009, \apj, 696, 797

\bibitem[{{Cristallo} {et~al.}(2015){Cristallo}, {Straniero}, {Piersanti}, \&
  {Gobrecht}}]{Cristallo2015}
{Cristallo}, S., {Straniero}, O., {Piersanti}, L., \& {Gobrecht}, D. 2015,
  \apjs, 219, 40

\bibitem[{{Cristallo} {et~al.}(2011){Cristallo}, {Piersanti}, {Straniero},
  {Gallino}, {Dom{\'\i}nguez}, {Abia}, {Di Rico}, {Quintini}, \&
  {Bisterzo}}]{Cristallo2011}
{Cristallo}, S., {Piersanti}, L., {Straniero}, O., {et~al.} 2011, \apjs, 197,
  17

\bibitem[{{Cunha} {et~al.}(2017){Cunha}, {Smith}, {Hasselquist}, {Souto},
  {Shetrone}, {Allende Prieto}, {Bizyaev}, {Frinchaboy},
  {Garc{\'\i}a-Hern{\'a}ndez}, {Holtzman}, {Johnson}, {J{\H{o}}nsson},
  {Majewski}, {M{\'e}sz{\'a}ros}, {Nidever}, {Pinsonneault}, {Schiavon},
  {Sobeck}, {Skrutskie}, {Zamora}, {Zasowski}, \&
  {Fern{\'a}ndez-Trincado}}]{Cunha2017}
{Cunha}, K., {Smith}, V.~V., {Hasselquist}, S., {et~al.} 2017, \apj, 844, 145

\bibitem[{{da Silva} {et~al.}(2012){da Silva}, {Porto de Mello}, {Milone}, {da
  Silva}, {Ribeiro}, \& {Rocha-Pinto}}]{DaSilva2012}
{da Silva}, R., {Porto de Mello}, G.~F., {Milone}, A.~C., {et~al.} 2012, \aap,
  542, A84

\bibitem[{{De Silva} {et~al.}(2015){De Silva}, {Freeman}, {Bland-Hawthorn},
  {Martell}, {de Boer}, {Asplund}, {Keller}, {Sharma}, {Zucker}, {Zwitter},
  {Anguiano}, {Bacigalupo}, {Bayliss}, {Beavis}, {Bergemann}, {Campbell},
  {Cannon}, {Carollo}, {Casagrande}, {Casey}, {Da Costa}, {D'Orazi}, {Dotter},
  {Duong}, {Heger}, {Ireland}, {Kafle}, {Kos}, {Lattanzio}, {Lewis}, {Lin},
  {Lind}, {Munari}, {Nataf}, {O'Toole}, {Parker}, {Reid}, {Schlesinger},
  {Sheinis}, {Simpson}, {Stello}, {Ting}, {Traven}, {Watson}, {Wittenmyer},
  {Yong}, \& {{\v{Z}}erjal}}]{DaSilva2015}
{De Silva}, G.~M., {Freeman}, K.~C., {Bland-Hawthorn}, J., {et~al.} 2015,
  \mnras, 449, 2604

\bibitem[{{Delgado Mena} {et~al.}(2019){Delgado Mena}, {Moya}, {Adibekyan},
  {Tsantaki}, {Gonz{\'a}lez Hern{\'a}ndez}, {Israelian}, {Davies}, {Chaplin},
  {Sousa}, {Ferreira}, \& {Santos}}]{Delgado-Mena2019}
{Delgado Mena}, E., {Moya}, A., {Adibekyan}, V., {et~al.} 2019, \aap, 624, A78

\bibitem[{{Djordjevic} {et~al.}(2019){Djordjevic}, {Thompson}, {Urquhart}, \&
  {Forbrich}}]{Djordjevic2019}
{Djordjevic}, J.~O., {Thompson}, M.~A., {Urquhart}, J.~S., \& {Forbrich}, J.
  2019, \mnras, 487, 1057

\bibitem[{{Donor} {et~al.}(2020){Donor}, {Frinchaboy}, {Cunha}, {O'Connell},
  {Allende Prieto}, {Almeida}, {Anders}, {Beaton}, {Bizyaev}, {Brownstein},
  {Carrera}, {Chiappini}, {Cohen}, {Garc{\'\i}a-Hern{\'a}ndez}, {Geisler},
  {Hasselquist}, {J{\"o}nsson}, {Lane}, {Majewski}, {Minniti}, {Bidin}, {Pan},
  {Roman-Lopes}, {Sobeck}, \& {Zasowski}}]{Donor2020}
{Donor}, J., {Frinchaboy}, P.~M., {Cunha}, K., {et~al.} 2020, \aj, 159, 199

\bibitem[{{D'Orazi} {et~al.}(2009){D'Orazi}, {Magrini}, {Randich}, {Galli},
  {Busso}, \& {Sestito}}]{DOrazi2009}
{D'Orazi}, V., {Magrini}, L., {Randich}, S., {et~al.} 2009, \apjl, 693, L31

\bibitem[{{Feltzing} {et~al.}(2017){Feltzing}, {Howes}, {McMillan}, \&
  {Stonkut{\.{e}}}}]{Feltzing2017}
{Feltzing}, S., {Howes}, L.~M., {McMillan}, P.~J., \& {Stonkut{\.{e}}}, E.
  2017, \mnras, 465, L109

\bibitem[{{Fishlock} {et~al.}(2017){Fishlock}, {Yong}, {Karakas},
  {Alves-Brito}, {Mel{\'e}ndez}, {Nissen}, {Kobayashi}, \&
  {Casey}}]{Fishlock2017}
{Fishlock}, C.~K., {Yong}, D., {Karakas}, A.~I., {et~al.} 2017, \mnras, 466,
  4672

\bibitem[{{Foreman-Mackey} {et~al.}(2013){Foreman-Mackey}, {Hogg}, {Lang}, \&
  {Goodman}}]{Foreman-Mackey2013}
{Foreman-Mackey}, D., {Hogg}, D.~W., {Lang}, D., \& {Goodman}, J. 2013, \pasp,
  125, 306

\bibitem[{{Forsberg} {et~al.}(2019){Forsberg}, {J{\"o}nsson}, {Ryde}, \&
  {Matteucci}}]{Forsberg2019}
{Forsberg}, R., {J{\"o}nsson}, H., {Ryde}, N., \& {Matteucci}, F. 2019, \aap,
  631, A113

\bibitem[{{Gallino} {et~al.}(1998){Gallino}, {Arlandini}, {Busso}, {Lugaro},
  {Travaglio}, {Straniero}, {Chieffi}, \& {Limongi}}]{Gallino1998}
{Gallino}, R., {Arlandini}, C., {Busso}, M., {et~al.} 1998, \apj, 497, 388

\bibitem[{{Gallino} {et~al.}(2006){Gallino}, {Bisterzo}, {Straniero}, {Ivans},
  \& {K{\"a}ppeler}}]{Gallino2006}
{Gallino}, R., {Bisterzo}, S., {Straniero}, O., {Ivans}, I.~I., \&
  {K{\"a}ppeler}, F. 2006, \memsai, 77, 786

\bibitem[{{Garc{\'\i}a-Hern{\'a}ndez}
  {et~al.}(2006){Garc{\'\i}a-Hern{\'a}ndez}, {Garc{\'\i}a-Lario}, {Plez},
  {D'Antona}, {Manchado}, \& {Trigo-Rodr{\'\i}guez}}]{Garcia-Hernandez2006}
{Garc{\'\i}a-Hern{\'a}ndez}, D.~A., {Garc{\'\i}a-Lario}, P., {Plez}, B.,
  {et~al.} 2006, Science, 314, 1751

\bibitem[{{Garc{\'\i}a-Hern{\'a}ndez}
  {et~al.}(2013){Garc{\'\i}a-Hern{\'a}ndez}, {Zamora}, {Yag{\"u}e},
  {Uttenthaler}, {Karakas}, {Lugaro}, {Ventura}, \&
  {Lambert}}]{Garcia-Hernandez2013}
{Garc{\'\i}a-Hern{\'a}ndez}, D.~A., {Zamora}, O., {Yag{\"u}e}, A., {et~al.}
  2013, \aap, 555, L3

\bibitem[{{Garc{\'\i}a P{\'e}rez} {et~al.}(2016){Garc{\'\i}a P{\'e}rez},
  {Allende Prieto}, {Holtzman}, {Shetrone}, {M{\'e}sz{\'a}ros}, {Bizyaev},
  {Carrera}, {Cunha}, {Garc{\'\i}a-Hern{\'a}ndez}, {Johnson}, {Majewski},
  {Nidever}, {Schiavon}, {Shane}, {Smith}, {Sobeck}, {Troup}, {Zamora},
  {Weinberg}, {Bovy}, {Eisenstein}, {Feuillet}, {Frinchaboy}, {Hayden},
  {Hearty}, {Nguyen}, {O'Connell}, {Pinsonneault}, {Wilson}, \&
  {Zasowski}}]{GarciaPerez2016}
{Garc{\'\i}a P{\'e}rez}, A.~E., {Allende Prieto}, C., {Holtzman}, J.~A.,
  {et~al.} 2016, \aj, 151, 144

\bibitem[{{Gilmore} {et~al.}(2012){Gilmore}, {Randich}, {Asplund}, {Binney},
  {Bonifacio}, {Drew}, {Feltzing}, {Ferguson}, {Jeffries}, {Micela},
  {Negueruela}, {Prusti}, {Rix}, {Vallenari}, {Alfaro}, {Allende-Prieto},
  {Babusiaux}, {Bensby}, {Blomme}, {Bragaglia}, {Flaccomio}, {Fran{\c{c}}ois},
  {Irwin}, {Koposov}, {Korn}, {Lanzafame}, {Pancino}, {Paunzen},
  {Recio-Blanco}, {Sacco}, {Smiljanic}, {Van Eck}, {Walton}, {Aden}, {Aerts},
  {Affer}, {Alcala}, {Altavilla}, {Alves}, {Antoja}, {Arenou}, {Argiroffi},
  {Asensio Ramos}, {Bailer-Jones}, {Balaguer-Nunez}, {Bayo}, {Barbuy},
  {Barisevicius}, {Barrado y Navascues}, {Battistini}, {Bellas Velidis},
  {Bellazzini}, {Belokurov}, {Bergemann}, {Bertelli}, {Biazzo}, {Bienayme},
  {Bland-Hawthorn}, {Boeche}, {Bonito}, {Boudreault}, {Bouvier}, {Brandao},
  {Brown}, {de Bruijne}, {Burleigh}, {Caballero}, {Caffau}, {Calura},
  {Capuzzo-Dolcetta}, {Caramazza}, {Carraro}, {Casagrande}, {Casewell},
  {Chapman}, {Chiappini}, {Chorniy}, {Christlieb}, {Cignoni}, {Cocozza},
  {Colless}, {Collet}, {Collins}, {Correnti}, {Covino}, {Crnojevic}, {Cropper},
  {Cunha}, {Damiani}, {David}, {Delgado}, {Duffau}, {Edvardsson}, {Eldridge},
  {Enke}, {Eriksson}, {Evans}, {Eyer}, {Famaey}, {Fellhauer}, {Ferreras},
  {Figueras}, {Fiorentino}, {Flynn}, {Folha}, {Franciosini}, {Frasca},
  {Freeman}, {Fremat}, {Friel}, {Gaensicke}, {Gameiro}, {Garzon}, {Geier},
  {Geisler}, {Gerhard}, {Gibson}, {Gomboc}, {Gomez}, {Gonzalez-Fernandez},
  {Gonzalez Hernandez}, {Gosset}, {Grebel}, {Greimel}, {Groenewegen},
  {Grundahl}, {Guarcello}, {Gustafsson}, {Hadrava}, {Hatzidimitriou}, {Hambly},
  {Hammersley}, {Hansen}, {Haywood}, {Heber}, {Heiter}, {Held}, {Helmi},
  {Hensler}, {Herrero}, {Hill}, {Hodgkin}, {Huelamo}, {Huxor}, {Ibata},
  {Jackson}, {de Jong}, {Jonker}, {Jordan}, {Jordi}, {Jorissen}, {Katz},
  {Kawata}, {Keller}, {Kharchenko}, {Klement}, {Klutsch}, {Knude}, {Koch},
  {Kochukhov}, {Kontizas}, {Koubsky}, {Lallement}, {de Laverny}, {van Leeuwen},
  {Lemasle}, {Lewis}, {Lind}, {Lindstrom}, {Lobel}, {Lopez Santiago}, {Lucas},
  {Ludwig}, {Lueftinger}, {Magrini}, {Maiz Apellaniz}, {Maldonado}, {Marconi},
  {Marino}, {Martayan}, {Martinez-Valpuesta}, {Matijevic}, {McMahon},
  {Messina}, {Meyer}, {Miglio}, {Mikolaitis}, {Minchev}, {Minniti}, {Moitinho},
  {Momany}, {Monaco}, {Montalto}, {Monteiro}, {Monier}, {Montes}, {Mora},
  {Moraux}, {Morel}, {Mowlavi}, {Mucciarelli}, {Munari}, {Napiwotzki},
  {Nardetto}, {Naylor}, {Naze}, {Nelemans}, {Okamoto}, {Ortolani}, {Pace},
  {Palla}, {Palous}, {Parker}, {Penarrubia}, {Pillitteri}, {Piotto}, {Posbic},
  {Prisinzano}, {Puzeras}, {Quirrenbach}, {Ragaini}, {Read}, {Read}, {Reyle},
  {De Ridder}, {Robichon}, {Robin}, {Roeser}, {Romano}, {Royer}, {Ruchti},
  {Ruzicka}, {Ryan}, {Ryde}, {Santos}, {Sanz Forcada}, {Sarro Baro},
  {Sbordone}, {Schilbach}, {Schmeja}, {Schnurr}, {Schoenrich}, {Scholz},
  {Seabroke}, {Sharma}, {De Silva}, {Smith}, {Solano}, {Sordo}, {Soubiran},
  {Sousa}, {Spagna}, {Steffen}, {Steinmetz}, {Stelzer}, {Stempels},
  {Tabernero}, {Tautvaisiene}, {Thevenin}, {Torra}, {Tosi}, {Tolstoy}, {Turon},
  {Walker}, {Wambsganss}, {Worley}, {Venn}, {Vink}, {Wyse}, {Zaggia},
  {Zeilinger}, {Zoccali}, {Zorec}, {Zucker}, {Zwitter}, \& {Gaia-ESO Survey
  Team}}]{Gilmore2012}
{Gilmore}, G., {Randich}, S., {Asplund}, M., {et~al.} 2012, The Messenger, 147,
  25

\bibitem[{{Grevesse} {et~al.}(2007){Grevesse}, {Asplund}, \&
  {Sauval}}]{Grevesse2007}
{Grevesse}, N., {Asplund}, M., \& {Sauval}, A.~J. 2007, \ssr, 130, 105

\bibitem[{{Gunn} {et~al.}(2006){Gunn}, {Siegmund}, {Mannery}, {Owen}, {Hull},
  {Leger}, {Carey}, {Knapp}, {York}, {Boroski}, {Kent}, {Lupton}, {Rockosi},
  {Evans}, {Waddell}, {Anderson}, {Annis}, {Barentine}, {Bartoszek}, {Bastian},
  {Bracker}, {Brewington}, {Briegel}, {Brinkmann}, {Brown}, {Carr},
  {Czarapata}, {Drennan}, {Dombeck}, {Federwitz}, {Gillespie}, {Gonzales},
  {Hansen}, {Harvanek}, {Hayes}, {Jordan}, {Kinney}, {Klaene}, {Kleinman},
  {Kron}, {Kresinski}, {Lee}, {Limmongkol}, {Lindenmeyer}, {Long}, {Loomis},
  {McGehee}, {Mantsch}, {Neilsen}, {Neswold}, {Newman}, {Nitta}, {Peoples},
  {Pier}, {Prieto}, {Prosapio}, {Rivetta}, {Schneider}, {Snedden}, \&
  {Wang}}]{Gunn2006}
{Gunn}, J.~E., {Siegmund}, W.~A., {Mannery}, E.~J., {et~al.} 2006, \aj, 131,
  2332

\bibitem[{{Gustafsson} {et~al.}(2008){Gustafsson}, {Edvardsson}, {Eriksson},
  {J{\o}rgensen}, {Nordlund}, \& {Plez}}]{Gustafsson2008}
{Gustafsson}, B., {Edvardsson}, B., {Eriksson}, K., {et~al.} 2008, \aap, 486,
  951

\bibitem[{Harris {et~al.}(2020)Harris, Millman, van~der Walt, Gommers,
  Virtanen, Cournapeau, Wieser, Taylor, Berg, Smith, Kern, Picus, Hoyer, van
  Kerkwijk, Brett, Haldane, Fernández~del Río, Wiebe, Peterson,
  Gérard-Marchant, Sheppard, Reddy, Weckesser, Abbasi, Gohlke, \&
  Oliphant}]{Harris2020}
Harris, C.~R., Millman, K.~J., van~der Walt, S.~J., {et~al.} 2020, Nature, 585,
  357–362

\bibitem[{{Holtzman} {et~al.}(2015){Holtzman}, {Shetrone}, {Johnson}, {Allende
  Prieto}, {Anders}, {Andrews}, {Beers}, {Bizyaev}, {Blanton}, {Bovy},
  {Carrera}, {Chojnowski}, {Cunha}, {Eisenstein}, {Feuillet}, {Frinchaboy},
  {Galbraith-Frew}, {Garc{\'\i}a P{\'e}rez}, {Garc{\'\i}a-Hern{\'a}ndez},
  {Hasselquist}, {Hayden}, {Hearty}, {Ivans}, {Majewski}, {Martell},
  {Meszaros}, {Muna}, {Nidever}, {Nguyen}, {O'Connell}, {Pan}, {Pinsonneault},
  {Robin}, {Schiavon}, {Shane}, {Sobeck}, {Smith}, {Troup}, {Weinberg},
  {Wilson}, {Wood-Vasey}, {Zamora}, \& {Zasowski}}]{Holtzman2015}
{Holtzman}, J.~A., {Shetrone}, M., {Johnson}, J.~A., {et~al.} 2015, \aj, 150,
  148

\bibitem[{{Hunter}(2007)}]{Hunter2007}
{Hunter}, J.~D. 2007, Computing in Science and Engineering, 9, 90

\bibitem[{{Jacobson} \& {Friel}(2013)}]{JacobsonFriel2013}
{Jacobson}, H.~R., \& {Friel}, E.~D. 2013, \aj, 145, 107

\bibitem[{{J{\'\i}lkov{\'a}} {et~al.}(2012){J{\'\i}lkov{\'a}}, {Carraro},
  {Jungwiert}, \& {Minchev}}]{Jilkova2012}
{J{\'\i}lkov{\'a}}, L., {Carraro}, G., {Jungwiert}, B., \& {Minchev}, I. 2012,
  \aap, 541, A64

\bibitem[{{Jofr{\'e}} {et~al.}(2020){Jofr{\'e}}, {Jackson}, \& {Tucci
  Maia}}]{Jofre2020}
{Jofr{\'e}}, P., {Jackson}, H., \& {Tucci Maia}, M. 2020, \aap, 633, L9

\bibitem[{{J{\"o}nsson} {et~al.}(2020){J{\"o}nsson}, {Holtzman}, {Allende
  Prieto}, {Cunha}, {Garc{\'\i}a-Hern{\'a}ndez}, {Hasselquist}, {Masseron},
  {Osorio}, {Shetrone}, {Smith}, {Stringfellow}, {Bizyaev}, {Edvardsson},
  {Majewski}, {M{\'e}sz{\'a}ros}, {Souto}, {Zamora}, {Beaton}, {Bovy}, {Donor},
  {Pinsonneault}, {Poovelil}, \& {Sobeck}}]{Jonsson2020}
{J{\"o}nsson}, H., {Holtzman}, J.~A., {Allende Prieto}, C., {et~al.} 2020, \aj,
  160, 120

\bibitem[{{K{\"a}ppeler} {et~al.}(2011){K{\"a}ppeler}, {Gallino}, {Bisterzo},
  \& {Aoki}}]{Kappeler2011}
{K{\"a}ppeler}, F., {Gallino}, R., {Bisterzo}, S., \& {Aoki}, W. 2011, Reviews
  of Modern Physics, 83, 157

\bibitem[{{Karakas} \& {Lattanzio}(2014)}]{Karakas2014}
{Karakas}, A.~I., \& {Lattanzio}, J.~C. 2014, \pasa, 31, e030

\bibitem[{{Karakas} \& {Lugaro}(2016)}]{KarakasLugaro2016}
{Karakas}, A.~I., \& {Lugaro}, M. 2016, \apj, 825, 26

\bibitem[{{Kharchenko} {et~al.}(2013){Kharchenko}, {Piskunov}, {Schilbach},
  {R{\"o}ser}, \& {Scholz}}]{Kharchenko2013}
{Kharchenko}, N.~V., {Piskunov}, A.~E., {Schilbach}, E., {R{\"o}ser}, S., \&
  {Scholz}, R.~D. 2013, \aap, 558, A53

\bibitem[{{Lugaro} {et~al.}(2003){Lugaro}, {Herwig}, {Lattanzio}, {Gallino}, \&
  {Straniero}}]{Lugaro2003}
{Lugaro}, M., {Herwig}, F., {Lattanzio}, J.~C., {Gallino}, R., \& {Straniero},
  O. 2003, \apj, 586, 1305

\bibitem[{{Magrini} {et~al.}(2018){Magrini}, {Spina}, {Randich}, {Friel},
  {Kordopatis}, {Worley}, {Pancino}, {Bragaglia}, {Donati},
  {Tautvai{\v{s}}ien{\.{e}}}, {Bagdonas}, {Delgado-Mena}, {Adibekyan}, {Sousa},
  {Jim{\'e}nez-Esteban}, {Sanna}, {Roccatagliata}, {Bonito}, {Sbordone},
  {Duffau}, {Gilmore}, {Feltzing}, {Jeffries}, {Vallenari}, {Alfaro}, {Bensby},
  {Francois}, {Koposov}, {Korn}, {Recio-Blanco}, {Smiljanic}, {Bayo},
  {Carraro}, {Casey}, {Costado}, {Damiani}, {Franciosini}, {Frasca},
  {Hourihane}, {Jofr{\'e}}, {de Laverny}, {Lewis}, {Masseron}, {Monaco},
  {Morbidelli}, {Prisinzano}, {Sacco}, \& {Zaggia}}]{Magrini2018}
{Magrini}, L., {Spina}, L., {Randich}, S., {et~al.} 2018, \aap, 617, A106

\bibitem[{{Magrini} {et~al.}(2021){Magrini}, {Vescovi}, {Casali}, {Cristallo},
  {Viscasillas V{\'a}zquez}, {Cescutti}, {Spina}, {Van Der Swaelmen}, \&
  {Randich}}]{Magrini2021}
{Magrini}, L., {Vescovi}, D., {Casali}, G., {et~al.} 2021, \aap, 646, L2

\bibitem[{{Maiorca} {et~al.}(2012){Maiorca}, {Magrini}, {Busso}, {Randich},
  {Palmerini}, \& {Trippella}}]{Maiorca2012}
{Maiorca}, E., {Magrini}, L., {Busso}, M., {et~al.} 2012, \apj, 747, 53

\bibitem[{{Maiorca} {et~al.}(2011){Maiorca}, {Randich}, {Busso}, {Magrini}, \&
  {Palmerini}}]{Maiorca2011}
{Maiorca}, E., {Randich}, S., {Busso}, M., {Magrini}, L., \& {Palmerini}, S.
  2011, \apj, 736, 120

\bibitem[{{Majewski} {et~al.}(2017){Majewski}, {Schiavon}, {Frinchaboy},
  {Allende Prieto}, {Barkhouser}, {Bizyaev}, {Blank}, {Brunner}, {Burton},
  {Carrera}, {Chojnowski}, {Cunha}, {Epstein}, {Fitzgerald}, {Garc{\'\i}a
  P{\'e}rez}, {Hearty}, {Henderson}, {Holtzman}, {Johnson}, {Lam}, {Lawler},
  {Maseman}, {M{\'e}sz{\'a}ros}, {Nelson}, {Nguyen}, {Nidever}, {Pinsonneault},
  {Shetrone}, {Smee}, {Smith}, {Stolberg}, {Skrutskie}, {Walker}, {Wilson},
  {Zasowski}, {Anders}, {Basu}, {Beland}, {Blanton}, {Bovy}, {Brownstein},
  {Carlberg}, {Chaplin}, {Chiappini}, {Eisenstein}, {Elsworth}, {Feuillet},
  {Fleming}, {Galbraith-Frew}, {Garc{\'\i}a}, {Garc{\'\i}a-Hern{\'a}ndez},
  {Gillespie}, {Girardi}, {Gunn}, {Hasselquist}, {Hayden}, {Hekker}, {Ivans},
  {Kinemuchi}, {Klaene}, {Mahadevan}, {Mathur}, {Mosser}, {Muna}, {Munn},
  {Nichol}, {O'Connell}, {Parejko}, {Robin}, {Rocha-Pinto}, {Schultheis},
  {Serenelli}, {Shane}, {Silva Aguirre}, {Sobeck}, {Thompson}, {Troup},
  {Weinberg}, \& {Zamora}}]{Majewski2017}
{Majewski}, S.~R., {Schiavon}, R.~P., {Frinchaboy}, P.~M., {et~al.} 2017, \aj,
  154, 94

\bibitem[{{Martinez-Medina} {et~al.}(2018){Martinez-Medina}, {Gieles},
  {Pichardo}, \& {Peimbert}}]{Martinez-Medina2018}
{Martinez-Medina}, L.~A., {Gieles}, M., {Pichardo}, B., \& {Peimbert}, A. 2018,
  \mnras, 474, 32

\bibitem[{{Masseron} {et~al.}(2016){Masseron}, {Merle}, \&
  {Hawkins}}]{Masseron2016}
{Masseron}, T., {Merle}, T., \& {Hawkins}, K. 2016, {BACCHUS: Brussels
  Automatic Code for Characterizing High accUracy Spectra}, , , ascl:1605.004

\bibitem[{{Miglio} {et~al.}(2021){Miglio}, {Chiappini}, {Mackereth}, {Davies},
  {Brogaard}, {Casagrande}, {Chaplin}, {Girardi}, {Kawata}, {Khan}, {Izzard},
  {Montalb{\'a}n}, {Mosser}, {Vincenzo}, {Bossini}, {Noels}, {Rodrigues},
  {Valentini}, \& {Mandel}}]{Miglio2021}
{Miglio}, A., {Chiappini}, C., {Mackereth}, J.~T., {et~al.} 2021, \aap, 645,
  A85

\bibitem[{{Mishenina} {et~al.}(2015){Mishenina}, {Pignatari}, {Carraro},
  {Kovtyukh}, {Monaco}, {Korotin}, {Shereta}, {Yegorova}, \&
  {Herwig}}]{Mishenina2015}
{Mishenina}, T., {Pignatari}, M., {Carraro}, G., {et~al.} 2015, \mnras, 446,
  3651

\bibitem[{{Mishenina} {et~al.}(2013){Mishenina}, {Pignatari}, {Korotin},
  {Soubiran}, {Charbonnel}, {Thielemann}, {Gorbaneva}, \&
  {Basak}}]{Mishenina2013}
{Mishenina}, T.~V., {Pignatari}, M., {Korotin}, S.~A., {et~al.} 2013, \aap,
  552, A128

\bibitem[{{Misiriotis} {et~al.}(2006){Misiriotis}, {Xilouris},
  {Papamastorakis}, {Boumis}, \& {Goudis}}]{Misiriotis2006}
{Misiriotis}, A., {Xilouris}, E.~M., {Papamastorakis}, J., {Boumis}, P., \&
  {Goudis}, C.~D. 2006, \aap, 459, 113

\bibitem[{{Nidever} {et~al.}(2015){Nidever}, {Holtzman}, {Allende Prieto},
  {Beland}, {Bender}, {Bizyaev}, {Burton}, {Desphande}, {Fleming}, {Garc{\'\i}a
  P{\'e}rez}, {Hearty}, {Majewski}, {M{\'e}sz{\'a}ros}, {Muna}, {Nguyen},
  {Schiavon}, {Shetrone}, {Skrutskie}, {Sobeck}, \& {Wilson}}]{Nidever2015}
{Nidever}, D.~L., {Holtzman}, J.~A., {Allende Prieto}, C., {et~al.} 2015, \aj,
  150, 173

\bibitem[{{Nissen}(2015)}]{Nissen2015}
{Nissen}, P.~E. 2015, \aap, 579, A52

\bibitem[{{Pe{\~n}a Su{\'a}rez} {et~al.}(2018){Pe{\~n}a Su{\'a}rez}, {Sales
  Silva}, {Katime Santrich}, {Drake}, \& {Pereira}}]{PenaSuarez2018}
{Pe{\~n}a Su{\'a}rez}, V.~J., {Sales Silva}, J.~V., {Katime Santrich}, O.~J.,
  {Drake}, N.~A., \& {Pereira}, C.~B. 2018, \apj, 854, 184

\bibitem[{{Piersanti} {et~al.}(2013){Piersanti}, {Cristallo}, \&
  {Straniero}}]{Piersanti2013}
{Piersanti}, L., {Cristallo}, S., \& {Straniero}, O. 2013, \apj, 774, 98

\bibitem[{{Pignatari} {et~al.}(2010){Pignatari}, {Gallino}, {Heil}, {Wiescher},
  {K{\"a}ppeler}, {Herwig}, \& {Bisterzo}}]{Pignatari2010}
{Pignatari}, M., {Gallino}, R., {Heil}, M., {et~al.} 2010, \apj, 710, 1557

\bibitem[{{Plez}(2012)}]{Plez2012}
{Plez}, B. 2012, {Turbospectrum: Code for spectral synthesis}, , ,
  ascl:1205.004

\bibitem[{{Prantzos} {et~al.}(2018){Prantzos}, {Abia}, {Limongi}, {Chieffi}, \&
  {Cristallo}}]{Prantzos2018}
{Prantzos}, N., {Abia}, C., {Limongi}, M., {Chieffi}, A., \& {Cristallo}, S.
  2018, \mnras, 476, 3432

\bibitem[{{Reddy} {et~al.}(2012){Reddy}, {Giridhar}, \& {Lambert}}]{Reddy2012}
{Reddy}, A. B.~S., {Giridhar}, S., \& {Lambert}, D.~L. 2012, \mnras, 419, 1350

\bibitem[{{Reddy} {et~al.}(2013){Reddy}, {Giridhar}, \& {Lambert}}]{Reddy2013}
---. 2013, \mnras, 431, 3338

\bibitem[{{Reddy} {et~al.}(2006){Reddy}, {Lambert}, \& {Allende
  Prieto}}]{Reddy2006}
{Reddy}, B.~E., {Lambert}, D.~L., \& {Allende Prieto}, C. 2006, \mnras, 367,
  1329

\bibitem[{{Reddy} {et~al.}(2003){Reddy}, {Tomkin}, {Lambert}, \& {Allende
  Prieto}}]{Reddy2003}
{Reddy}, B.~E., {Tomkin}, J., {Lambert}, D.~L., \& {Allende Prieto}, C. 2003,
  \mnras, 340, 304

\bibitem[{{Santrich} {et~al.}(2013){Santrich}, {Pereira}, \&
  {Drake}}]{Santrich2013}
{Santrich}, O.~J.~K., {Pereira}, C.~B., \& {Drake}, N.~A. 2013, \aap, 554, A2

\bibitem[{{Smith} {et~al.}(2021){Smith}, {Bizyaev}, {Cunha}, {Shetrone},
  {Souto}, {Allende Prieto}, {Masseron}, {M{\'e}sz{\'a}ros}, {J{\"o}nsson},
  {Hasselquist}, {Osorio}, {Garc{\'\i}a-Hern{\'a}ndez}, {Plez}, {Beaton},
  {Holtzman}, {Majewski}, {Stringfellow}, \& {Sobeck}}]{Smith2021}
{Smith}, V.~V., {Bizyaev}, D., {Cunha}, K., {et~al.} 2021, \aj, 161, 254

\bibitem[{{Souto} {et~al.}(2018){Souto}, {Cunha}, {Smith}, {Allende Prieto},
  {Garc{\'\i}a-Hern{\'a}ndez}, {Pinsonneault}, {Holzer}, {Frinchaboy},
  {Holtzman}, {Johnson}, {J{\"o}nsson}, {Majewski}, {Shetrone}, {Sobeck},
  {Stringfellow}, {Teske}, {Zamora}, {Zasowski}, {Carrera}, {Stassun},
  {Fernandez-Trincado}, {Villanova}, {Minniti}, \& {Santana}}]{Souto2018}
{Souto}, D., {Cunha}, K., {Smith}, V.~V., {et~al.} 2018, \apj, 857, 14

\bibitem[{{Souto} {et~al.}(2019){Souto}, {Allende Prieto}, {Cunha},
  {Pinsonneault}, {Smith}, {Garcia-Dias}, {Bovy}, {Garc{\'\i}a-Hern{\'a}ndez},
  {Holtzman}, {Johnson}, {J{\"o}nsson}, {Majewski}, {Shetrone}, {Sobeck},
  {Zamora}, {Pan}, \& {Nitschelm}}]{Souto2019}
{Souto}, D., {Allende Prieto}, C., {Cunha}, K., {et~al.} 2019, \apj, 874, 97

\bibitem[{{Spina} {et~al.}(2018){Spina}, {Mel{\'e}ndez}, {Karakas}, {dos
  Santos}, {Bedell}, {Asplund}, {Ram{\'\i}rez}, {Yong}, {Alves-Brito}, {Bean},
  \& {Dreizler}}]{Spina2018}
{Spina}, L., {Mel{\'e}ndez}, J., {Karakas}, A.~I., {et~al.} 2018, \mnras, 474,
  2580

\bibitem[{{Spina} {et~al.}(2020){Spina}, {Nordlander}, {Casey}, {Bedell},
  {D'Orazi}, {Mel{\'e}ndez}, {Karakas}, {Desidera}, {Baratella}, {Yana
  Galarza}, \& {Casali}}]{Spina2020}
{Spina}, L., {Nordlander}, T., {Casey}, A.~R., {et~al.} 2020, \apj, 895, 52

\bibitem[{{Spina} {et~al.}(2021){Spina}, {Ting}, {De Silva}, {Frankel},
  {Sharma}, {Cantat-Gaudin}, {Joyce}, {Stello}, {Karakas}, {Asplund},
  {Nordlander}, {Casagrande}, {D'Orazi}, {Casey}, {Cottrell},
  {Tepper-Garc{\'\i}a}, {Baratella}, {Kos}, {{\v{C}}otar}, {Bland-Hawthorn},
  {Buder}, {Freeman}, {Hayden}, {Lewis}, {Lin}, {Lind}, {Martell},
  {Schlesinger}, {Simpson}, {Zucker}, \& {Zwitter}}]{Spina2021}
{Spina}, L., {Ting}, Y.~S., {De Silva}, G.~M., {et~al.} 2021, \mnras, 503, 3279

\bibitem[{{Tautvai{\v{s}}ien{\.{e}}} {et~al.}(2021){Tautvai{\v{s}}ien{\.{e}}},
  {Viscasillas V{\'a}zquez}, {Mikolaitis}, {Stonkut{\.{e}}},
  {Minkevi{\v{c}}i{\={u}}t{\.{e}}}, {Drazdauskas}, \&
  {Bagdonas}}]{Tautvaisiene2021}
{Tautvai{\v{s}}ien{\.{e}}}, G., {Viscasillas V{\'a}zquez}, C., {Mikolaitis},
  {\v{S}}., {et~al.} 2021, \aap, 649, A126

\bibitem[{{Thielemann} {et~al.}(2017){Thielemann}, {Eichler}, {Panov}, \&
  {Wehmeyer}}]{Thielemann2017}
{Thielemann}, F.~K., {Eichler}, M., {Panov}, I.~V., \& {Wehmeyer}, B. 2017,
  Annual Review of Nuclear and Particle Science, 67, 253

\bibitem[{{van Raai} {et~al.}(2012){van Raai}, {Lugaro}, {Karakas},
  {Garc{\'\i}a-Hern{\'a}ndez}, \& {Yong}}]{van-Raai2012}
{van Raai}, M.~A., {Lugaro}, M., {Karakas}, A.~I., {Garc{\'\i}a-Hern{\'a}ndez},
  D.~A., \& {Yong}, D. 2012, \aap, 540, A44

\bibitem[{{Vescovi} {et~al.}(2020){Vescovi}, {Cristallo}, {Busso}, \&
  {Liu}}]{Vescovi2020}
{Vescovi}, D., {Cristallo}, S., {Busso}, M., \& {Liu}, N. 2020, \apjl, 897, L25

\bibitem[{{Villanova} {et~al.}(2018){Villanova}, {Carraro}, {Geisler},
  {Monaco}, \& {Assmann}}]{Villanova2018}
{Villanova}, S., {Carraro}, G., {Geisler}, D., {Monaco}, L., \& {Assmann}, P.
  2018, \apj, 867, 34

\bibitem[{Virtanen {et~al.}(2020)Virtanen, Gommers, Oliphant, Haberland, Reddy,
  Cournapeau, Burovski, Peterson, Weckesser, Bright, {van der Walt}, Brett,
  Wilson, Millman, Mayorov, Nelson, Jones, Kern, Larson, Carey, Polat, Feng,
  Moore, {VanderPlas}, Laxalde, Perktold, Cimrman, Henriksen, Quintero, Harris,
  Archibald, Ribeiro, Pedregosa, {van Mulbregt}, \& {SciPy 1.0
  Contributors}}]{Virtanen2020}
Virtanen, P., Gommers, R., Oliphant, T.~E., {et~al.} 2020, Nature Methods, 17,
  261

\bibitem[{{Wilson} {et~al.}(2019){Wilson}, {Hearty}, {Skrutskie}, {Majewski},
  {Holtzman}, {Eisenstein}, {Gunn}, {Blank}, {Henderson}, {Smee}, {Nelson},
  {Nidever}, {Arns}, {Barkhouser}, {Barr}, {Beland}, {Bershady}, {Blanton},
  {Brunner}, {Burton}, {Carey}, {Carr}, {Colque}, {Crane}, {Damke}, {Davidson},
  {Dean}, {Di Mille}, {Don}, {Ebelke}, {Evans}, {Fitzgerald}, {Gillespie},
  {Hall}, {Harding}, {Harding}, {Hammond}, {Hancock}, {Harrison}, {Hope},
  {Horne}, {Karakla}, {Lam}, {Leger}, {MacDonald}, {Maseman}, {Matsunari},
  {Melton}, {Mitcheltree}, {O'Brien}, {O'Connell}, {Patten}, {Richardson},
  {Rieke}, {Rieke}, {Roman-Lopes}, {Schiavon}, {Sobeck}, {Stolberg}, {Stoll},
  {Tembe}, {Trujillo}, {Uomoto}, {Vernieri}, {Walker}, {Weinberg}, {Young},
  {Anthony-Brumfield}, {Bizyaev}, {Breslauer}, {De Lee}, {Downey}, {Halverson},
  {Huehnerhoff}, {Klaene}, {Leon}, {Long}, {Mahadevan}, {Malanushenko},
  {Nguyen}, {Owen}, {S{\'a}nchez-Gallego}, {Sayres}, {Shane}, {Shectman},
  {Shetrone}, {Skinner}, {Stauffer}, \& {Zhao}}]{Wilson2019}
{Wilson}, J.~C., {Hearty}, F.~R., {Skrutskie}, M.~F., {et~al.} 2019, \pasp,
  131, 055001

\bibitem[{{Yong} {et~al.}(2012){Yong}, {Carney}, \& {Friel}}]{Yong2012}
{Yong}, D., {Carney}, B.~W., \& {Friel}, E.~D. 2012, \aj, 144, 95

\bibitem[{{Zasowski} {et~al.}(2017){Zasowski}, {Cohen}, {Chojnowski},
  {Santana}, {Oelkers}, {Andrews}, {Beaton}, {Bender}, {Bird}, {Bovy},
  {Carlberg}, {Covey}, {Cunha}, {Dell'Agli}, {Fleming}, {Frinchaboy},
  {Garc{\'\i}a-Hern{\'a}ndez}, {Harding}, {Holtzman}, {Johnson}, {Kollmeier},
  {Majewski}, {M{\'e}sz{\'a}ros}, {Munn}, {Mu{\~n}oz}, {Ness}, {Nidever},
  {Poleski}, {Rom{\'a}n-Z{\'u}{\~n}iga}, {Shetrone}, {Simon}, {Smith},
  {Sobeck}, {Stringfellow}, {Szigeti{\'a}ros}, {Tayar}, \&
  {Troup}}]{Zasowski2017}
{Zasowski}, G., {Cohen}, R.~E., {Chojnowski}, S.~D., {et~al.} 2017, \aj, 154,
  198

\end{thebibliography}




\end{document}